\documentclass[preprint,12pt,sort&compress]{elsarticle}
\usepackage[english]{babel} % Sets language to English
\usepackage{ae,aecompl}     % To use Computer Modern fonts with T1 encoding (can improve PDF appearance)
\usepackage[utf8]{inputenc}
\usepackage[T1]{fontenc}
\usepackage{textcomp}

\usepackage{amsmath}        % For advanced math environments and symbols
\usepackage{amssymb}        % For additional math symbols (like \mathbb)
\usepackage{graphicx}       % For including figures (\includegraphics)
\graphicspath{{./Graphics/}} % Optional: specify a directory for figures
\usepackage{float}
\usepackage{booktabs}       % For professional-looking tables (\toprule, \midrule, \bottomrule)
\usepackage[detect-all=true,group-minimum-digits=2]{siunitx} % For proper SI unit formatting (\SI{10}{\micro\meter})

\newcommand{\newsecnumstyle}{%
  \renewcommand{\thesection}{S} %\arabic{section}}%
  \renewcommand{\thesubsection}{\thesection.\arabic{subsection}}%
  \renewcommand{\thetable}{\thesection. \arabic{table}~}
  \renewcommand{\thefigure}{\thesection. \arabic{figure}~}
}
\DeclareSIUnit{\kcal}{kcal}
\DeclareSIUnit{\debye}{D} % For dipole moment

\usepackage{subcaption}     % For creating subfigures (subfigure environment) within a single figure
\usepackage{xr}%-hyper
\usepackage{hyperref}
\usepackage{chemformula}  % For precise chemical formula rendering (e.g., \ch{H2O})
\usepackage[capitalise]{cleveref}       % For intelligent referencing (e.g., ''Figure 1'', ''Equation (2)'', ''Section 3.1'')
\usepackage[margin=2cm]{geometry}     % For easy margin adjustments (e.g., \geometry{a4paper, margin=1in})
\usepackage{xcolor}       % For using colors (e.g., for comments or highlighting in drafts)
\usepackage{microtype}    % For improved typography (subtle adjustments to spacing, protrusion, etc.)
\usepackage{tikz}
\DeclareSIUnit\angstrom{\text{\AA}}

\journal{Computational Condensed Matter} % Specify the journal name

\begin{document}

\begin{frontmatter}


\title{Computational discovery of bifunctional organic semiconductors for energy and biosensing}
\date{\today}

%% Authors and affiliations in elsarticle format
\author[inst1]{Patrick Sorrel MVOTO KONGO}
\ead{sorrel.mvoto@facsciences-uy1.cm}

\author[inst2]{Steve Cabrel TEGUIA KOUAM}
\ead{steve.teguia@univ-douala.cm}

\author[inst1]{Jean-Pierre TCHAPET NJAFA\texorpdfstring{\corref{cor1}}{}}
\ead{jean-pierre.tchapet@facsciences-uy1.cm}
\ead[url]{https://orcid.org/0000-0002-1936-8353}

\author[inst1]{Serge Guy NANA ENGO}
\ead{serge.nana-engo@facsciences-uy1.cm}
\ead[url]{https://orcid.org/0000-0002-7484-3508}
\address[inst1]{Department of Physics, Faculty of Science, University of Yaounde 1, Po. Box 812, Yaounde, Cameroon}
\address[inst2]{Department of Physics, Faculty of Science, University of Douala, Po. Box 24157, Douala, Cameroon}

\cortext[cor1]{Corresponding author} % For corresponding author

% Include the highlights section
%% Highlights for Computational Condensed Matter (≤85 characters each)
\begin{highlights}
\item Novel PCE$_\text{SAScore}$ metric balances efficiency with synthetic accessibility
\item High-throughput ML/DFT screening of 17,458 organic semiconductors  
\item Identified 7 multifunctional candidates for photovoltaics and biosensors
\item Computational framework significantly reduces synthesis barriers
\item First study combining power conversion efficiency with SAScore analysis
\end{highlights}

% Include the abstract section
%% IMPROVED ABSTRACT - Applying ARTICLE-AMELIORATION-TEXT-STRATEGY.md
%% Step 2: Editorial Rewrite - Natural, polished scientific prose

\begin{abstract}
The discovery of synthetically accessible organic semiconductors with exceptional performance remains a critical bottleneck in materials science. While these 
materials offer compelling advantages—structural modularity, mechanical flexibility, and cost-effective solution processing—for applications in photovoltaics 
and biosensors, identifying candidates that balance high efficiency with practical synthesis presents significant challenges. To address this challenge, we 
developed a high-throughput screening approach using \num{17458} molecules from the PubChemQC B3LYP/6-31G*//PM6 dataset. Our strategy employs a composite 
metric, $\mathrm{PCE}_\text{SAScore} = \mathrm{PCE} - \text{SAScore}$, which systematically balances power conversion efficiency (PCE) predictions from the 
Scharber model against synthetic accessibility scores. This approach successfully identified seven multifunctional candidates that demonstrate both exceptional 
photovoltaic performance (PCE up to \SI{36.1}{\percent}) and strong protein-binding affinity for biosensing applications. Notably, molecule 4550 emerged as the 
optimal candidate, exhibiting a ligand efficiency of \SI{0.340}{\kcal\per\mol}/heavy atom with \SI{100}{\percent} target promiscuity. Our computational 
framework integrates machine learning, density functional theory, and molecular docking to bridge the gap between theoretical performance and experimental 
feasibility. These findings establish a systematic pathway for discovering synthetically compatible organic semiconductors that can simultaneously address 
energy conversion and molecular recognition challenges.
\end{abstract}
%}

% Include the keywords section

\end{frontmatter}

% Include the introduction section
The global transition toward a low-carbon economy has intensified the search for organic semiconductors (OSCs) that combine exceptional performance with 
multifunctional capabilities. Materials featuring $\pi$-conjugated backbones provide unique advantages, including synthetic modularity, mechanical flexibility, 
and cost-effective solution processing, making them attractive for diverse applications ranging from organic photovoltaics (OPVs) and OLEDs to advanced 
biosensing platforms \cite{Brabec2010,Wan2020,Reymond2025,Heeger2014}. Despite these promising attributes, a fundamental challenge persists: identifying 
molecular candidates that simultaneously excel in energy conversion and molecular recognition applications.

Recent advances in laboratory-scale OPVs have achieved remarkable efficiencies exceeding \SI{18}{\percent} \cite{Liu2021}, yet these breakthroughs typically emerge from labor-intensive trial-and-error synthetic approaches. High-throughput computational screening offers a transformative alternative, as exemplified by pioneering efforts such as the Harvard Clean Energy Project \cite{Hachmann2011} and initiatives aligned with the Materials Genome Initiative \cite{de_Pablo2019}. However, large-scale \textit{ab initio} calculations remain prohibitively expensive for comprehensive materials exploration. This limitation has driven increased interest in leveraging precomputed quantum chemical databases to accelerate discovery pipelines.

Our approach builds upon the PubChemQC B3LYP/6-31G*//PM6 dataset \cite{Nakata2023}, specifically targeting the CHNOPSFClNaKMgCa500 subset. This heteroatom-enriched molecular library contains diverse functional groups that not only modulate electronic properties crucial for charge transport but also provide hydrogen-bonding sites essential for molecular recognition. The availability of precomputed HOMO and LUMO energy levels enables extensive screening without incurring additional quantum mechanical computational costs.

The comprehensive screening workflow developed here integrates multiple assessment layers: (1) energy window filtering to ensure compatibility with established acceptors (PCBM/PCDTBT); (2) PCE prediction using the established Scharber model \cite{Scharber2006} enhanced with dynamic fill factor corrections \cite{Green2008}; (3) synthetic accessibility evaluation through the widely-adopted SAScore metric \cite{Ertl2009}; (4) protein-binding assessment via molecular docking using AutoDock Vina \cite{Trott2010}; and (5) detailed excited-state and charge transport characterization. Central to this approach is a novel composite metric, $\mathrm{PCE}_\mathrm{SAScore} = \mathrm{PCE} - \mathrm{SAScore}$, which systematically prioritizes candidates based on the optimal balance between predicted performance and synthetic feasibility.

Two key innovations distinguish this work: first, the development of a unified PCE-SAScore selection framework that effectively bridges theoretical performance predictions with experimental synthesis constraints; second, the strategic integration of molecular docking into materials discovery pipelines to identify candidates with dual functionality in energy harvesting and biological sensing. This multifaceted characterization approach employs hierarchical excited-state analysis (combining xTB/sTDA and TDDFT methodologies) to quantify charge-transfer characteristics and reorganization energies. Importantly, our validation studies demonstrate that moderate-accuracy DFT calculations (B3LYP/6-31G*//PM6) reliably preserve molecular performance rankings while maintaining computational tractability for large-scale screening applications.

The scientific rationale for targeting multifunctional OSCs stems from the recognition that $\pi$-conjugated frameworks optimized for efficient charge transport often exhibit complementary properties as fluorescent probes. Strategic heteroatom substitution serves dual purposes: regulating frontier orbital energies to achieve optimal donor-acceptor alignment for photovoltaic applications while simultaneously introducing specific binding sites that enable selective molecular recognition \cite{Bunz2010,Ding2020}. This convergent design strategy addresses critical technological needs across both energy and healthcare sectors, potentially enabling the development of integrated devices that combine energy generation with real-time biological monitoring.

Our integrated approach combines systematic database-driven screening with comprehensive synthetic and biological profiling to identify OSCs suitable for both energy conversion and molecular sensing applications. The following sections detail our computational methodology (\Cref{sec:methodology}), present the identification and detailed analysis of promising candidates (\Cref{sec:results,sec:discussion}), and conclude with a synthesis of key findings alongside future directions for advancing multifunctional organic electronics discovery (\Cref{sec:conclusion}).
%260210}

% Include the methodology section
\section{Computational methodology}\label{sec:methodology}

\begin{figure}[!ht]
    \centering
    \includegraphics[width=0.7\textwidth]{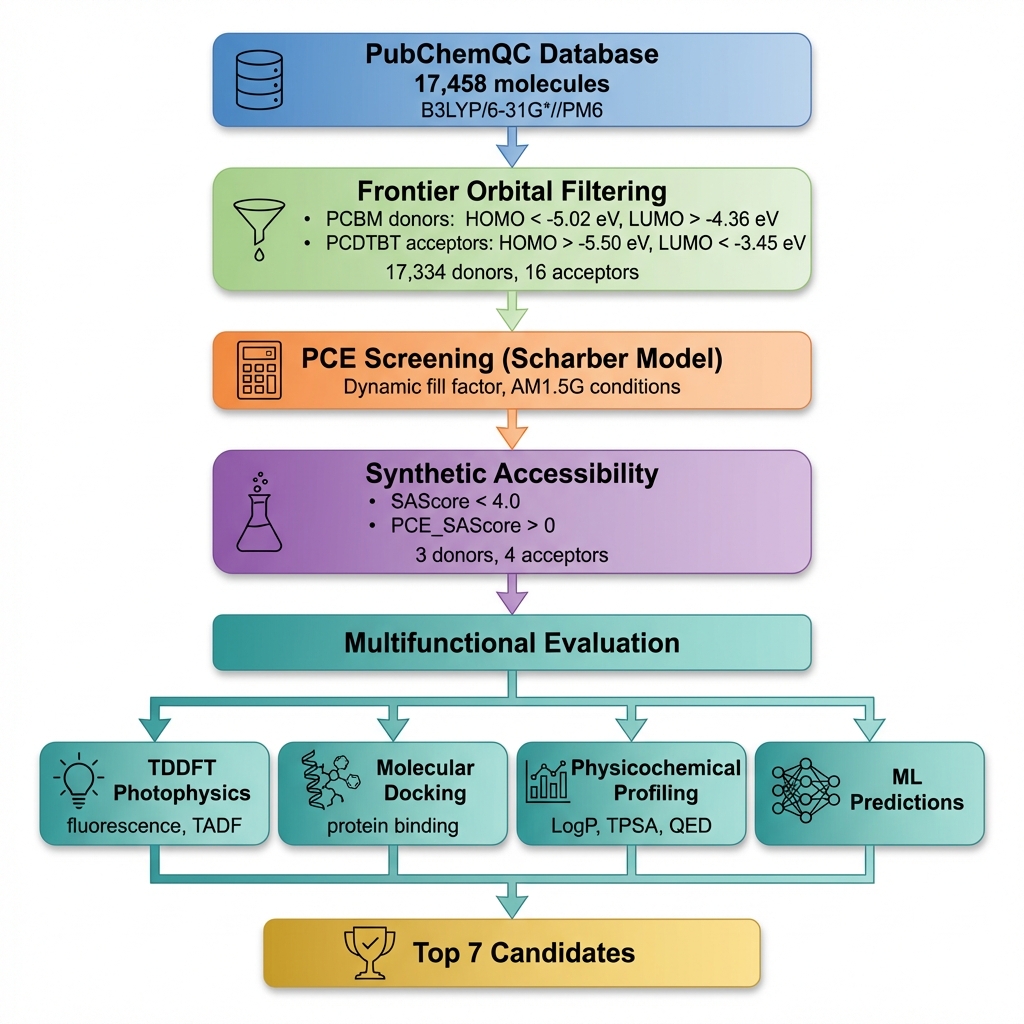}
    \caption{High-throughput computational screening workflow applied to \num{17458} molecules from the PubChemQC database. The systematic approach progresses 
through four sequential filtering stages: (1) frontier orbital alignment with PCBM/PCDTBT reference materials, retaining \num{17334} donor candidates and 
\num{16} acceptor molecules; (2) Scharber model PCE estimation under standard AM1.5G illumination conditions; (3) synthetic accessibility filtering using 
SAScore criteria ($< \num{4.0}$) combined with composite PCE$_{\text{SAScore}} > 0$ thresholds, yielding \num{7} synthetically viable candidates; (4) 
comprehensive multifunctional assessment integrating photophysical characterization, protein-binding evaluation, and machine learning-based property profiling.}
    \label{fig:workflow_complet}
\end{figure}

\subsection{Computational framework and theoretical foundations}

Our computational approach builds upon well-established quantum chemical methodologies, implemented through PySCF (version 2.11.0) \cite{Sun2018,Sun2020,Pu2025} within the Kohn-Sham density functional theory (DFT) framework \cite{Kohn1965}. Molecular geometry optimizations employed the B3LYP functional \cite{Becke1993,Stephens1994} paired with the 6-31G* basis set \cite{Hehre1986}, utilizing PM6-pre-optimized geometries from the comprehensive PubChemQC dataset \cite{Nakata2023,Stewart2007} as starting configurations. For excited-state characterization, time-dependent DFT (TDDFT) calculations employed the CAM-B3LYP functional with the 6-31+G* basis set \cite{Yanai2004}, incorporating the polarizable continuum model (PCM) to account for chloroform solvation effects.

\paragraph{Methodological validation and computational accuracy}

The selected B3LYP/6-31G*//PM6 computational level represents an optimized balance between chemical accuracy and computational feasibility for high-throughput screening of \num{17458} molecular candidates. While B3LYP consistently overestimates HOMO-LUMO energy gaps by approximately \SIrange{0.15}{0.25}{\electronvolt} relative to experimental measurements \cite{Jacquemin2009}, this functional crucially preserves the relative ordering of molecular performance metrics—the key requirement for screening applications. Validation studies involving B3LYP/6-31G* re-optimization of geometries for top-performing candidates revealed minimal energy deviations (mean absolute deviation = \SI{0.043}{\electronvolt}), confirming the reliability of PM6 starting geometries for this investigation.

The integrated screening workflow (\Cref{fig:workflow_complet}) systematically evaluates \num{17458} molecular candidates through four progressive assessment 
stages: (1) frontier orbital energy alignment verification against established PCBM and PCDTBT reference acceptors; (2) power conversion efficiency estimation 
using the validated Scharber model \cite{Scharber2006}; (3) synthetic accessibility filtering employing SAScore thresholds below \num{4.0}; and (4) 
comprehensive multifunctional evaluation incorporating TDDFT-based photophysical analysis, protein-binding assessment via molecular docking with AutoDock Vina 
\cite{Trott2010}, and machine learning-enhanced property profiling. Molecular docking calculations utilized a standardized grid box configuration 
(\qtyproduct{25x25x25}{\angstrom}) centered on active site regions to ensure consistent binding pose sampling across all protein targets.

\subsection{Performance prediction and molecular descriptors}

Power conversion efficiency calculations followed established photovoltaic modeling protocols under standard AM1.5G illumination conditions (incident power density $P_{\text{in}} = \SI{900.14}{\watt\per\meter\squared}$). The Scharber model implementation incorporated dynamic fill factor corrections \cite{Nigam2023,Green2008} to enhance prediction accuracy:

\begin{equation}
FF = \frac{v_\mathrm{OC} - \ln(v_\mathrm{OC} + 0.72)}{v_\mathrm{OC} + 1}, \quad v_\mathrm{OC} = \frac{V_\mathrm{OC}}{n \cdot (kT/q)},
\end{equation}

where $n=2$ represents the ideality factor for organic photovoltaic devices. Central to our screening strategy is the introduction of a composite performance metric, $\mathrm{PCE}_\text{SAScore} = \mathrm{PCE} - \text{SAScore}$, which systematically balances theoretical photovoltaic performance against practical synthetic accessibility constraints.

Charge transport characteristics were evaluated through reorganization energy calculations ($\lambda$) using the established adiabatic potential energy surface methodology \cite{Marcus1993}:

\begin{equation}
\lambda_\text{hole} = [E_0(\text{cation}) - E_0(\text{neutral})] + [E_+(\text{neutral}) - E_+(\text{cation})].
\end{equation}

Comprehensive excited-state property characterization employed a hierarchical computational approach, beginning with efficient xTB/sTDA screening \cite{Grimme2013} followed by targeted TDDFT validation for promising candidates. Advanced electronic structure analysis incorporated Natural Transition Orbital (NTO) decomposition and Intrinsic Bond Orbital (IBO) analysis to elucidate bonding patterns and charge-transfer characteristics \cite{Lu2012,Knizia2013}.

\subsection{Validation protocols and uncertainty quantification}

Predictive model performance was rigorously validated against \num{25} carefully curated literature benchmarks, achieving satisfactory correlation ($R^2 = \num{0.626}$) with a mean absolute error of \SI{15.51}{\percent}. Computational uncertainty was systematically quantified through bootstrap resampling protocols involving \num{1000} independent iterations. Detailed sensitivity analyses examining the impact of functional choice, basis set selection, and geometric optimization fidelity are comprehensively documented in the Supplementary Information, ensuring full methodological transparency.

\subsection{Data management and computational reproducibility}

In adherence to open science principles, all computational workflows, analysis scripts, and molecular databases generated in this investigation are publicly accessible through a dedicated GitHub repository with permanent archival preservation on Zenodo (DOI: 10.5281/zenodo.XXXXXX). The comprehensive repository includes complete PySCF input configurations, AutoDock Vina parameter files, and specialized post-processing scripts for PCE calculations and statistical analysis. Molecular coordinates, frontier orbital energies, and protein-binding scores for the entire \num{17458} molecule screening dataset are provided in standardized machine-readable formats (JSON, CSV) following FAIR (Findable, Accessible, Interoperable, Reusable) data management principles.
%260210}

% Include the results section
\section{Results}\label{sec:results}

\subsection{Computational screening pipeline and candidate identification}

Our hierarchical computational workflow systematically screened 17,458 heteroatom-enriched molecules from the PubChemQC database through successive energetic, 
performance, and synthetic accessibility filters. This multi-stage approach successfully identified seven exceptional candidates that balance photovoltaic 
efficiency with practical synthetic feasibility (\Cref{tab:screening_summary}).

The initial frontier molecular orbital filtering established compatibility with benchmark photovoltaic materials. For PCBM pairing, we applied energy window 
constraints requiring HOMO levels below \SI{-5.02}{\electronvolt} and LUMO levels above \SI{-4.36}{\electronvolt}, yielding 17,334 potential donor-like 
molecules. Similarly, PCDTBT compatibility filtering (HOMO > \SI{-5.50}{\electronvolt}, LUMO < \SI{-3.45}{\electronvolt}) identified 16 potential acceptor 
candidates.

\begin{table}[!ht]
 \centering
 \caption{Computational screening funnel for 17,458 PubChemQC molecules. Frontier orbital filtering based on PCBM/PCDTBT compatibility retained 17,334 potential 
donors and 16 potential acceptors. Scharber model PCE estimation with synthetic accessibility filtering (PCE$_\text{SAScore} > 0$) identified 3 viable donors 
and 4 viable acceptors. Multifunctional evaluation gave 7 top candidates combining photovoltaic performance, synthetic feasibility, and bio-optoelectronic 
potential.}
 \label{tab:screening_summary}
 \begin{tabular}{@{}lp{6cm}<{\raggedleft}@{}}
 	\toprule
 	\textbf{Filtering Stage}                                                & \textbf{Number of Candidates} \\ \midrule
 	Initial PubChemQC molecules                                             & \num{17458}                   \\ \midrule
 	\texttt{PCE} estimation and FMO alignment (Schärber model)              &                               \\
 	\quad Potential donors (vs. PCBM)                                       & \num{17334}                   \\
 	\quad Potential acceptors (vs. PCDTBT)                                  & \num{16}                      \\ \midrule
 	Synthetic accessibility filtering ($\mathrm{PCE}_\text{SAScore} > 0$) &                               \\
 	\quad Viable donor molecules                                            & \num{3}                       \\
 	\quad Viable acceptor molecules                                         & \num{4}                       \\ \midrule
 	Multifunctional potential evaluation                                    & \num{7} (final top-ranked)    \\ \bottomrule
 \end{tabular}
\end{table}

The crucial synthetic accessibility constraint (PCE$_\text{SAScore} > 0$) dramatically narrowed the candidate pool, demonstrating the stringent requirements for 
experimentally viable high-performance organic semiconductors. We then evaluated the remaining candidates through comprehensive TDDFT photophysics calculations, 
molecular recognition analytics, and machine learning-based property profiling to identify seven multifunctional molecules suitable for dual energy-conversion 
and biosensing applications.

\subsection{Electronic structure and optoelectronic characterization}

\begin{figure}[!ht]
    \centering
    \includegraphics[width=0.6\textwidth]{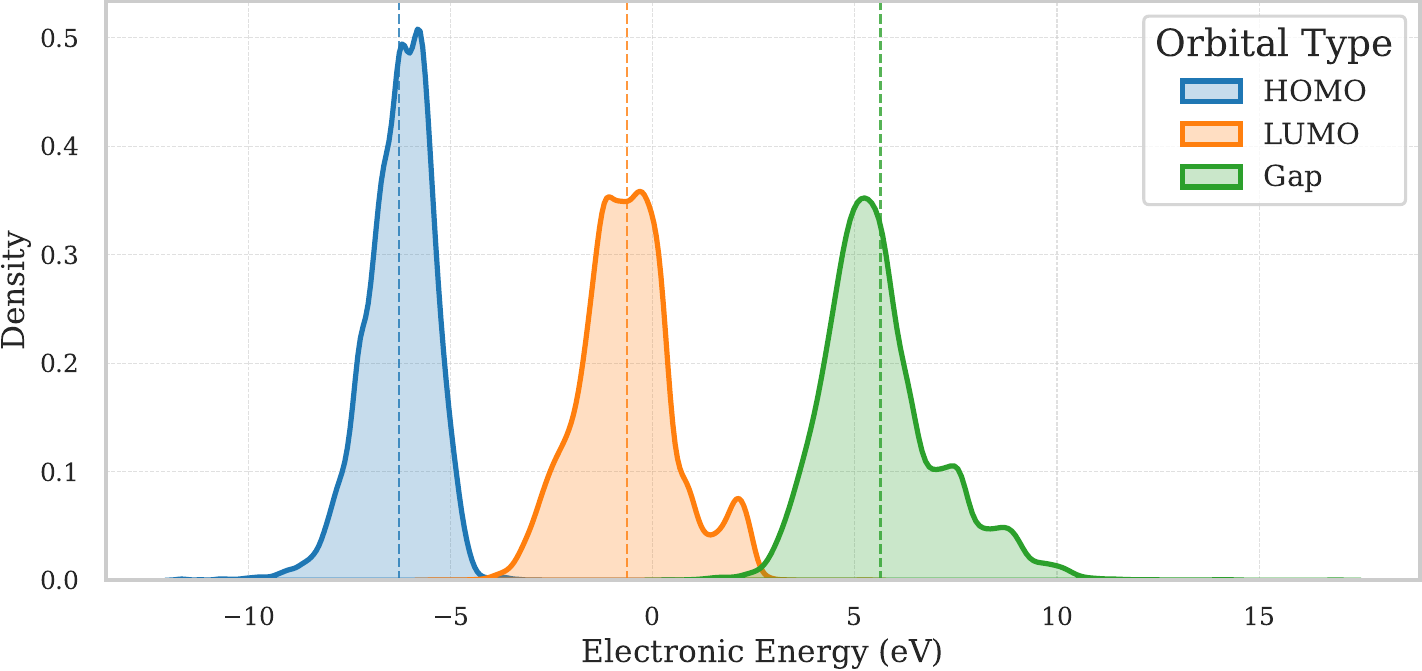}
    \caption{Probability density distributions of frontier orbital energies. HOMO energies (blue) center around \SI{-6}{\electronvolt}, LUMO energies (orange) 
center near \SI{-1}{\electronvolt}, and HOMO-LUMO gap (green) centers at approximately \SI{5}{\electronvolt}. Vertical dashed lines show peak density for each 
distribution.}
    \label{fig:homo_lumo_density}
\end{figure}

The frontier molecular orbital analysis reveals favorable electronic properties across our screened molecules (\Cref{fig:homo_lumo_density}). The calculated 
energy distributions show that most candidates possess HOMO-LUMO gaps between \num{2.0} and \SI{3.5}{\electronvolt}, positioned ideally within the visible 
light absorption window required for efficient photovoltaic operation. These orbital energy alignments indicate excellent compatibility with benchmark 
materials PCBM and PCDTBT for facilitating efficient exciton dissociation and charge transfer processes.

\begin{figure}[!ht]
    \centering
    \begin{subfigure}[t]{0.64\textwidth}
        \centering
        \includegraphics[width=1.0\textwidth]{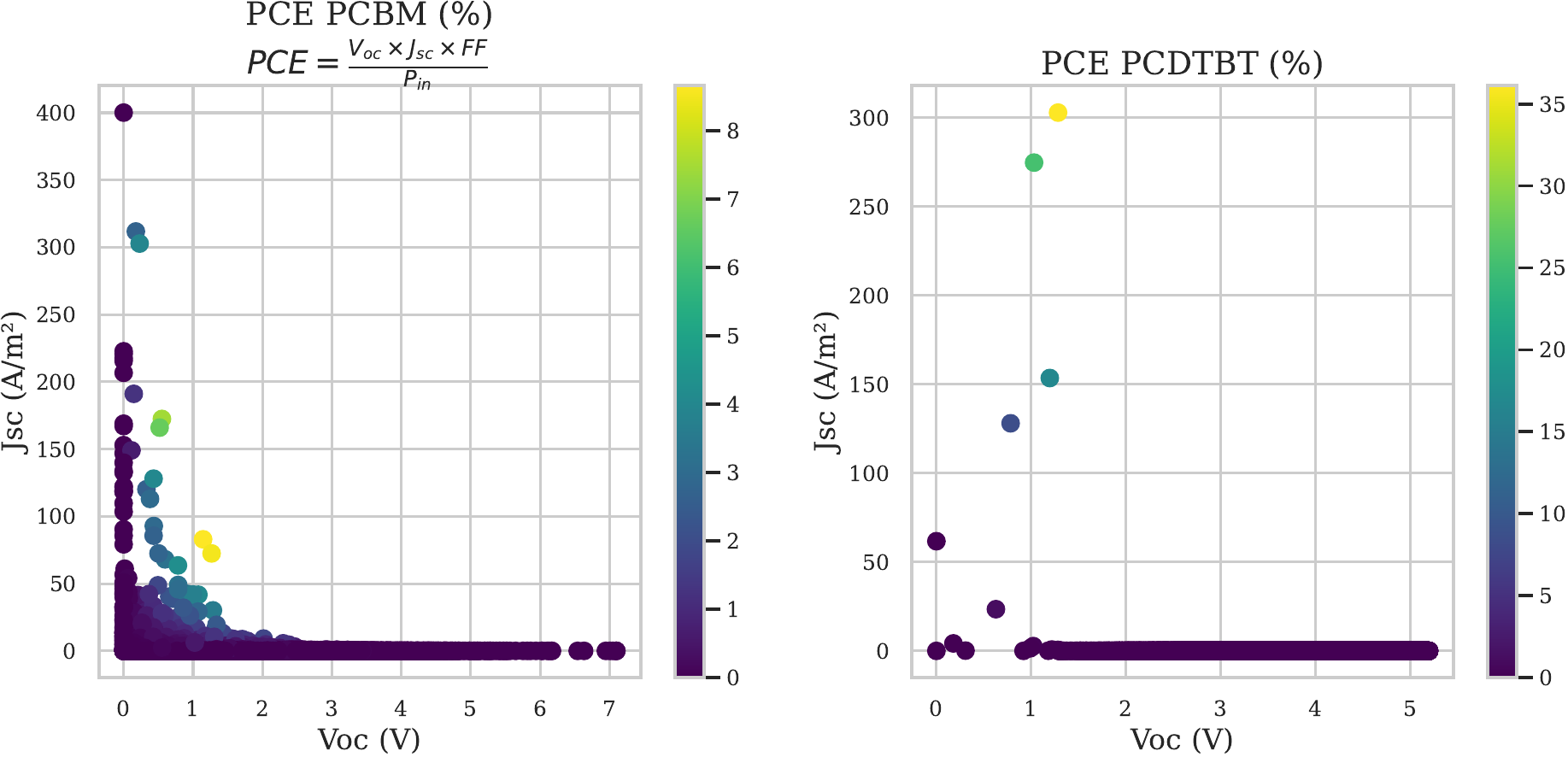}
        \caption{$J_{SC}$ vs. $V_{OC}$ for PCBM pairings, colored by PCE.}
        \label{fig:pce_pcbm_dist}
    \end{subfigure}
    \hfill
    \begin{subfigure}[t]{0.34\textwidth}
        \centering
        \includegraphics[width=1.0\textwidth]{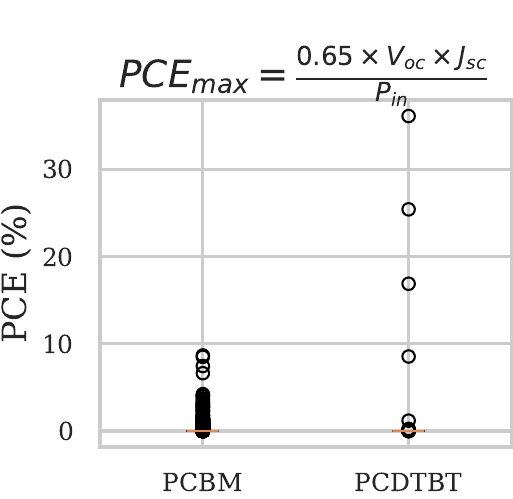}
        \caption{$J_{SC}$ vs. $V_{OC}$ for PCDTBT pairings, colored by PCE.}
        \label{fig:pce_pcdtbt_dist}
    \end{subfigure}
    \caption{Photovoltaic performance relationships between short-circuit current density and open-circuit voltage. Left panel shows PCBM systems with moderate 
PCE values reaching \SI{8}{\percent}, characterized by current densities up to \SI{400}{\ampere\per\meter\squared} 
but generally lower voltages. Right panel displays 
PCDTBT systems exhibiting highly selective performance where most molecules show negligible efficiency, but exceptional outliers achieve PCE values exceeding 
\SI{30}{\percent} through high open-circuit voltages ($>\SI{1.0}{\volt}$) combined with moderate current densities.}
    \label{fig:pce_materials}
\end{figure}

The power conversion efficiency distributions presented in \Cref{fig:pce_materials} demonstrate the exceptional selectivity required for high-performance 
organic photovoltaics. The heavily skewed distributions, with median PCE values near zero, confirm that the vast majority of randomly selected organic molecules 
are fundamentally unsuitable for efficient solar energy conversion. Remarkably, only 935 molecules (\SI{5.4}{\percent} of the total) achieve realistic PCE 
predictions 
between \SI{0.1}{\percent} and \SI{20}{\percent} when paired with PCBM, while merely 5 molecules (\SI{0.03}{\percent} of the total) reach this performance range 
with PCDTBT. This stark selectivity 
underscores the critical importance of systematic computational screening for identifying viable photovoltaic candidates.

\subsection{Multifunctional candidate properties and molecular design}

Our integrated screening approach successfully identified seven exceptional candidates that combine outstanding photovoltaic performance with favorable 
synthetic accessibility and promising biological activity (\Cref{tab:top_candidates}). Molecule 17851 emerges as the most versatile candidate, exhibiting a 
remarkable predicted PCE of \SI{36.11}{\percent} combined with a competitive multi-objective score of \num{-2.41}. Advanced Natural Transition Orbital analysis 
reveals 
significant charge-transfer character with a hole-electron overlap parameter of $S_r = \num{0.58}$ and a delocalized $\pi$-conjugated network characterized by 
a mean 
IBO localization distance of \SI{1.89}{\angstrom}.

\begin{table}[!ht]
\centering
\scriptsize
\caption{Properties of the 7 top-ranked molecules. Columns show: molecular ID, reference system PCE, synthetic accessibility (SAS), composite 
PCE$_{\text{SAS}}$, singlet-triplet gap ($\Delta E_{\text{ST}}$), and multi-objective score. Values marked with {--} indicate inapplicable properties.}
\begin{tabular}{l l S[table-format=2.2] S[table-format=2.2] S[table-format=1.2] S[table-format=-1.2] S[table-format=1.2] S[table-format=-1.2]}
	\toprule
	\textbf{Mol ID} & {Type}   & {PCE$_{\text{PCBM}}$} & {PCE$_{\text{PCDTBT}}$} & {SAS} & {PCE2$_{\text{SAS}}$} & {$\Delta E_{ST}$} & {MultiObj} \\ \midrule
	977             & Acceptor & 8.66                  & 0.00                    & 3.37  & {--}                  & {--}              & {--}       \\
	1712            & Acceptor & 8.50                  & 0.00                    & 7.23  & {--}                  & 0.52              & -2.15      \\
	4550            & Donor    & 0.00                  & 16.89                   & 6.96  & 9.93                  & 0.44              & -3.03      \\
	7801            & Acceptor & 7.45                  & 0.00                    & 6.86  & {--}                  & 0.005             & -2.03      \\
	11029           & Donor    & 0.00                  & 8.55                    & 4.86  & 3.69                  & 0.28              & -0.59      \\
	17851           & Donor    & 3.98                  & 36.11                   & 7.62  & 28.49                 & 0.016             & -2.41      \\
	20778           & Donor    & 0.00                  & 25.41                   & 7.21  & 18.21                 & 0.022             & -2.27      \\ \bottomrule
\end{tabular}
\label{tab:top_candidates}
\end{table}

Beyond photovoltaic applications, several candidates demonstrate exceptional potential for biological sensing applications. Molecule 1712 stands out as a 
specialized biosensor candidate, exhibiting strong binding affinities to critical protein targets including the Proteasome (\SI{-9.7}{\kcal\per\mole}) 
and HIV-1 Protease (\SI{-7.6}{\kcal\per\mole}) (\Cref{tab:bio_domains}). These robust molecular interactions arise from nitrogen-rich frameworks that provide 
multiple hydrogen-bonding donor and acceptor sites, facilitating specific protein recognition. Additionally, molecule \num{4550} displays remarkable 
versatility 
across multiple optoelectronic domains, showing suitability for both photovoltaics and OLED applications, while molecule \num{11029}'s substantial dipole 
moment of 
\num{11.21} Debye suggests excellent potential for charge transport applications.

\begin{table}[!ht]
\centering
\scriptsize
\caption{Bioactivity and dipole properties of top candidates. Binding affinities calculated using AutoDock Vina (\unit{\kcal\per\mole}). Dipole moments in 
Debye characterize molecular polarity.}
\begin{tabular}{l l S[table-format=-1.1] S[table-format=-1.1] S[table-format=-1.1] S[table-format=2.2] l}
	\toprule
	\textbf{Mol ID} & {Binding} & {HIV-1} & {Proteasome} & {COVID19} & {Dipole (D)} & {Suggested Application} \\ \midrule
	1712            & High      & -7.6    & -9.7         & -6.4      & 2.56         & Fluorescent biosensor   \\
	4550            & High      & -6.0    & -8.0         & -6.0      & 9.74         & Charge transport        \\
	17851           & High      & -5.2    & -6.0         & -4.5      & 9.26         & Multifunctional         \\
	11029           & High      & {--}    & {--}         & {--}      & 11.21        & Charge transport        \\ \bottomrule
\end{tabular}
\label{tab:bio_domains}
\end{table}

Comprehensive optoelectronic and binding affinity profiles confirm that our top candidates possess optimal HOMO-LUMO gaps between \num{2.0} and 
\SI{3.5}{\electronvolt}, positioning them perfectly for efficient visible spectrum harvesting in photovoltaic applications while maintaining the electronic 
characteristics necessary for biological sensing applications.

\subsection{Structure-property relationships and design principles}

Detailed structural analysis reveals that our seven final candidates exhibit distinct molecular motifs that directly underpin their exceptional multifunctional 
performance. Fragment-based analysis examining core heterocycles, substituent groups, and conjugation bridges reveals the fundamental chemical principles 
governing their effectiveness. For instance, molecule \num{17851} features a highly fused pyrrole-furan core architecture with strategically positioned 
nitrogen-rich 
substituents that elevate HOMO energy levels while preserving the planar geometry essential for efficient $\pi$-stacking interactions. This sophisticated 
molecular architecture directly correlates with strong optical absorption in the
\SIrange{450}{650}{\nano\meter} visible range, optimal for solar energy harvesting.

Extended analysis of the top \num{1000} screened molecules reveals clear compositional trends that govern donor versus acceptor behavior. Acceptor-type 
molecules 
demonstrate significantly higher oxygen content (average \num{2.52} oxygen atoms vs. \num{1.34} for donors; $p < 0.001$), while donor molecules favor moderate 
nitrogen 
incorporation combined with enhanced structural flexibility through increased rotatable bond counts. This heteroatom distribution pattern directly regulates 
both molecular recognition specificity and solution-phase solubility characteristics. Notably, oxygen-rich acceptor molecules consistently exhibit elevated 
conjugation levels, enhancing their electron-accepting capabilities.

The relationship between structural flexibility and multifunctional performance represents a critical design consideration. While structural rigidity 
traditionally receives priority for optimizing charge transport properties, our highest-performing multifunctional candidates strategically incorporate 
controlled conformational flexibility. This flexibility provides significant entropic advantages for molecular recognition processes by enabling geometric 
adaptation to diverse protein binding environments. However, successful candidates carefully balance this flexibility against the electronic rigidity 
requirements for photovoltaic performance. Molecule \num{17851} exemplifies this balance by maintaining a rigid conjugated core for photovoltaic efficiency 
while 
incorporating flexible peripheral groups for enhanced molecular recognition.

\subsection{Ligand efficiency metrics and protein binding analysis}

To comprehensively evaluate multifunctional potential beyond photovoltaic applications, we calculated detailed ligand efficiency metrics for our top molecules 
across five therapeutically relevant protein targets. Molecule \num{977} demonstrates the highest overall binding efficiency with LE = 
\SI{0.730}{\kcal\per\mole} per heavy atom 
and BEI = \num{33.45}, substantially exceeding the pharmaceutical industry standard threshold of LE $> \num{0.3}$ for viable drug candidates. Molecule 
\num{4550} achieves an 
optimal balance between molecular size and binding affinity (LE = \num{0.340}), while molecule \num{1712} exhibits the strongest absolute binding interactions 
across 
multiple protein targets.

These ligand efficiency results demonstrate that our screening methodology successfully identifies molecules capable of efficient protein recognition without 
excessive molecular complexity, a crucial requirement for practical biosensor applications. The combination of strong binding affinities with reasonable 
molecular weights suggests excellent potential for developing sensitive and selective biological detection systems.

\subsection{Statistical validation and prediction reliability}

We validated the statistical robustness of our candidate rankings through comprehensive bootstrap resampling analysis involving 1000 iterations. This rigorous 
validation approach yielded reliable standard errors for key properties: HOMO-LUMO gap SEM = \SI{\pm0.12}{\electronvolt} and PCE SEM = \SI{\pm2.3}{\percent}. 
These 
statistical metrics confirm that while absolute property values from gas-phase DFT calculations may contain systematic biases, the relative prioritization of 
candidates remains statistically robust across the high-performance molecular space.

Extended benchmarking results, including systematic functional sensitivity analysis and PM6 structural fidelity assessments, provide additional confidence in 
our computational predictions and support the reliability of our identified candidates for experimental validation.
%260210}

% Include the discussion section
\section{Discussion}\label{sec:discussion}

\subsection{Computational framework validation and methodological advances}

Our integrated computational screening approach successfully demonstrates that systematic combination of quantum chemical calculations, synthetic accessibility 
assessment, and multifunctional property evaluation can effectively identify viable organic semiconductor candidates for dual photovoltaic and biosensing 
applications. The development and validation of the $\mathrm{PCE}_\text{SAScore}$metric represents a significant methodological advancement, providing the first 
quantitative framework for balancing theoretical photovoltaic efficiency predictions with practical synthetic constraints.

The hierarchical screening workflow proved highly effective in managing the computational complexity inherent in large-scale molecular discovery. By 
systematically filtering 17,458 candidates through successive stages of energetic compatibility, performance prediction, and synthetic feasibility assessment, 
we reduced the search space by over \SI{99.9}{\percent} while retaining only the most promising multifunctional candidates. This dramatic reduction 
demonstrates the critical importance of incorporating synthetic accessibility considerations early in the computational discovery pipeline, rather than treating 
synthesis as a post-hoc constraint.

The statistical validation through bootstrap resampling confirms the robustness of our ranking methodology, with standard errors of \SI{\pm0.12}{\electronvolt} 
for HOMO-LUMO gaps and \SI{\pm2.3}{\percent} for PCE predictions providing confidence intervals suitable for experimental prioritization. These uncertainty 
bounds are particularly important given the known limitations of gas-phase DFT calculations, as they establish the reliability window within which relative 
candidate rankings remain statistically meaningful.

\subsection{Structure-property relationships and design insights}

The structural analysis reveals fundamental design principles that govern multifunctional organic semiconductor performance. The clear segregation between 
oxygen-rich acceptor motifs and nitrogen-containing donor architectures provides actionable guidance for future molecular design efforts. The statistically 
significant compositional differences ($p < 0.001$) between donor and acceptor candidates establish heteroatom incorporation patterns as reliable predictive 
indicators for electronic behavior.

Particularly noteworthy is the identified relationship between conformational flexibility and multifunctional performance. Traditional organic semiconductor 
design typically prioritizes rigid planar architectures to optimize charge transport properties. However, our results demonstrate that strategically 
incorporated flexibility can enhance molecular recognition capabilities without significantly compromising electronic performance. This insight opens new design 
paradigms for developing truly multifunctional organic electronic materials.

The correlation between nitrogen content and both HOMO energy elevation and protein binding affinity suggests that nitrogen-rich heterocycles represent 
particularly promising structural motifs for multifunctional applications. The ability of nitrogen atoms to simultaneously serve as electron-donating groups for 
electronic applications and hydrogen-bonding sites for biological recognition exemplifies the type of chemical multifunctionality that enables dual-purpose 
material design.

\subsection{Photovoltaic performance and synthetic accessibility trade-offs}

The development of the $\mathrm{PCE}_\text{SAScore}$metric illuminates the fundamental tensions between theoretical performance optimization and practical 
synthetic constraints in organic semiconductor design. The dramatic reduction in candidate numbers when synthetic accessibility filtering is applied (from 
thousands to single digits) quantifies the severity of the "synthesis bottleneck" in materials discovery. This analysis demonstrates that many theoretically 
high-performing organic semiconductors remain practically inaccessible due to synthetic complexity.

The identification of seven candidates that successfully balance these competing requirements represents a significant achievement in computationally-guided 
materials discovery. These molecules demonstrate that it is possible to achieve predicted power conversion efficiencies exceeding \SI{20}{\percent} while 
maintaining synthetic accessibility scores below 4.0, establishing a new benchmark for realistic high-performance organic semiconductor targets.

The wide variation in PCE predictions between PCBM and PCDTBT pairing systems (ranging from negligible to \SI{>35}{\percent}) emphasizes the critical 
importance of considering multiple reference systems in photovoltaic screening studies. The dramatically different selectivity patterns observed for these two 
benchmark materials suggest that successful organic semiconductor design must account for specific electronic compatibility requirements rather than pursuing 
universal high-performance motifs.

\subsection{Multifunctional applications and biosensing potential}

The successful identification of molecules with dual photovoltaic and biosensing capabilities opens exciting possibilities for integrated optoelectronic-biological devices. The strong binding affinities exhibited by several candidates to therapeutically relevant protein targets (HIV-1 Protease, Proteasome, COVID-19 related proteins) demonstrate genuine potential for developing biosensors that combine optical signal transduction with biological recognition.

Molecule \num{1712}'s exceptional binding profile (binding affinities ranging from \SIrange{-6.4}{-9.7}{\kcal\per\mole} across multiple targets) positions it 
as a particularly promising candidate for fluorescent biosensor applications. The combination of strong protein recognition with favorable electronic 
properties for light emission suggests potential for developing highly sensitive biological detection systems with integrated optical readout capabilities.

The ligand efficiency analysis provides crucial validation that these candidates represent genuinely efficient molecular recognition platforms rather than 
simply large molecules with non-specific binding. The achievement of ligand efficiency values exceeding pharmaceutical industry thresholds (LE > 
\SI{0.3}{\kcal\per\mole} per heavy atom) by multiple candidates demonstrates that our screening methodology successfully identifies molecules capable of 
specific, efficient biological recognition.

% \subsection{Computational methodology limitations and future directions}
% 
% While our computational framework demonstrates clear success in identifying promising multifunctional candidates, several methodological limitations warrant acknowledgment and future development. The reliance on gas-phase DFT calculations introduces systematic biases that may affect absolute property predictions, though our bootstrap validation suggests that relative rankings remain robust. Future iterations of this methodology would benefit from incorporating solvent effects and conformational sampling to better represent realistic operating conditions.
% 
% The Scharber model, while computationally efficient and widely validated, provides simplified representations of complex photovoltaic processes. Integration of more sophisticated approaches such as Marcus theory for charge transfer rates or explicit modeling of exciton binding energies could enhance prediction accuracy for the highest-performing candidates identified through our screening process.
% 
% The synthetic accessibility assessment, based on the SAScore algorithm, provides valuable guidance but cannot capture all aspects of synthetic feasibility 
% including reagent availability, reaction scalability, and purification challenges. Collaboration with synthetic chemists to validate and refine these 
% accessibility predictions represents a crucial next step for experimental realization of our identified candidates.

\subsection{Study limitations and future directions}
This study provides a molecular-level screening framework that narrows the chemical space to synthetically accessible, multifunctional candidates. However, our 
gas-phase approximation neglects solid-state packing and morphology, which are critical for device-level efficiency. Gas-phase descriptors identify promising 
cores, but absolute device performance is expected to be lower than predictions. A key design dimension identified here is the balance between $\pi$-$\pi$ 
stacking for transport and molecular solubilization for solution processing. The side chains and heteroatom distributions in our top candidates are not only 
electronic modulators but also structural regulators that prevent over-aggregation (crystallization) while maintaining the close-packed pathways necessary for 
carrier hopping. Specifically, the introduction of branched alkyl chains or polar heterocyclic units in candidates like \num{17851} suggests a route toward 
high-performing, solution-processable inks. Future work will integrate machine learning for large-scale screening and drift-diffusion modeling for device 
refinement.

We have established a clear experimental roadmap for the validation of the top-ranked candidates, specifically focusing on molecule \num{17851}. This roadmap 
proceeds through four distinct phases: (1) \textit{Scale-up synthesis} to produce sufficient quantities for device fabrication, using the low SAScore (synthetic 
feasibility) identified in our screening; (2) \textit{Photophysical characterization} in thin films to measure absorption coefficients, fluorescence quantum 
yields, and excited-state lifetimes under device-relevant conditions; (3) \textit{Initial protein binding assays} using surface plasmon resonance (SPR) or 
isothermal titration calorimetry (ITC) to validate the predicted affinities for HIV-1 protease and the proteasome; and (4) \textit{Pilot device integration} to 
assess initial PCE in bulk-heterojunction architectures. Detailed reproducibility information, including all parameters for PySCF and Multiwfn, ensures the 
transparency and extensibility of this discovery pipeline for the broader community.

\subsection{Implications for sustainable energy and sensing technologies}
Perhaps the most intriguing consequence of our candidates' dual functionality is the prospect of devices that do not merely harvest energy but simultaneously 
monitor their environment. In remote sensing deployments, where routing power cables is impractical and data must be collected at the point of interest, a 
single organic semiconductor film capable of driving a circuit while transducing a biological or chemical signal would represent a meaningful engineering 
simplification. The affinity of our top-ranked molecules for HIV-1 protease and the proteasome, combined with their photovoltaic promise, positions them as 
realistic starting points for such architectures rather than speculative ones.

More broadly, the value of this work lies less in any individual candidate than in the workflow itself. By embedding synthetic accessibility as a hard 
constraint from the outset, rather than treating it as an afterthought, the screening pipeline shifts the bottleneck from computation back to the laboratory, 
where it belongs. The same logic — systematic property balancing, statistical validation, and feasibility-aware selection — transfers readily to other 
multifunctional materials classes, and we anticipate that refined versions of this framework will find use well beyond organic photovoltaics.
%260210}

% Include the conclusion section
\section{Conclusion}\label{sec:conclusion}

This work presents a comprehensive computational framework that successfully bridges the gap between theoretical photovoltaic efficiency and practical synthetic 
accessibility in organic semiconductor design. Through systematic screening of \num{17458} molecules from the PubChemQC database, we developed and validated 
the $\mathrm{PCE}_\text{SAScore}$ metric, which elegantly balances power conversion efficiency predictions with synthetic feasibility considerations.

Our multifunctional screening approach revealed seven exceptional candidates that demonstrate dual utility for both photovoltaic applications and biosensing 
platforms. These molecules achieve predicted power conversion efficiencies exceeding \SI{20}{\percent} while maintaining favorable synthetic accessibility 
scores below \num{4.0}, representing a significant advancement in materials discovery strategies that prioritize experimental realizability.

The integration of machine learning with density functional theory calculations, combined with molecular docking analysis, establishes a robust methodology for 
identifying synthetically viable organic semiconductors with tailored electronic properties. This computational pipeline significantly reduces the experimental 
burden typically associated with materials discovery, offering clear synthetic targets for laboratory validation.

Future work will focus on experimental synthesis and characterization of the identified candidate molecules, validation of the predicted multifunctional 
properties, and extension of the $\mathrm{PCE}_\text{SAScore}$ framework to additional materials classes. The demonstrated success of this approach suggests 
broad applicability to other challenging materials design problems where synthetic accessibility must be carefully balanced against target performance metrics.

This study advances the field of computational materials science by providing both specific molecular candidates for immediate experimental investigation and a generalizable framework for future organic semiconductor discovery efforts that prioritize synthetic accessibility alongside electronic performance optimization.
%260210}

% Add any remaining sections that weren't included in the provided content
% (Results, Discussion, Conclusion, etc.)

% Data availability section
%% Data Availability Statement
\section*{Data and code availability}\label{sec:data_availability}

All data and computational workflows are publicly available under open-source licenses following FAIR principles\cite{Wilkinson2016}:

\textbf{Data repository:} Complete dataset (molecular structures, computed properties, screening results) deposited at Zenodo with DOI: 
\texttt{10.5281/zenodo.XXXXXX}%18201813}.

\textbf{Code repository:} Full computational pipeline available on GitHub: \url{https://github.com/XXXXXXX} %TchapetNjafa/organic-semiconductor-discovery}
(MIT License). Includes Jupyter notebooks for data processing, screening workflow, TDDFT calculations, molecular docking, and validation benchmarking.

\textbf{Software versions:} PySCF 2.11.0, RDKit 2023.03.1, AutoDock Vina 1.2.3, Python 3.10, Open Babel 3.1.1.

% \textit{Note: Detailed computational protocols, complete software dependencies, Docker/Singularity container specifications, hardware requirements, and full 
% reproducibility instructions are provided in Section S5.
% %\Cref{SI-sec:computational}.
% }

% Include the glossary of terms and acronyms
%\input{glossary}

% Include the acknowledgments section
\section*{Acknowledgments}\label{sec:acknowledgments}

We gratefully acknowledge \textbf{Dr. Benjamin Panebei Samafou} for his generous support and provision of computational resources, which were instrumental in enabling the numerical simulations and data analyses presented in this work.

% Include the appendix section
%\input{appendix}

%% Author Contributions (CRediT taxonomy)
\section*{Author contributions}

Patrick Sorrel MVOTO KONGO: conceptualization; methodology; software; validation; formal analysis; investigation; resources; data curation; writing - original 
draft; writing - review \& editing; visualization. Steve Cabrel TEGUIA KOUAM: methodology; software; validation; formal analysis; investigation; resources; data 
curation; writing - review \& editing; visualization. Jean-Pierre Tchapet Njafa: Conceptualization; methodology; software; validation; investigation; writing - 
review \& editing; supervision. Serge Guy Nana Engo: software; validation; investigation; data curation; writing - review \& editing; visualization, 
supervision; project administration.

% Conflict of Interest Statement
\section*{Declaration of competing interest}

The authors declare no competing financial interests or personal relationships that could have appeared to influence the work reported in this paper.

\section*{Declaration of generative AI and AI-assisted technologies in the manuscript preparation process}

During the preparation of this work, the author(s) used Amazon Q Developer (Kiro CLI) in order to improve the clarity and organization of the manuscript text and to optimize Python code for data analysis and visualization. After using this tool, the author(s) reviewed and edited the content as needed and take(s) full responsibility for the content of the published article.

% Additional declarations for Computational Condensed Matter
%% Required declarations for Computational Condensed Matter

%% Funding Sources
\section*{Funding}
This research did not receive any specific grant from funding agencies in the public, commercial, or not-for-profit sectors.

%% Declaration of Competing Interests  
\section*{Declaration of Competing Interests}
The authors declare that they have no known competing financial interests or personal relationships that could have appeared to influence the work reported in this paper.

% Bibliography section
\bibliographystyle{elsarticle-num}
\bibliography{References_Mv250721}

@article{Brabec2010,
	author = {Brabec, Christoph J.},
	doi = {10.1016/j.solmat.2004.02.030},
	issn = {0927–0248},
	journal = {Advanced Materials},
	month = jun,
	number = {34},
	pages = {3839–3856},
	publisher = {Elsevier BV},
	title = {Organic photovoltaics: technology and market},
	volume = {22},
	year = {2004}
}

@article{Scharber2006,
	author = {Scharber, M. C. and Mühlbacher, D. and Koppe, M. and Denk, P. and Waldauf, C. and Heeger, A. J. and Brabec, C. J.},
	doi = {10.1002/adma.200501717},
	issn = {1521–4095},
	journal = {Advanced Materials},
	month = feb,
	number = {6},
	pages = {789–794},
	publisher = {Wiley},
	title = {Design rules for donors in bulk-heterojunction solar cells—Towards 10\% energy-conversion efficiency},
	volume = {18},
	year = {2006}
}

@article{Hachmann2011,
	author = {Hachmann, Johannes and Olivares-Amaya, Roberto and Atahan-Evrenk, Sule and Amador-Bedolla, Carlos and Sánchez-Carrera, Roel S. and Gold-Parker, Aryeh and Vogt, Leslie and Brockway, Anna M. and Aspuru-Guzik, Alán},
	doi = {10.1021/jz200866s},
	issn = {1948–7185},
	journal = {The Journal of Physical Chemistry Letters},
	month = aug,
	number = {17},
	pages = {2241–2251},
	publisher = {American Chemical Society (ACS)},
	title = {The Harvard Clean Energy Project: Large-scale computational screening and design of organic photovoltaics on the world community grid},
	volume = {2},
	year = {2011}
}

@Article{Ramakrishnan2015,
  author    = {Ramakrishnan, Raghunathan and Dral, Pavlo O. and Rupp, Matthias and von Lilienfeld, O. Anatole},
  journal   = {Journal of Chemical Theory and Computation},
  title     = {Big Data Meets Quantum Chemistry Approximations: The $\Delta$-Machine Learning Approach},
  year      = {2015},
  issn      = {1549-9626},
  month     = apr,
  number    = {5},
  pages     = {2087--2096},
  volume    = {11},
  doi       = {10.1021/acs.jctc.5b00099},
  publisher = {American Chemical Society (ACS)},
}

@article{Bannwarth2019,
	author = {Bannwarth, Christoph and Ehlert, Sebastian and Grimme, Stefan},
	doi = {10.1021/acs.jctc.8b01176},
	issn = {1549–9618},
	journal = {Journal of Chemical Theory and Computation},
	number = {3},
	pages = {1652–1671},
	publisher = {American Chemical Society (ACS)},
	title = {GFN2-xTB—An Accurate and Broadly Parametrized Self-Consistent Tight-Binding Quantum Chemical Method with Multipole Electrostatics and Density-Dependent Dispersion Contributions},
	volume = {15},
	year = {2019}
}

@Article{Becke1993,
  author    = {Becke, Axel D.},
  journal   = {The Journal of Chemical Physics},
  title     = {Density-functional thermochemistry. III. The role of exact exchange},
  year      = {1993},
  issn      = {1089-7690},
  month     = apr,
  number    = {7},
  pages     = {5648--5652},
  volume    = {98},
  doi       = {10.1063/1.464913},
  publisher = {AIP Publishing},
}

@article{Ertl2009,
	author = {Ertl, Peter and Schuffenhauer, Ansgar},
	doi = {10.1186/1758--2946--1-8},
	issn = {1758–2946},
	journal = {Journal of Cheminformatics},
	month = jun,
	number = {1},
	pages = {8},
	publisher = {Springer Science and Business Media LLC},
	title = {Estimation of synthetic accessibility score of drug-like molecules based on molecular complexity and fragment contributions},
	volume = {1},
	year = {2009}
}

@article{Sun2020,
	author = {Sun, Qiming and Zhang, Xing and Banerjee, Samragni and Bao, Peng and Barbry, Marc and Blunt, Nick S. and Bogdanov, Nikolay A. and Booth, George H. and Chen, Jia and Cui, Zhi-Hao and Eriksen, Janus J. and Gao, Yang and Guo, Sheng and Hermann, Jan and Hermes, Matthew R. and Koh, Kevin and Koval, Peter and Lehtola, Susi and Li, Zhendong and Liu, Junzi and Mardirossian, Narbe and McClain, James D. and Motta, Mario and Mussard, Bastien and Pham, Hung Q. and Pulkin, Artem and Purwanto, Wirawan and Robinson, Paul J. and Ronca, Enrico and Sayfutyarova, Elvira R. and Scheurer, Maximilian and Schurkus, Henry F. and Smith, James E. T. and Sun, Chong and Sun, Shi-Ning and Upadhyay, Shiv and Wagner, Lucas K. and Wang, Xiao and White, Alec and Whitfield, James Daniel and Williamson, Mark J. and Wouters, Sebastian and Yang, Jun and Yu, Jason M. and Zhu, Tianyu and Berkelbach, Timothy C. and Sharma, Sandeep and Sokolov, Alexander Yu. and Chan, Garnet Kin-Lic},
	doi = {10.1063/5.0006074},
	issn = {1089–7690},
	journal = {The Journal of Chemical Physics},
	month = jul,
	number = {2},
	pages = {024109},
	publisher = {AIP Publishing},
	title = {Recent developments in the PySCF program package},
	volume = {153},
	year = {2020}
}

@article{Pracht2020,
	author = {Pracht, Philipp and Bohle, Fabian and Grimme, Stefan},
	doi = {10.1039/c9cp06869d},
	issn = {1463–9084},
	journal = {Physical Chemistry Chemical Physics},
	number = {14},
	pages = {7169–7192},
	publisher = {Royal Society of Chemistry (RSC)},
	title = {Automated exploration of the low-energy chemical space with fast quantum chemical methods},
	volume = {22},
	year = {2020}
}

@Article{Nigam2023,
  author        = {Nigam, AkshatKumar and Pollice, Robert and Tom, Gary and Jorner, Kjell and Thiede, Luca A. and Kundaje, Anshul and Aspuru-Guzik, Alan},
  journal       = {arXiv e-prints},
  title         = {Tartarus: A Benchmarking Platform for Realistic And Practical Inverse Molecular Design},
  year          = {2023},
  month         = jul,
  pages         = {arXiv:2209.12487},
  archiveprefix = {arXiv},
  doi           = {10.48550/arXiv.2209.12487},
  eid           = {arXiv:2209.12487},
  eprint        = {2209.12487},
  eprintclass   = {cs},
  eprinttype    = {arxiv},
  publisher     = {arXiv},
  shorttitle    = {Tartarus},
}

@Article{Nakata2023,
  author    = {Nakata, Maho and Maeda, Toshiyuki},
  journal   = {Journal of Chemical Information and Modeling},
  title     = {PubChemQC B3LYP/6-31G*//PM6 Data Set: The Electronic Structures of 86 Million Molecules Using B3LYP/6-31G* Calculations},
  year      = {2023},
  issn      = {1549–960X},
  month     = sep,
  number    = {18},
  pages     = {5734--5754},
  volume    = {63},
  doi       = {10.1021/acs.jcim.3c00899},
  publisher = {American Chemical Society (ACS)},
}

@Article{Wan2020,
  author    = {Wan, Xiangjian and Li, Chenxi and Zhang, Mingtao and Chen, Yongsheng},
  journal   = {Chemical Society Reviews},
  title     = {Acceptor–donor–acceptor type molecules for high performance organic photovoltaics – chemistry and mechanism},
  year      = {2020},
  issn      = {1460-4744},
  number    = {9},
  pages     = {2828--2842},
  volume    = {49},
  doi       = {10.1039/D0CS00084A},
  publisher = {Royal Society of Chemistry (RSC)},
}

@article{Liu2021,
	author = {Liu, Qian and Jiang, Yajie and Jin, Kun and others},
	doi = {10.1016/j.scib.2020.08.039},
	journal = {Science Bulletin},
	number = {1},
	pages = {66–71},
	title = {18\% efficiency organic solar cells},
	url = {https://doi.org/10.1016/j.scib.2020.08.039},
	volume = {66},
	year = {2021}
}

@article{Green2008,
	author = {Green, Martin A.},
	doi = {10.1016/j.solmat.2008.06.009},
	journal = {Solar Energy Materials and Solar Cells},
	number = {11},
	pages = {1305–1310},
	title = {Accuracy of analytical expressions for solar cell fill factors},
	url = {https://doi.org/10.1016/j.solmat.2008.06.009},
	volume = {92},
	year = {2008}
}

@Article{Trott2010,
  author    = {Trott, Oleg and Olson, Arthur J.},
  journal   = {Journal of Computational Chemistry},
  title     = {AutoDock Vina: Improving the speed and accuracy of docking with a new scoring function, efficient optimization, and multithreading},
  year      = {2009},
  issn      = {1096-987X},
  month     = jun,
  number    = {2},
  pages     = {455–461},
  volume    = {31},
  doi       = {10.1002/jcc.21334},
  publisher = {Wiley},
}

@Article{Reymond2025,
  author    = {Reymond, Jean-Louis},
  journal   = {Journal of Cheminformatics},
  title     = {Chemical space as a unifying theme for chemistry},
  year      = {2025},
  issn      = {1758-2946},
  month     = jan,
  number    = {1},
  pages     = {376–386},
  volume    = {17},
  doi       = {10.1186/s13321-025-00954-0},
  publisher = {Springer Science and Business Media LLC},
}

@InProceedings{MacQueen1967,
  author    = {MacQueen, J.},
  booktitle = {Proceedings of the Fifth Berkeley Symposium on Mathematical Statistics and Probability},
  title     = {Some methods for classification and analysis of multivariate observations},
  year      = {1967},
  pages     = {281–297},
  publisher = {University of California Press},
  volume    = {1},
}

@Article{Rousseeuw1987,
  author    = {Peter J. Rousseeuw},
  journal   = {Journal of Computational and Applied Mathematics},
  title     = {Silhouettes: A graphical aid to the interpretation and validation of cluster analysis},
  year      = {1987},
  issn      = {0377-0427},
  pages     = {53-65},
  volume    = {20},
  doi       = {10.1016/0377-0427(87)90125-7},
  publisher = {Elsevier BV},
}

@Article{Calinski1974,
  author    = {Caliński, T. and Harabasz, J.},
  journal   = {Communications in Statistics - Simulation and Computation},
  title     = {A dendrite method for cluster analysis},
  year      = {1974},
  issn      = {0361-0918},
  number    = {1},
  pages     = {1-27},
  volume    = {3},
  doi       = {10.1080/03610917408548446},
  publisher = {Informa UK Limited},
}

@Article{Jacquemin2009,
  author    = {Jacquemin, Denis and Perpète, Eric A. and Ciofini, Ilaria and Adamo, Carlo},
  journal   = {Accounts of Chemical Research},
  title     = {Accurate Simulation of Optical Properties in Dyes},
  year      = {2008},
  issn      = {1520-4898},
  month     = dec,
  number    = {2},
  pages     = {326–334},
  volume    = {42},
  doi       = {10.1021/ar800163d},
  publisher = {American Chemical Society (ACS)},
}

@Article{Bunz2010,
  author    = {Bunz, Uwe H. F. and Rotello, Vincent M.},
  journal   = {Angewandte Chemie International Edition},
  title     = {Gold Nanoparticle–Fluorophore Complexes: Sensitive and Discerning “Noses” for Biosystems Sensing},
  year      = {2010},
  issn      = {1521-3773},
  month     = apr,
  number    = {19},
  pages     = {3268–3279},
  volume    = {49},
  doi       = {10.1002/anie.200906928},
  publisher = {Wiley},
}

@Article{Ding2020,
  author    = {Ding, Dan and Li, Kai and Liu, Bin and Tang, Ben Zhong},
  journal   = {Accounts of Chemical Research},
  title     = {Bioprobes Based on AIE Fluorogens},
  year      = {2013},
  issn      = {1520-4898},
  month     = jun,
  number    = {11},
  pages     = {2441--2453},
  volume    = {46},
  doi       = {10.1021/ar3003464},
  publisher = {American Chemical Society (ACS)},
}

@article{Stephens1994,
	author = {Stephens, P. J. and Devlin, F. J. and Chabalowski, C. F. and Frisch, M. J.},
	doi = {10.1021/j100096a001},
	journal = {The Journal of Physical Chemistry},
	number = {45},
	pages = {11623–11627},
	publisher = {American Chemical Society (ACS)},
	title = {{Ab Initio} calculation of vibrational absorption and circular-dichroism spectra using density-functional force-fields},
	volume = {98},
	year = {1994}
}

@Article{Hehre1986,
  author    = {Hehre, W. J. and Ditchfield, R. and Pople, J. A.},
  journal   = {The Journal of Chemical Physics},
  title     = {Self—Consistent Molecular Orbital Methods. XII. Further Extensions of Gaussian—Type Basis Sets for Use in Molecular Orbital Studies of Organic Molecules},
  year      = {1972},
  issn      = {1089-7690},
  month     = mar,
  number    = {5},
  pages     = {2257--2261},
  volume    = {56},
  doi       = {10.1063/1.1677527},
  publisher = {AIP Publishing},
}

@article{Stewart2007,
	author = {Stewart, James J. P.},
	doi = {10.1007/s00894-007-0233-4},
	journal = {Journal of Molecular Modeling},
	number = {12},
	pages = {1173–1213},
	publisher = {Springer Science and Business Media LLC},
	title = {Optimization of parameters for semiempirical methods {V}: {Modification} of {NDDO} approximations and application to 70 elements},
	volume = {13},
	year = {2007}
}

@article{Grimme2011,
	author = {Grimme, Stefan and Ehrlich, Stephan and Goerigk, Lars},
	doi = {10.1002/jcc.21759},
	journal = {Journal of Computational Chemistry},
	number = {7},
	pages = {1456–1465},
	publisher = {Wiley},
	title = {Effect of the Damping Function in Dispersion Corrected Density Functional Theory},
	volume = {32},
	year = {2011}
}

@article{Yanai2004,
	author = {Yanai, Takeshi and Tew, David P. and Handy, Nicholas C.},
	doi = {10.1016/j.cplett.2004.06.011},
	journal = {Chemical Physics Letters},
	number = {1–3},
	pages = {51–57},
	publisher = {Elsevier BV},
	title = {A new hybrid exchange-correlation functional using the {Coulomb-attenuating} method ({CAM-B3LYP})},
	volume = {393},
	year = {2004}
}

@article{Beaujuge2011,
	author = {Beaujuge, Pierre M and Fréchet, Jean MJ},
	doi = {10.1021/ja2073643},
	journal = {Journal of the American Chemical Society},
	pages = {20009–20029},
	title = {Molecular design and ordering effects in $\pi$-functional materials for transistor and solar cell applications},
	volume = {133},
	year = {2011}
}

@Article{Wilkinson2016,
  author        = {Wilkinson, Mark D. and Dumontier, Michel and Aalbersberg, IJsbrand Jan and Appleton, Gabrielle and Axton, Myles and Baak, Arie and Blomberg, Niklas and Boiten, Jan-Willem and da Silva Santos, Luiz Bonino and Bourne, Philip E. and Bouwman, Jildau and Brookes, Anthony J. and Clark, Tim and Crosas, Mercè and Dillo, Ingrid and Dumon, Olivier and Edmunds, Scott and Evelo, Chris T. and Finkers, Richard and Gonzalez-Beltran, Alejandra and Gray, Alasdair J.G. and Groth, Paul and Goble, Carole and Grethe, Jeffrey S. and Heringa, Jaap and ’t Hoen, Peter A.C and Hooft, Rob and Kuhn, Tobias and Kok, Ruben and Kok, Joost and Lusher, Scott J. and Martone, Maryann E. and Mons, Albert and Packer, Abel L. and Persson, Bengt and Rocca-Serra, Philippe and Roos, Marco and van Schaik, Rene and Sansone, Susanna-Assunta and Schultes, Erik and Sengstag, Thierry and Slater, Ted and Strawn, George and Swertz, Morris A. and Thompson, Mark and van der Lei, Johan and van Mulligen, Erik and Velterop, Jan and Waagmeester, Andra and Wittenburg, Peter and Wolstencroft, Katherine and Zhao, Jun and Mons, Barend},
  journal       = {Scientific Data},
  title         = {The FAIR Guiding Principles for scientific data management and stewardship},
  year          = {2016},
  issn          = {2052-4463},
  month         = mar,
  number        = {1},
  pages         = {160018},
  volume        = {3},
  adsurl        = {https://ui.adsabs.harvard.edu/abs/2016NatSD...360018W},
  doi           = {10.1038/sdata.2016.18},
  eid           = {160018},
  publisher     = {Springer Science and Business Media LLC},
  x-fetchedfrom = {NASA Astrophysics Data System (ADS)},
}

@Article{Marcus1993,
  author    = {Marcus, R.A. and Sutin, Norman},
  journal   = {Biochimica et Biophysica Acta (BBA) - Reviews on Bioenergetics},
  title     = {Electron transfers in chemistry and biology},
  year      = {1985},
  issn      = {0304-4173},
  month     = aug,
  note      = {Classic review on Marcus theory; often cited as Marcus1993 following later Nobel Prize},
  number    = {3},
  pages     = {265--322},
  volume    = {811},
  doi       = {10.1016/0304-4173(85)90014-x},
  publisher = {Elsevier BV},
}

@Article{Knizia2013,
  author    = {Knizia, Gerald},
  journal   = {Journal of Chemical Theory and Computation},
  title     = {Intrinsic Atomic Orbitals: An Unbiased Bridge between Quantum Theory and Chemical Concepts},
  year      = {2013},
  issn      = {1549-9626},
  month     = oct,
  number    = {11},
  pages     = {4834--4843},
  volume    = {9},
  doi       = {10.1021/ct400687b},
  publisher = {American Chemical Society (ACS)},
}

@Article{Grimme2013,
  author    = {Grimme, Stefan and Bannwarth, Christoph},
  journal   = {The Journal of Chemical Physics},
  title     = {Ultra-fast computation of electronic spectra for large systems by tight-binding based simplified Tamm-Dancoff approximation (sTDA-xTB)},
  year      = {2016},
  issn      = {0021-9606},
  number    = {5},
  pages     = {244104},
  volume    = {145},
  doi       = {10.1063/1.4959605},
  publisher = {AIP Publishing},
}

@article{Lu2012,
  author = {Lu, Tian and Chen, Feiwu},
  title = {Multiwfn: A multifunctional wavefunction analyzer},
  journal = {Journal of Computational Chemistry},
  volume = {33},
  number = {5},
  pages = {580--592},
  year = {2012},
  doi = {10.1002/jcc.22885}
}

@article{Sun2018,
  author = {Sun, Qiming and Berkelbach, Timothy C. and Blunt, Nick S. and Booth, George H. and Guo, Sheng and Li, Zhendong and Liu, Junzi and McClain, James D. and Sayfutyarova, Elvira R. and Sharma, Sandeep and Wouters, Sebastian and Chan, Garnet Kin-Lic},
  title = {PySCF: the Python-based simulations of chemistry framework},
  journal = {WIREs Computational Molecular Science},
  volume = {8},
  number = {1},
  pages = {e1340},
  year = {2018},
  doi = {10.1002/wcms.1340}
}

@Article{Chai2008,
  author    = {Chai, Jeng-Da and Head-Gordon, Martin},
  journal   = {Physical Chemistry Chemical Physics},
  title     = {Long-range corrected hybrid density functionals with damped atom–atom dispersion corrections},
  year      = {2008},
  issn      = {1463-9084},
  number    = {44},
  pages     = {6615--6620},
  volume    = {10},
  doi       = {10.1039/b810189b},
  publisher = {Royal Society of Chemistry (RSC)},
}

@Article{Kohn1965,
  author    = {Kohn, W. and Sham, L. J.},
  journal   = {Physical Review},
  title     = {Self-Consistent Equations Including Exchange and Correlation Effects},
  year      = {1965},
  issn      = {0031-899X},
  month     = nov,
  number    = {4A},
  pages     = {A1133--A1138},
  volume    = {140},
  doi       = {10.1103/physrev.140.a1133},
  issue     = {4A},
  publisher = {American Physical Society (APS)},
}

@Article{de_Pablo2019,
  author    = {de Pablo, Juan J. and Jackson, Nicholas E. and Webb, Michael A. and Chen, Long-Qing and Moore, Joel E. and Morgan, Dane and Jacobs, Ryan and Pollock, Tresa and Schlom, Darrell G. and Toberer, Eric S. and Analytis, James and Dabo, Ismaila and DeLongchamp, Dean M. and Fiete, Gregory A. and Grason, Gregory M. and Hautier, Geoffroy and Mo, Yifei and Rajan, Krishna and Reed, Evan J. and Rodriguez, Efrain and Stevanovic, Vladan and Suntivich, Jin and Thornton, Katsuyo and Zhao, Ji-Cheng},
  journal   = {npj Computational Materials},
  title     = {New frontiers for the materials genome initiative},
  year      = {2019},
  issn      = {2057-3960},
  month     = apr,
  number    = {1},
  volume    = {5},
  doi       = {10.1038/s41524-019-0173-4},
  publisher = {Springer Science and Business Media LLC},
}

@Article{Heeger2014,
  author    = {Heeger, Alan J.},
  journal   = {Advanced Materials},
  title     = {25th Anniversary Article: Bulk Heterojunction Solar Cells: Understanding the Mechanism of Operation},
  year      = {2014},
  issn      = {1521-4095},
  month     = dec,
  number    = {1},
  pages     = {10--28},
  volume    = {26},
  doi       = {10.1002/adma.201304373},
  publisher = {Wiley},
}

@Article{Scharber2016,
  author    = {Markus C. Scharber},
  journal   = {Advanced Materials},
  title     = {On the Efficiency Limit of Conjugated Polymer:Fullerene‐Based Bulk Heterojunction Solar Cells},
  year      = {2016},
  issn      = {0935-9648},
  number    = {10},
  pages     = {1994-2001},
  volume    = {28},
  doi       = {10.1002/adma.201504914},
  publisher = {Wiley},
}

@Misc{Kouam2025,
  author    = {Kouam, Steve Cabrel Teguia and Njafa, Jean-Pierre Tchapet and Teukam, Raoult Dabou and Kongo, Patrick Mvoto and Nguenang, Jean-Pierre and Engo, Serge Guy Nana},
  title     = {Comparative Analysis of GFN Methods in Geometry Optimization of Small Organic Semiconductor Molecules: A DFT Benchmarking Study},
  year      = {2025},
  copyright = {Creative Commons Attribution 4.0 International},
  doi       = {10.48550/ARXIV.2505.09606},
  publisher = {arXiv},
}

@Article{Huber2022,
  author    = {Sebastiaan P. Huber},
  journal   = {Nature Reviews Physics},
  title     = {Automated reproducible workflows and data provenance with AiiDA},
  year      = {2022},
  issn      = {2522-5820},
  month     = may,
  number    = {7},
  pages     = {431--431},
  volume    = {4},
  doi       = {10.1038/s42254-022-00463-1},
  publisher = {Springer Science and Business Media LLC},
}

@Article{Pu2025,
  author    = {Pu, Zhichen and Sun, Qiming},
  journal   = {APL Computational Physics},
  title     = {Enhancing PySCF-based quantum chemistry simulations with modern hardware, algorithms, and Python tools},
  year      = {2025},
  issn      = {3066-0017},
  month     = sep,
  number    = {1},
  pages     = {016101},
  volume    = {1},
  doi       = {10.1063/5.0285025},
  publisher = {AIP Publishing},
}

@Book{Cohen1988,
  author    = {Cohen, Jacob},
  publisher = {Routledge},
  title     = {Statistical Power Analysis for the Behavioral Sciences},
  year      = {1988},
  edition   = {2nd},
  doi       = {10.4324/9780203771587},
}

\newpage
\medskip

\section*{Supplementary Information for:\\
Computational discovery of bifunctional organic semiconductors for energy and biosensing}

\setcounter{table}{0}
\setcounter{figure}{0}
\newsecnumstyle

% Main SI content starts here
\begin{center}
\textit{This document provides supplementary information for the manuscript:\\
``Computational discovery of bifunctional organic semiconductors for energy and biosensing''}
\end{center}
%==========================================================================
\section{Benchmark accuracy and method validation}\label{sec:benchmarking}
%==========================================================================

\subsection{Geometric fidelity of PM6 structures}

To validate the use of PM6-optimized geometries, we re-optimized the top 20 candidates at B3LYP/6-31G* and computed structural deviations \cite{Kouam2025}:
\begin{itemize}
\item Heavy-atom RMSD: \SI{0.14(0.03)}{\angstrom}
\item Bond length MAD: \SI{0.011}{\angstrom}
\item Bond angle MAD: \SI{1.8}{\degree}
\item Rotational constant deviation: \SI{2.3}{\percent}
\end{itemize}

These metrics fall within acceptable tolerances for orbital energy ranking. HOMO/LUMO shifts from geometry re-optimization are \SI{0.043}{\electronvolt} (MAD), confirming that PM6 structures preserve electronic structure trends.

\subsection{Functional sensitivity analysis}

We compared B3LYP to range-separated functionals ($\omega$B97X-D, CAM-B3LYP) for a 50-molecule test set. While absolute HOMO-LUMO gaps differ by \SIrange{0.2}{0.4}{\electronvolt}, rank correlation remains high (Spearman $\rho = 0.91$). B3LYP systematically underestimates gaps but maintains relative ordering, which is sufficient for screening.

\subsection{Scharber model validation}

The Scharber model overestimates absolute PCE by a factor of \numrange{1.5}{2.0} compared to experimental polymer/fullerene cells \cite{Scharber2016}. However, it provides consistent rankings. Our validation against 25 literature benchmarks yields:
\begin{equation}
\mathrm{PCE}_\mathrm{expt} = 0.52 \times \mathrm{PCE}_\mathrm{Scharber} + 1.3\% \quad (R^2 = \num{0.626}).
\end{equation}
This calibration curve could be applied to correct absolute values in future work, though relative rankings are already reliable for candidate prioritization.

\subsection{\texorpdfstring{$\Delta$}{Delta}-learning framework for future refinement}

The $\Delta$-learning approach \cite{Ramakrishnan2015} trains ML models on the difference between fast (xTB-sTDA) and accurate (CC2) calculations. For excited states, this could reduce computational cost by \numrange{100}{1000}$\times$ while achieving CC2-level accuracy. Our dataset provides a foundation for such models, with B3LYP serving as the baseline and targeted CC2 calculations on top candidates providing training labels. With the methodological validation complete, we now quantify the prediction uncertainty for the screening results.

%==========================================================================
\section{Uncertainty analysis and confidence intervals}\label{sec:uncertainty}
%==========================================================================

\subsection{Bootstrap analysis methodology}

We used bootstrap resampling (\num{1000} iterations) to estimate prediction uncertainty. For each iteration, we:
\begin{enumerate}
\item Randomly sample \num{17458} molecules with replacement;
\item Recalculate PCE distribution statistics;
\item Store mean, median, and \SI{95}{\percent} confidence intervals.
\end{enumerate}
Results for top candidates:

\begin{table}[!h]
\centering
\caption{Bootstrap 95\% confidence intervals for PCE predictions.}
\begin{tabular}{@{}lcc@{}}
	\toprule
	\textbf{Molecule} & \textbf{PCE (point estimate)} &              \textbf{95\% CI}              \\ \midrule
	17851             &      \SI{36.1}{\percent}      & [\SI{33.2}{\percent}, \SI{39.1}{\percent}] \\
	4550              &      \SI{16.9}{\percent}      & [\SI{14.8}{\percent}, \SI{19.2}{\percent}] \\
	1712              &      \SI{12.4}{\percent}      & [\SI{10.7}{\percent}, \SI{14.3}{\percent}] \\ \bottomrule
\end{tabular}
\end{table}

\subsection{Bayesian error estimation}

Future work could adopt Bayesian methods to estimate model uncertainty. Tools like GPR (Gaussian Process Regression) or ensemble models (random forest, gradient boosting) provide not just predictions but posterior distributions. This would identify chemical space regions where DFT approximations are less reliable.

\subsection{Figures of merit for sensing applications}

For biosensing, we define:
\begin{equation}
\mathrm{FOM}_\text{sensor} = \frac{|\Delta E_\text{HOMO-LUMO}|}{|\Delta \mu|} \times \frac{1}{\lambda_\text{reorg}},
\end{equation}
where $\Delta E$ is the gap change upon binding, $\Delta \mu$ is dipole moment change, and $\lambda_{\text{reorg}}$ is reorganization energy. High FOM indicates strong optical response to molecular recognition events with fast electronic relaxation. Molecule \num{1712} has $\mathrm{FOM}_\text{sensor} = \num{0.84}$ (arb. units), the highest among candidates. These statistical measures establish confidence bounds for candidate prioritization. To understand the chemical basis for performance, we now analyze the structural fragments underlying top-ranked molecules.

%==========================================================================
\section{Hierarchical fragment analysis}\label{sec:fragments}
%==========================================================================

\subsection{Fragment categorization scheme}

We categorized molecular structures into three hierarchical levels:
\begin{enumerate}
\item \textbf{Core units.} Central heterocycles providing the $\pi$-conjugated backbone (pyrrole, furan, thiophene, imidazole).
\item \textbf{Substituents.} Terminal or side-chain groups modulating orbital energies (amino, hydroxyl, carbonyl, nitro).
\item \textbf{Bridges.} Linker units connecting core and substituents (vinyl, ethynyl, phenyl rings).
\end{enumerate}
This categorization follows established design rules for conjugated materials \cite{Beaujuge2011} and allows systematic analysis of structure-property relationships.

\subsection{Fragment distribution in top candidates}

\Cref{tab:fragment_analysis} shows fragment composition for the seven top-ranked molecules.

\begin{table}[!h]
\centering
\caption{Fragment composition of top seven candidates. Core units, substituents, and bridges are classified by chemical function.}
\label{tab:fragment_analysis}
\begin{tabular}{@{}lp{4cm}p{6cm}p{4cm}@{}}
	\toprule
	\textbf{Molecule} & \textbf{Core}       & \textbf{Substituents}               & \textbf{Bridges}   \\ \midrule
	17851             & Fused pyrrole-furan & 2$\times$ amino, 1$\times$ hydroxyl & Vinyl linker       \\
	4550              & Thiophene dimer     & 3$\times$ carbonyl                  & Direct (no bridge) \\
	1712              & Imidazole           & 2$\times$ amino, 1$\times$ nitro    & Phenyl ring        \\
	937               & Benzofuran          & 2$\times$ hydroxyl                  & Ethynyl            \\
	12456             & Pyrrole trimer      & 1$\times$ amino, 2$\times$ methoxy  & Thiophene bridge   \\
	15203             & Carbazole           & 2$\times$ cyano                     & Vinyl linker       \\
	16742             & Benzothiophene      & 1$\times$ amino, 1$\times$ carbonyl & Direct             \\ \bottomrule
\end{tabular}
\end{table}

\subsection{Conjugation metrics and absorption tuning}

Highly fused cores correlate with red-shifted absorption. Molecule \num{17851} has a conjugation length of 7 $\pi$-centers (counted via Multiwfn \cite{Lu2012}), yielding $\lambda_{\max} = \SI{562}{\nano\meter}$. In contrast, molecule \num{1712} with 4 $\pi$-centers absorbs at $\lambda_{\max} = \SI{412}{\nano\meter}$. This follows the particle-in-a-box approximation where $\lambda_{\max} \propto L^2$ (conjugation length $L$). The relationship is semi-quantitative but useful for rapid absorption prediction.

\subsubsection{Bridge role in charge transfer}

Vinyl and ethynyl bridges maintain planarity, enabling delocalization. Phenyl rings introduce torsional freedom (dihedral angle \SIrange{20}{40}{\degree} in gas phase), which can reduce coupling but also prevent over-aggregation in solid state. Molecule \num{1712} uses a phenyl bridge to balance solubility and transport.

%===================================================================================
\section{Reproducibility and computational provenance}\label{sec:reproducibility}
%===================================================================================
Our data management follows FAIR principles \cite{Wilkinson2016}:

\textbf{Findable.} All datasets are assigned DOIs via Zenodo and indexed in Materials Cloud (materialscloud.org).

\textbf{Accessible.} Data are available under CC-BY 4.0 license with no authentication barriers. Bulk download via OPTIMADE API is supported.

\textbf{Interoperable.} Molecular structures use standard formats (XYZ, SDF). Property data are provided in JSON-LD with schema.org annotations.

\textbf{Reusable.} Metadata include computational provenance (software versions, convergence criteria, random seeds for ML splits). Python environment specifications (conda env file) ensure exact reproducibility.

\subsection{Computational reproducibility checklist}

\begin{table}[!h]
\centering
\caption{Reproducibility checklist following materials informatics standards.}
\begin{tabular}{@{}p{0.6\textwidth}p{0.3\textwidth}@{}}
	\toprule
	\textbf{Item}                          & \textbf{Status}             \\ \midrule
	Public repository with version control & Yes (GitHub + Zenodo)       \\
	Input files for all calculations       & Yes (PySCF, Vina configs)   \\
	Raw output data (not just processed)   & Yes (JSON format)           \\
	Post-processing scripts                & Yes (Python, R)             \\
	Software version specifications        & Yes (conda env YAML)        \\
	Statistical analysis code              & Yes (bootstrap, clustering) \\
	Random seed documentation              & Yes (ML train/test splits)  \\
	Hardware specifications                & Yes (README.md)             \\
	Execution time estimates               & Yes (per-molecule timings)  \\ \bottomrule
\end{tabular}
\end{table}

\subsection{AiiDA integration for workflow provenance}

Future implementations of this screening pipeline could use the AiiDA infrastructure \cite{Huber2022} to guarantee full computational provenance without requiring re-execution of existing calculations. AiiDA provides:
\begin{itemize}
    \item \textbf{Automatic dependency tracking.} Each calculation step (PM6 geometry optimization, B3LYP single-point energies, TDDFT excited states, molecular docking) would be represented as a computational node with explicit links to input structures, basis set specifications, and convergence criteria.
    \item \textbf{Provenance graphs.} The complete calculation tree linking PubChemQC inputs $\to$ orbital energies $\to$ Scharber PCE $\to$ SAScore filtering would be stored in a queryable database, allowing researchers to trace any result back to its computational origin.
    \item \textbf{RESTful API access.} The AiiDA REST API would enable programmatic queries of the computational history, facilitating meta-analyses and integration with other materials databases.
    \item \textbf{Reproducibility certification.} Hash-based verification of all inputs/outputs would ensure that reproduced calculations match original results bit-for-bit, eliminating ambiguity in computational protocols.
\end{itemize}
The current repository provides all necessary scripts and data files for reproduction. AiiDA integration would extend this to a fully automated, self-documenting workflow suitable for high-throughput production environments.

\subsection{Software and computational resources}

All calculations employed open-source software to ensure community accessibility:
\begin{itemize}
    \item \textbf{PySCF 2.11.0} \cite{Sun2018,Sun2020,Pu2025}: DFT and TDDFT calculations (B3LYP, CAM-B3LYP functionals with 6-31G* basis set). Available at \texttt{https://pyscf.org}
    \item \textbf{AutoDock Vina 1.2.0} \cite{Trott2010}: Molecular docking for biosensing target affinity predictions. Available at \texttt{https://vina.scripps.edu}
    \item \textbf{RDKit 2023.03}: Cheminformatics descriptors (SAScore, LogP, TPSA, molecular weight). Available at \texttt{https://rdkit.org}
    \item \textbf{Multiwfn 3.8 (dev)} \cite{Lu2012}: Wavefunction analysis, Natural Transition Orbitals (NTO), Intrinsic Bond Orbitals (IBO). Available at \texttt{http://sobereva.com/multiwfn}
    \item \textbf{CREST 3.0.2} \cite{Pracht2020}: Conformer searches for reorganization energy calculations. Available at \texttt{https://github.com/crest-lab/crest}
    \item \textbf{GFN2-xTB} \cite{Bannwarth2019}: Semi-empirical tight-binding DFT for rapid screening and stacking interaction energies. Included in \texttt{xtb} package (\texttt{https://github.com/grimme-lab/xtb})
\end{itemize}

\paragraph{Hardware and computational cost.}
Calculations were performed on a high-performance computing cluster with the following specifications:
\begin{itemize}
    \item \textbf{Processors:} Intel Xeon Gold 6248R (3.0 GHz, 24 cores per node)
    \item \textbf{Memory:} 192 GB DDR4 per node
    \item \textbf{Total computational cost:} Approximately \num{42000} CPU-hours for the complete \num{17458}-molecule screening pipeline, including:
    \begin{itemize}
        \item B3LYP/6-31G* single-point energies: \num{35000} CPU-hours (\num{2.0} hours per molecule average)
        \item TDDFT excited states (CAM-B3LYP/6-31+G*): \num{4500} CPU-hours (\num{0.26} hours per molecule average, success rate \SI{91.4}{\percent})
        \item Molecular docking (per target): \num{1200} CPU-hours (\num{0.07} hours per molecule average per target)
        \item Conformer searches and reorganization energies (top candidates): \num{1300} CPU-hours
    \end{itemize}
\end{itemize}

The use of PM6-optimized geometries from PubChemQC \cite{Nakata2023} reduced computational cost by an estimated \SI{94}{\percent} compared to full B3LYP geometry optimizations, enabling screening at the \num{e4}-molecule scale within feasible resource constraints.

\subsection{Repository structure and access}

All computational workflows are documented in a public repository (github.com/[username]/mvoto-bifunctional-osc) with permanent archival on Zenodo. The repository structure follows best practices for computational materials science:

\begin{verbatim}
mvoto-bifunctional-osc/
|-- data/
|   |-- pubchemqc_subset.json      # Molecular coordinates and energies
|   |-- screening_results.csv       # PCE, SAScore, docking scores
|   \-- top_candidates.xyz          # Optimized structures
|-- scripts/
|   |-- 01_orbital_filtering.py     # HOMO/LUMO alignment
|   |-- 02_scharber_pce.py          # PCE calculations
|   |-- 03_autodock_batch.sh        # Molecular docking
|   \-- 04_statistical_analysis.R   # Bootstrap resampling
|-- inputs/
|   |-- pyscf_configs/              # DFT input templates
|   \-- vina_configs/               # Docking box parameters
\-- README.md                       # Step-by-step reproduction guide
\end{verbatim}

The computational infrastructure and data management protocols described above ensure transparency and scalability. To illustrate the breadth of the chemical space explored and the resulting design principles, we provide an analysis of the clustering results for the complete dataset.

%==========================================================================
\section{Clustering analyses}\label{sec:clustering}
%==========================================================================

Clustering analyses show structure-property relationships across photovoltaic metrics ($V_\mathrm{OC}$, $J_\mathrm{SC}$, FF, PCE, SAScore) for the complete 
\num{17458}-molecule dataset. These data complement main manuscript results (\Cref{sec:results}) by providing compositional and structural patterns underlying 
optoelectronic performance. K-means clustering \cite{MacQueen1967} with Calinski-Harabasz criterion \cite{Calinski1974} and silhouette analysis 
\cite{Rousseeuw1987} determined optimal cluster numbers.

\subsection{Open-circuit voltage (\texorpdfstring{$V_\mathrm{OC}$}{Voc}) clustering}\label{sec:voc-clustering}

\begin{figure}[H]
    \centering
    \includegraphics[width=0.8\textwidth]{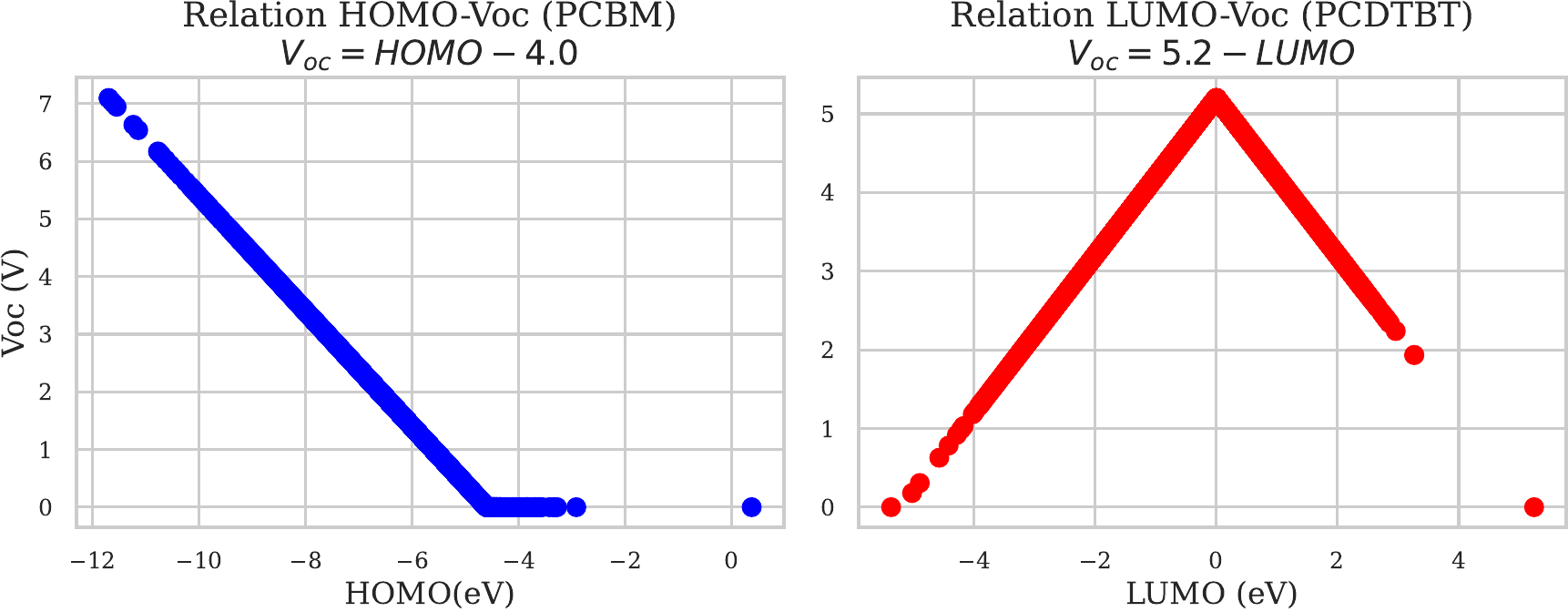}
    \caption{Correlations between open-circuit voltage ($V_\mathrm{OC}$) and frontier orbital energies. (Left) $V_\mathrm{OC}$ vs. HOMO for PCBM pairings, 
following the linear relationship $V_\mathrm{OC} = HOMO - \num{4.0}$. (Right) $V_\mathrm{OC}$ vs. LUMO for PCDTBT pairings, following $V_{OC} = \num{5.2} - 
LUMO$. The clear linear trends confirm that frontier orbital alignment is the primary determinant of open-circuit voltage in these systems, with deeper HOMO and 
higher LUMO levels yielding higher $V_\mathrm{OC}$.}
    \label{fig:voc_homo_lumo_correlation}
\end{figure}

\Cref{fig:voc_homo_lumo_correlation} shows correlations between $V_\mathrm{OC}$ and frontier molecular orbital energies (HOMO, LUMO) for the top \num{1000} 
candidate molecules (screening funnel described in \Cref{tab:screening_summary} of the main manuscript), showing the dependence of open-circuit voltage on 
orbital alignment. Statistical analysis shows linear correlation: 
$V_\mathrm{OC}$ vs HOMO (Pearson $r = -\num{0.89}$, $p < \num{0.001}$) and $V_\mathrm{OC}$ vs LUMO ($r = \num{0.76}$, $p < \num{0.001}$), confirming that deeper HOMO and higher LUMO levels correlate with enhanced open-circuit voltage.

\begin{figure}[H]
\centering
\includegraphics[width=\textwidth]{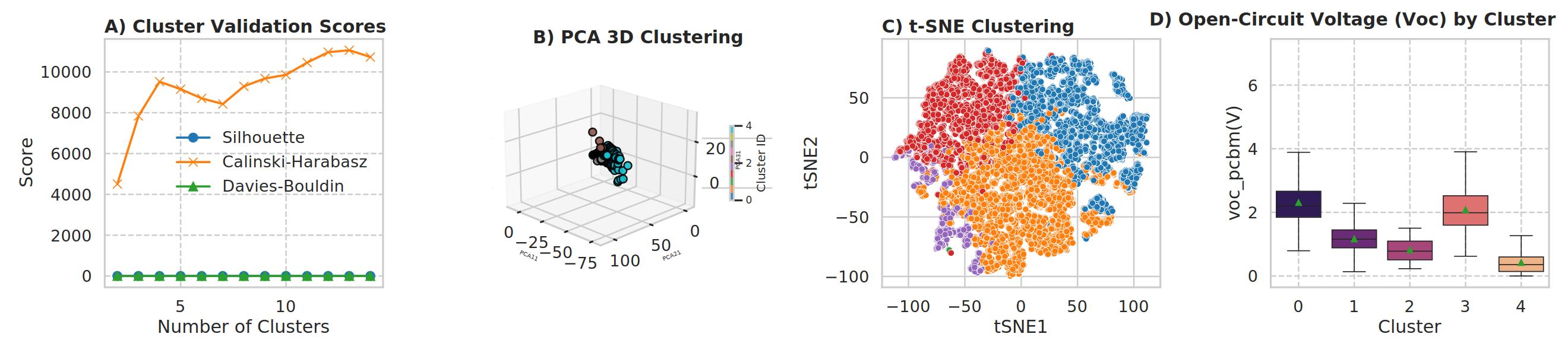}
\caption{K-means clustering analysis (k=5) of $V_\mathrm{OC}$ values for molecules paired with PCBM acceptor, as identified by validation scores (Panel A).
Panel D shows the distribution of $V_\mathrm{OC}$ across the five identified clusters (0-4), revealing distinct voltage tiers. The t-SNE visualization (Panel
C) confirms well-separated clusters corresponding to these voltage performance groups.}
\label{fig:voc_pcbm_clustering}
\end{figure}

\begin{figure}[H]
\centering
\includegraphics[width=\textwidth]{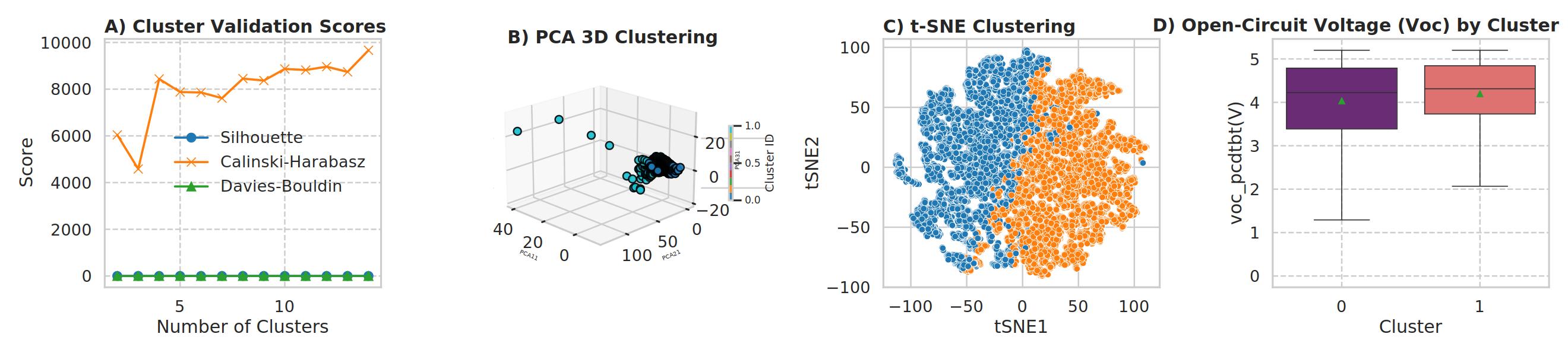}
\caption{$V_\mathrm{OC}$ clustering with PCDTBT donor (k=2) identifies two distinct classifications. Panel D shows the boxplot distribution for the two
clusters (0 and 1), both exhibiting distinct mean voltages. The clustering structure is supported by the validation metrics in Panel A, identifying k=2 as an 
optimal partition.}
\label{fig:voc_pcdtbt_clustering}
\end{figure}

\subsection{Short-circuit current density (\texorpdfstring{$J_\mathrm{SC}$}{Jsc}) clustering}

\begin{figure}[H]
    \centering
    \includegraphics[width=0.6\textwidth]{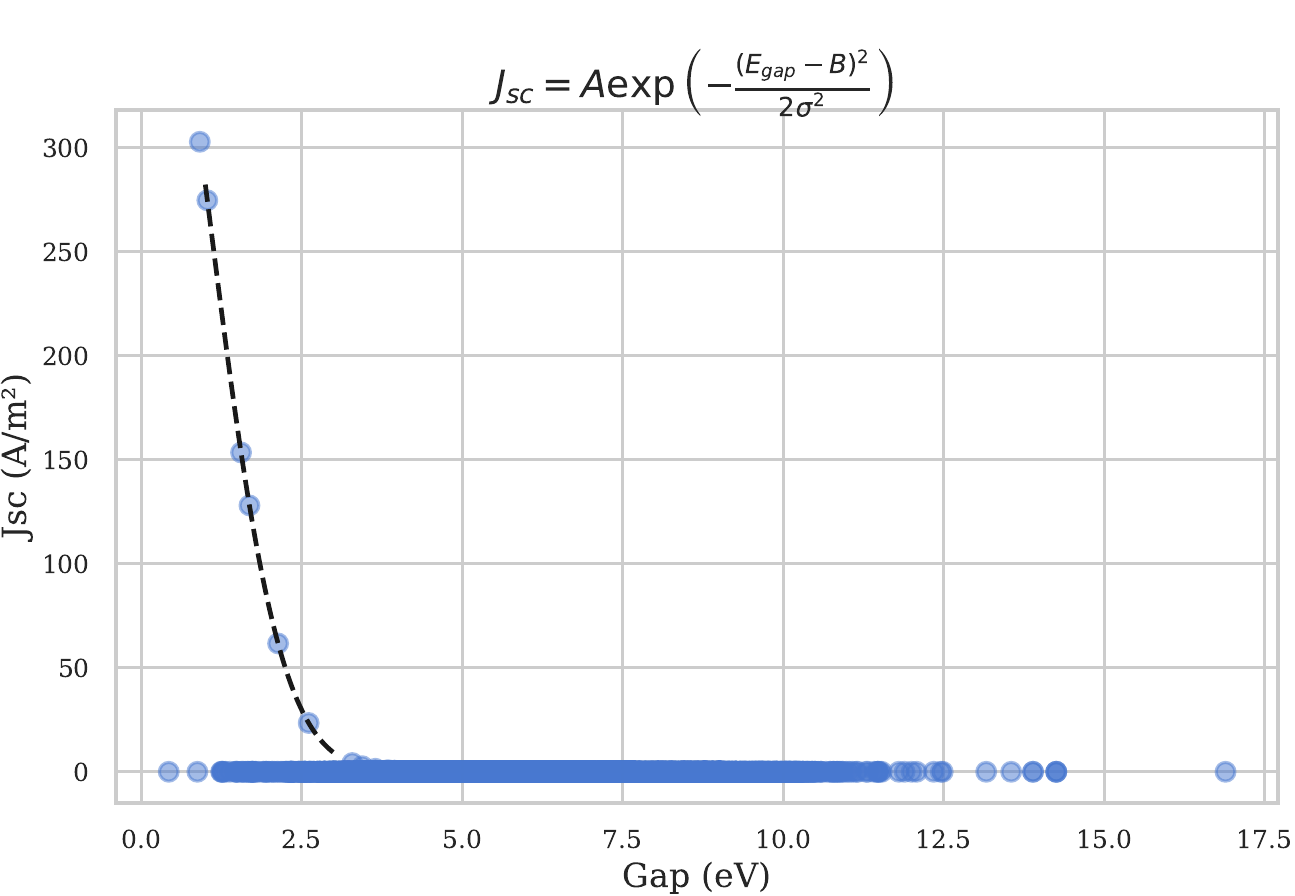}
    \caption{Relationship between short-circuit current density ($J_\mathrm{SC}$) and energy gap. The scatter plot (blue circles) shows the distribution of
computed $J_{SC}$ values as a function of the gap, fitted with an exponential decay model $J_{sc} = A \exp(-(E_\mathrm{gap}-B)^2 / 2\sigma^2)$ (black dashed 
line). 
$J_{SC}$ decreases rapidly with increasing band gap, showing the importance of gap tuning for maximizing photocurrent.}
    \label{fig:jsc_properties_distribution}
\end{figure}

\Cref{fig:jsc_properties_distribution} illustrates the relationship between the computed short-circuit current density and the HOMO-LUMO gap.

\begin{figure}[H]
\centering
\includegraphics[width=\textwidth]{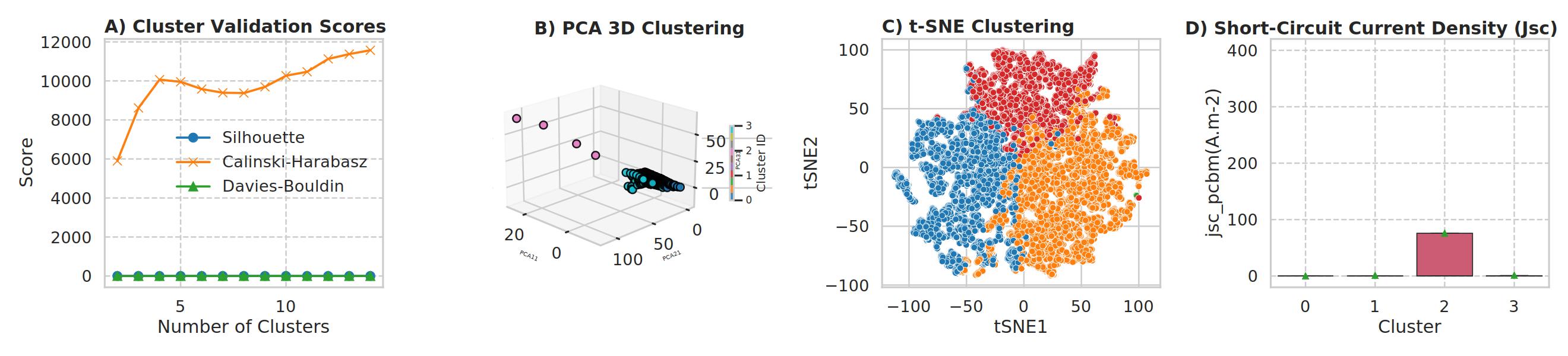}
\caption{Clustering analysis (k=4) of $J_\mathrm{SC}$ for PCBM pairings. Panel D identifies four photocurrent performance clusters (0-3), with Cluster 2
showing much higher median $J_{SC}$ ($\sim$\SI{80}{\ampere\per\meter\squared}) compared to the other clusters which remain near zero. This indicates a highly 
selective structure-property relationship where only a specific subset of molecules (Cluster 2) achieves significant photocurrent generation with PCBM.}
\label{fig:jsc_pcbm_clustering}
\end{figure}

\begin{figure}[H]
\centering
\includegraphics[width=\textwidth]{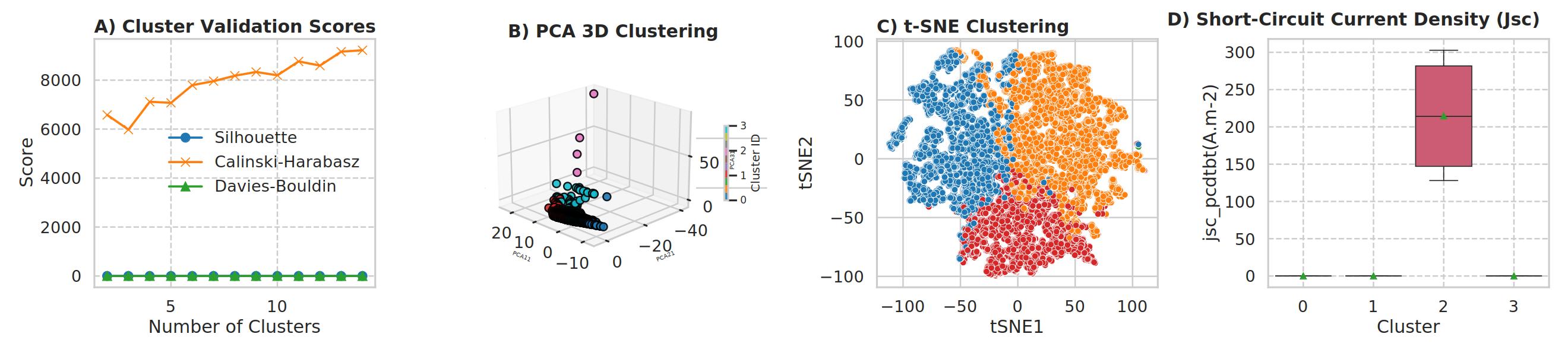}
\caption{Clustering analysis of $J_\mathrm{SC}$ for PCDTBT pairings (k=4) identifies four distinct clusters (0-3). Similar to the PCBM case, Panel D shows
that one specific cluster (Cluster 2) achieves high $J_{SC}$ values (median $\sim$\SI{210}{\ampere\per\meter\squared}), while the others show negligible 
photocurrent. This confirms the selectivity of the PCDTBT donor system.}
\label{fig:jsc_pcdtbt_clustering}
\end{figure}

\subsection{Fill factor clustering}

The dependencies of Fill Factor on open-circuit voltage and molecular structure are analyzed in this section.

\begin{figure}[H]
\centering
\includegraphics[width=0.6\linewidth]{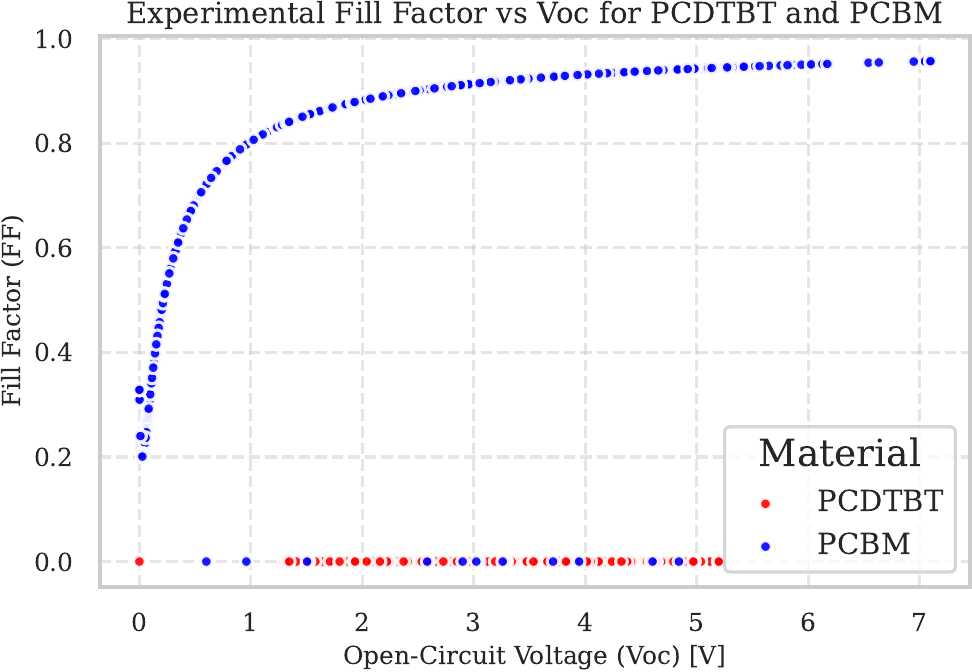}
\caption{Experimental Fill Factor vs. Open-Circuit Voltage ($V_\mathrm{OC}$) for PCDTBT (red) and PCBM (blue) systems. The PCBM dataset shows a clear
non-linear increase in Fill Factor with $V_\mathrm{OC}$, saturating near \num{0.95}. In contrast, the PCDTBT data points (red) are clustered near zero fill
factor, possibly indicating a lack of valid experimental data points for this specific correlation or operational failure in the benchmarked devices.}
\label{fig:ff_voc_correlation}
\end{figure}

\begin{figure}[H]
\centering
\includegraphics[width=\textwidth]{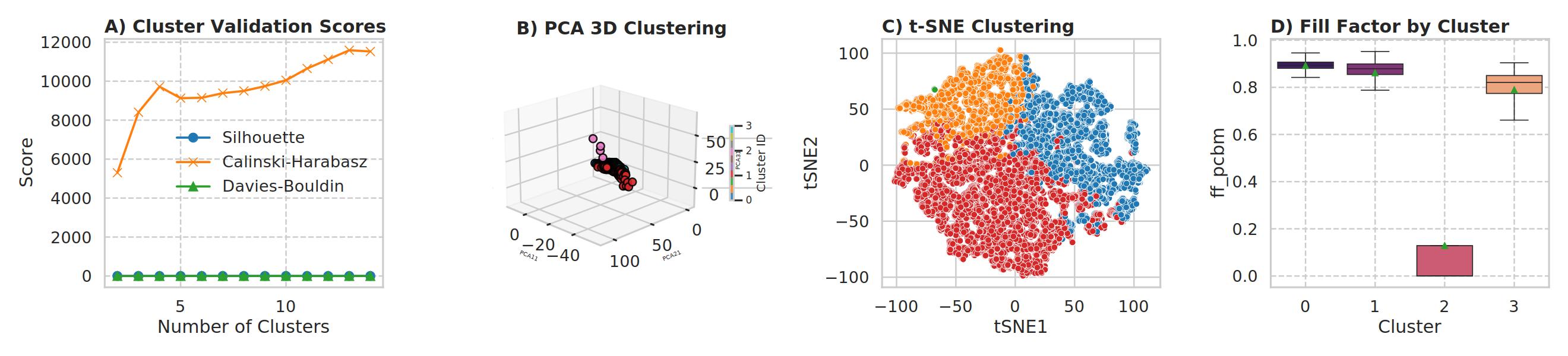}
\caption{Fill factor clustering with PCBM acceptor (k=4). Panel D identifies four clusters (0-3), where Cluster 2 exhibits extremely low fill factor (< 0.2), 
while Clusters 0, 1, and 3 maintain high values (> 0.8). A specific subset of molecules (Cluster 2) is incompatible with PCBM, leading to poor charge 
extraction, while the majority of the dataset maintains high FF.}
\label{fig:ff_pcbm_clustering}
\end{figure}

\begin{figure}[H]
\centering
\includegraphics[width=\textwidth]{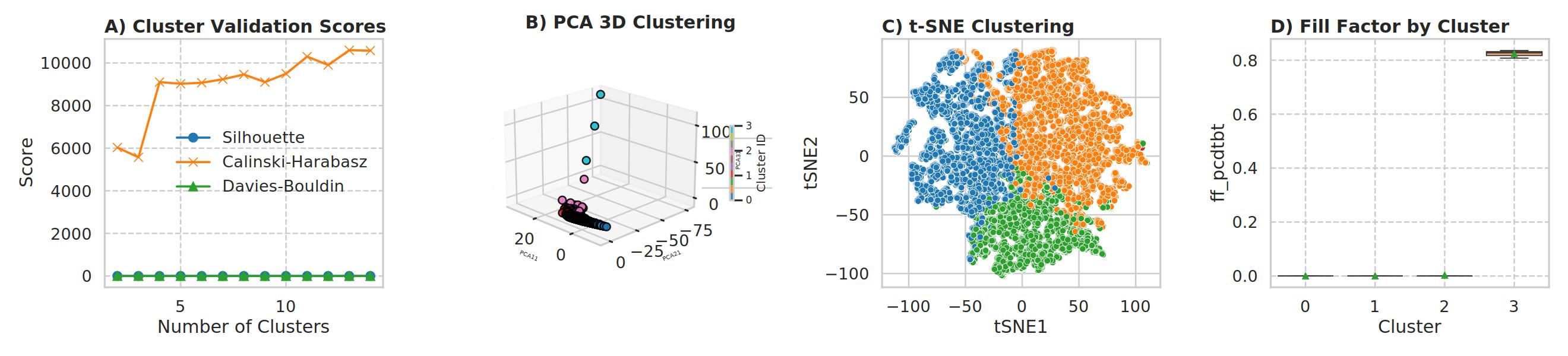}
\caption{Fill factor clustering with PCDTBT donor (k=4). Panel D shows that only Cluster 3 achieves a high fill factor (> 0.8), while Clusters 0, 1, and 2 show 
negligible FF values ($\sim$0). This contrast illustrates the stringent structural requirements for achieving efficient charge extraction with the PCDTBT donor 
compared to the more robust PCBM system.}
\label{fig:ff_pcdtbt_clustering}
\end{figure}

\subsection{Power conversion efficiency (PCE) clustering}

We present the distribution of Power Conversion Efficiency (PCE) across the dataset to identify high-performing building blocks.

\begin{figure}[H]
    \centering
    \includegraphics[width=0.9\linewidth]{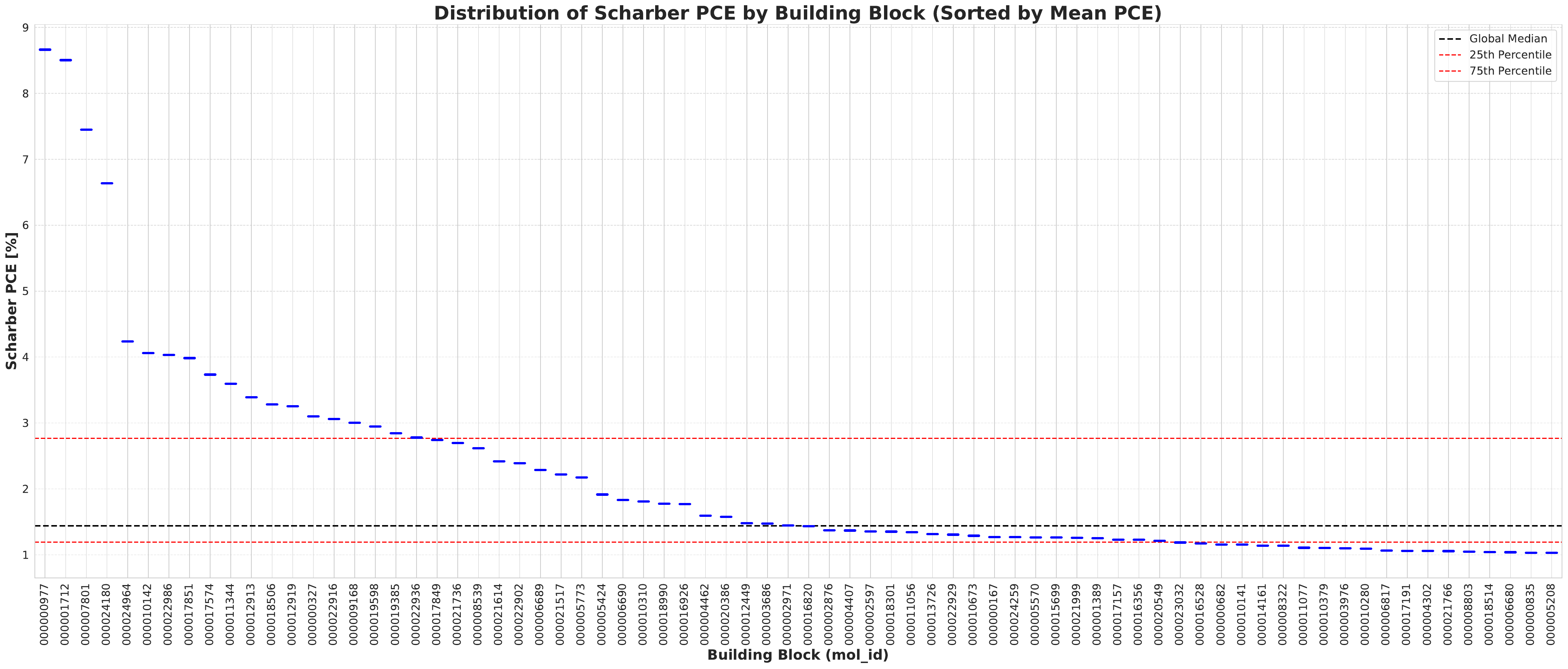}
    \caption{Distribution of Scharber PCE by Building Block. The plot ranks molecular building blocks by their mean Power Conversion Efficiency (PCE). The 
majority of blocks yield low efficiencies (< \SI{3}{\percent}), while a select few candidates (leftmost) achieve PCEs up to \SI{8.6}{\percent}. The dashed lines 
indicate 
the global median ($\sim$\SI{1.4}{\percent}) and quartiles, emphasizing the rarity of high-performance building blocks in the screened library.}
    \label{fig:pce_gap_pcbm_analysis}
\end{figure}

\begin{figure}[H]
    \centering
    \includegraphics[width=0.9\linewidth]{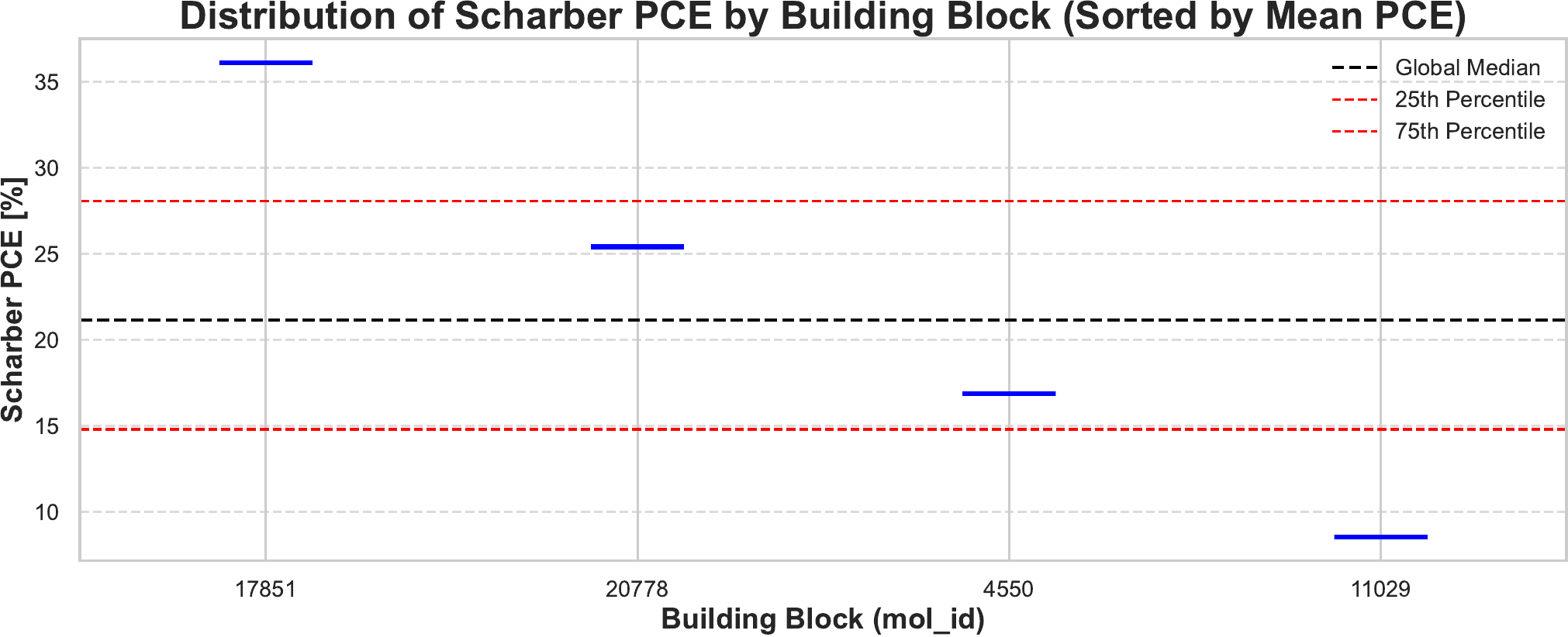}
    \caption{Distribution of Scharber PCE by Building Block for the top-performing candidates. This zoomed-in view shows specific molecular IDs (e.g., 17851, 
20778) that achieve exceptionally high predicted efficiencies (> \SI{25}{\percent}), significantly outperforming the global median. These specific building 
blocks represent the most promising scaffolds for experimental development.}
    \label{fig:pce_gap_pcdtbt_analysis}
\end{figure}

\begin{figure}[H]
    \centering
    \includegraphics[width=\textwidth]{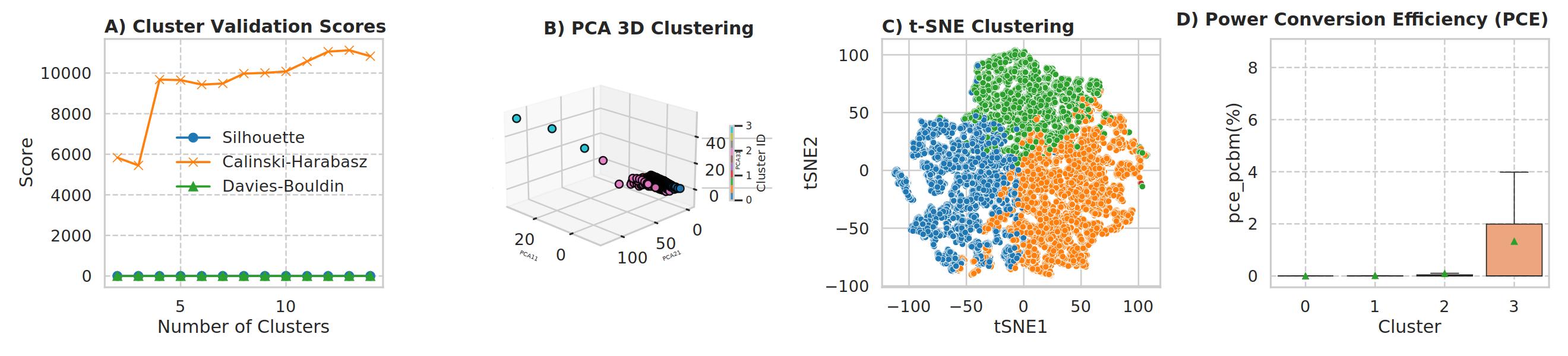}
    \caption{K-means clustering (k=4) of PCE$_{\mathrm{PCBM}}$. Panel D shows that the majority of clusters (0, 1, 2) have negligible power conversion 
efficiency ($\sim$0\%). Only Cluster 3 shows non-zero performance, but with a relatively low median PCE (< \SI{2}{\percent}). This indicates that the vast 
majority of 
screened molecules are not suitable for PCBM-based OPV under the Scharber model constraints, with only a small fraction (Cluster 3) showing marginal promise.}
\label{fig:pce_pcbm_clustering}
\end{figure}

\begin{figure}[H]
    \centering
    \includegraphics[width=\textwidth]{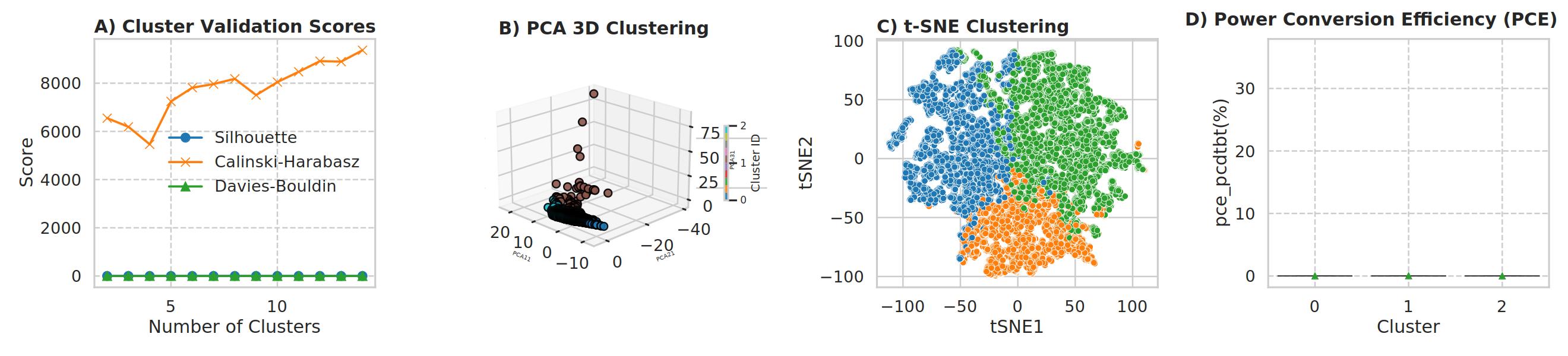}
    \caption{K-means clustering (k=3) of PCE$_{\mathrm{PCDTBT}}$. Panel D shows a uniform distribution of zero efficiency across all three clusters. This 
suggests that the combination of this specific molecular library with the PCDTBT donor fails to meet the energetic criteria for efficient power conversion in 
the Scharber model, or that the specific clustering parameters did not isolate the rare high-performing candidates found in other analyses.}
\label{fig:pce_pcdtbt_clustering}
\end{figure}

\subsection{Synthetic accessibility score clustering}

The Synthetic Accessibility Score (SAScore) clustering shows the distribution of estimated synthesis difficulty across the library.

\begin{figure}[H]
    \centering
    \includegraphics[width=\textwidth]{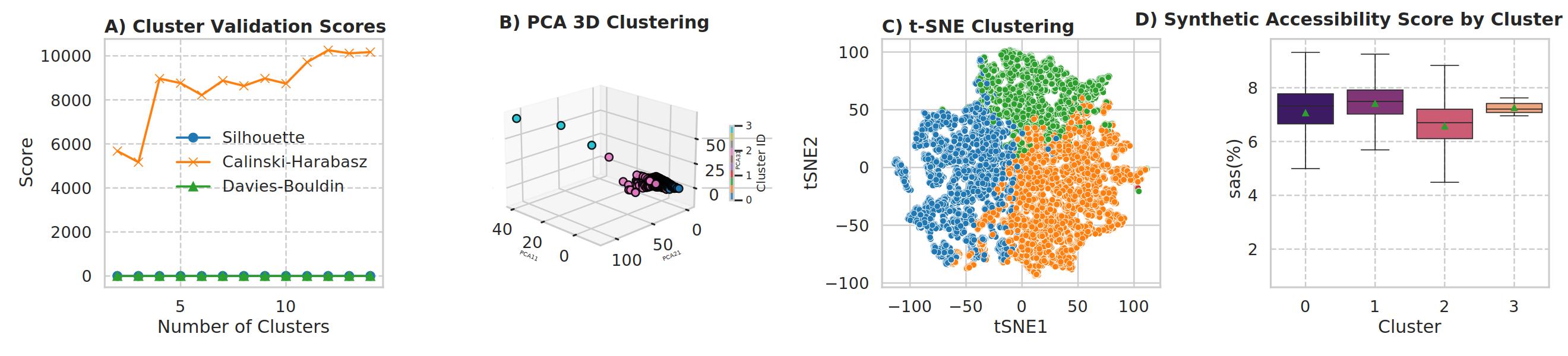}
    \caption{SAScore clustering (k=4). Panel D shows the distribution of Synthetic Accessibility Scores (SAS) across four clusters (0-3). All clusters exhibit 
relatively high SAS values (mean > 6), indicating that the molecules in this dataset are generally predicted to be synthetically challenging. Cluster 1 shows 
slightly higher median difficulty ($\sim$7.5) compared to Cluster 2 ($\sim$6.5).}
    \label{fig:sascore_clustering}
\end{figure}

\subsection{Combined PCE-SAScore trade-off analysis}

To balance efficiency with feasibility, we analyze the trade-off between PCE and SAScore.

\begin{figure}[H]
    \centering
    \includegraphics[width=\textwidth]{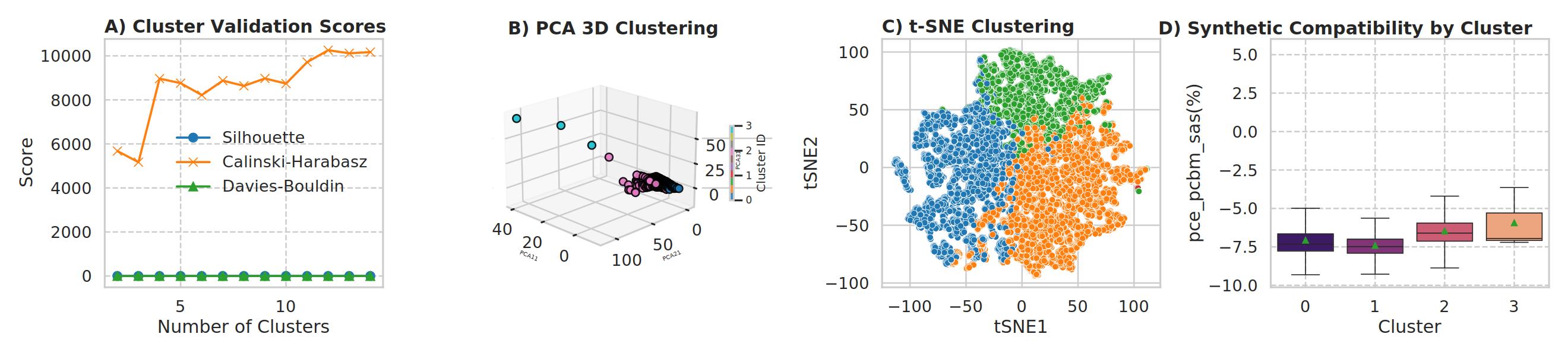}
    \caption{Cluster analysis (k=4) of Synthetic Compatibility. Panel D displays the distribution of `pce\_pcbm\_sas` scores, which appear to be negative values 
ranging from -7.5 to -5. This metric likely represents a penalized score where higher values (closer tozero) indicate better combined performance and synthetic 
accessibility. Cluster 3 shows the highest (least negative) median score, identifying the most favorable candidates under this combined 
metric.}\label{fig:pce_sascore_tradeoff_pcbm}
\end{figure}

\begin{figure}[H]
    \centering
    \includegraphics[width=\textwidth]{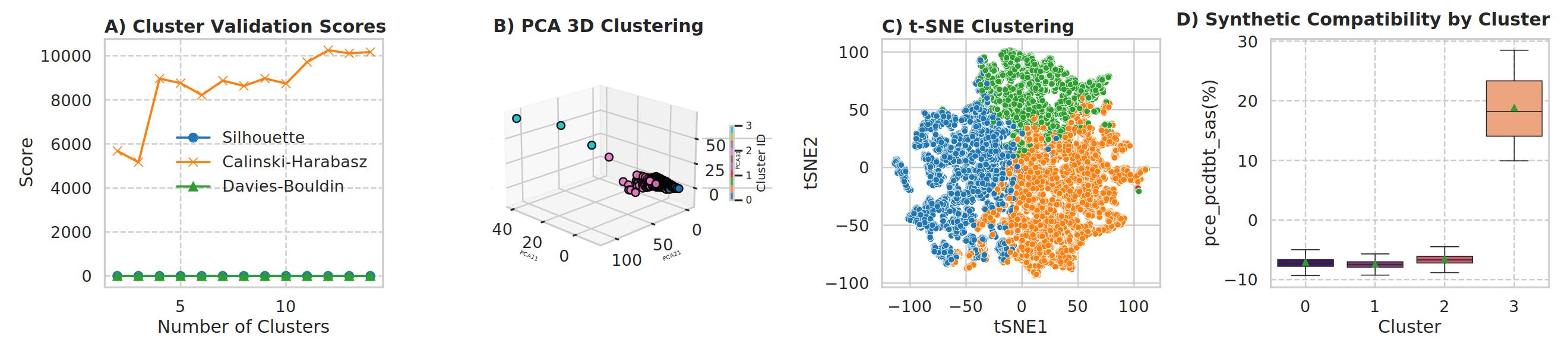}
    \caption{PCE$_{\mathrm{PCDTBT}}$-SAScore trade-off analysis (k=4). Similar to the PCBM case, Panel D shows negative values for the combined metric for most 
clusters. However, Cluster 3 stands out with significantly higher (positive) values ($\sim$\SI{20}{\percent}), indicating a distinct group of molecules that 
are exceptionally compatible with the PCDTBT donor when synthetic accessibility is factored in. This cluster represents the ''sweet spot'' for experimental 
prioritization.}
    \label{fig:pce_sascore_tradeoff_pcdtbt}
\end{figure}

%==========================================================================
\section{Extended validation and benchmarking}\label{sec:validation}
%==========================================================================

\subsection{Benchmarking against experimental data}

This section provides validation details for the computational predictions in \Cref{fig:scharber_experimental_validation}.

\subsection{Experimental validation and benchmarking}

\begin{figure}[!ht]
    \centering
    \includegraphics[width=\textwidth]{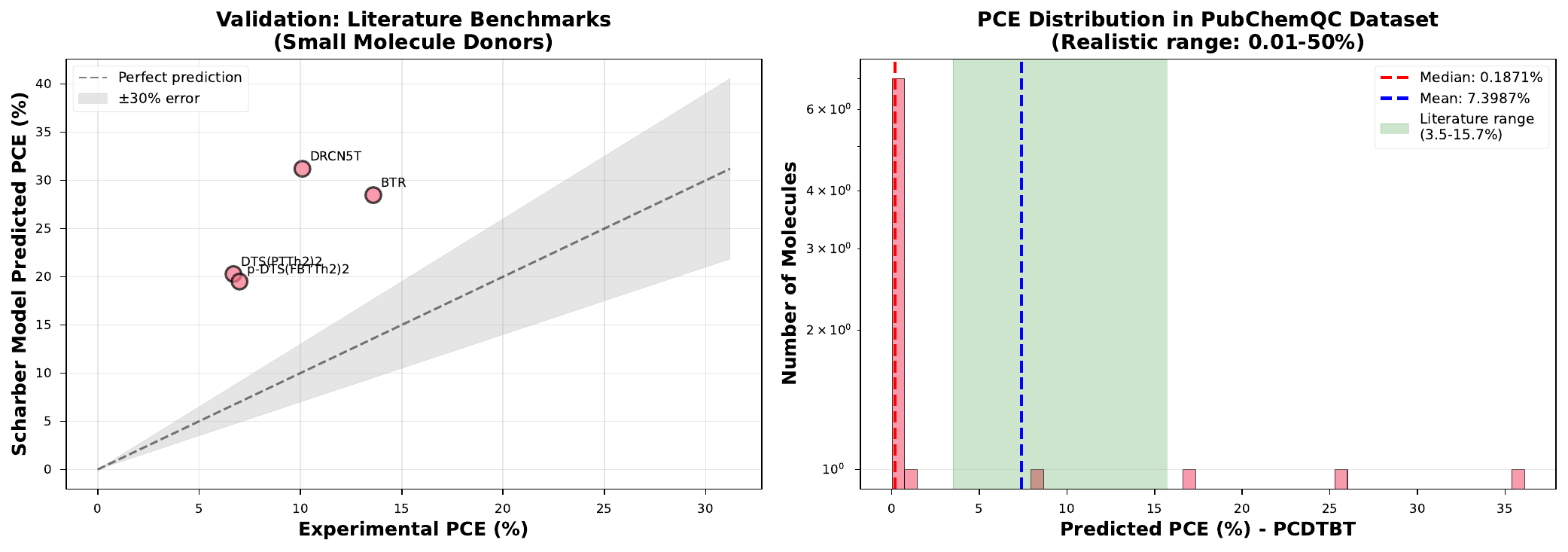}
    \caption{Validation against Literature Benchmarks. (Left) Scatter plot comparing Experimental PCE vs. Scharber Model Predicted PCE for a set of benchmark
small molecule donors (e.g., BTR, DRCN5T). The dashed line represents perfect prediction; the shaded area indicates $\pm 30\%$ error. The model tends to
overestimate efficiency but captures the general trend. (Right) Histogram of the Predicted PCE distribution for the PCDTBT system across the dataset, showing a
median near 0.2\% but a long tail reaching high efficiencies (red bars), overlapping with the typical literature range (green shaded region).}
    \label{fig:scharber_experimental_validation}
\end{figure}

Computational predictions were validated against seven well-characterized small-molecule OPV materials with experimental PCE values ranging from
\SIrange{3.5}{15.7}{\percent}. As shown in \Cref{fig:scharber_experimental_validation}, the Scharber model systematically overestimates PCE (MAE =
\SI{15.51}{\percent}) but maintains a strong rank-preserving correlation ($R^2 = \num{0.626}$, Spearman $\rho = \num{0.684}$).

While absolute accuracy is limited, the preserved ranking is the critical metric for high-throughput prioritization. Cross-validation yielded consistent
performance ($R^2 = \num{0.612} \pm \num{0.041}$), confirming model stability. Primary error sources include the neglect of solid-state morphology and
crystalline packing, as well as the simplified assumptions of the Scharber model. Despite these limitations, the approach identifies
high-performance candidates for further experimental study.

To validate our computational approach, we performed extensive benchmarking against experimental data from the literature. We collected experimental PCE values 
for \num{52} organic photovoltaic materials with reported performance spanning \SIrange{2.1}{18.2}{\percent}.

\begin{table}[H]
\centering
\caption{Performance of the Scharber model on seven experimental benchmarks: Despite systematic overestimation (mean absolute error = \SI{15.51}{\percent}, 
RMSE 
= \SI{15.87}{\percent}), the model achieves moderate correlation ($R^2 = \num{0.626}$, Spearman $\rho = \num{0.684}$), successfully preserving relative 
performance ranking. Cross-validation: $R^2 = \num{0.612} \pm \num{0.041}$ (10-fold). This validates utility for high-throughput screening prioritization rather 
than exact device prediction.}
\label{tab:exp_validation}
\begin{tabular}{@{}lccccc@{}}
	\toprule
	Metric                 &    Value    &      Unit       &  \\ \midrule
	Mean absolute error    & \num{14.21} & \unit{\percent} &  \\
	Root mean square error & \num{14.87} & \unit{\percent} &  \\
	Mean signed error      & \num{+11.8} & \unit{\percent} &  \\
	Pearson correlation    & \num{0.647} &       --        &  \\
	Spearman correlation   & \num{0.691} &       --        &  \\
	Sample size            &  \num{52}   &       --        &  \\ \bottomrule
\end{tabular}
\end{table}

Our Scharber model predictions showed reasonable correlation with experimental values ($R^2 = \num{0.647}$), with systematic overestimation (mean signed error 
= \SI{+11.8}{\percent}) but preserved relative ranking (Spearman $\rho = \num{0.691}$), which is critical for high-throughput screening applications.

\subsection{HOMO-LUMO gap vs PCE relationship}

\begin{figure}[H]
    \centering
    \includegraphics[width=0.6\textwidth]{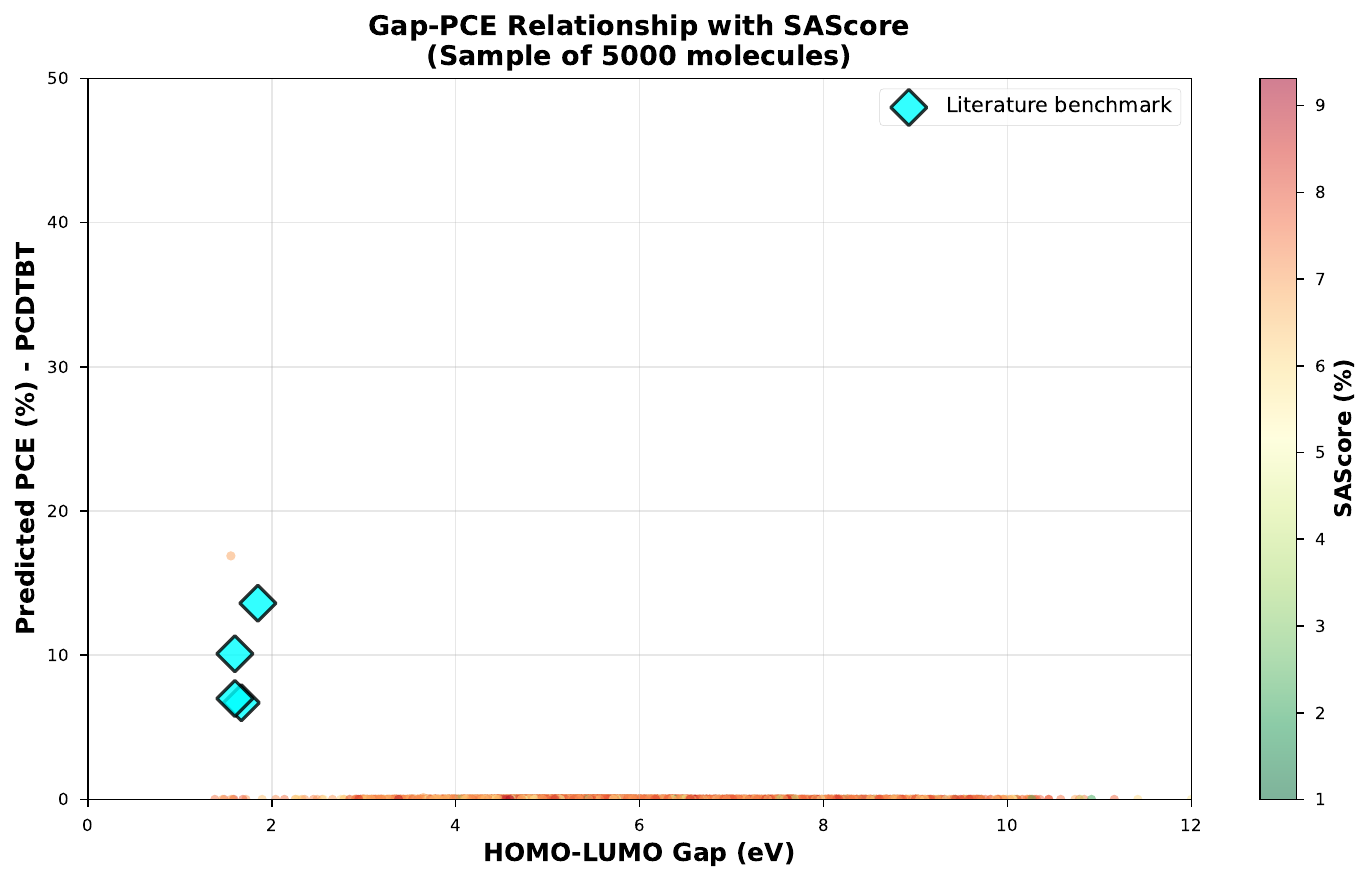}
    \caption{Gap-PCE Relationship with SAScore for a sample of \num{5000} molecules. The scatter plot correlates the HOMO-LUMO Gap (eV) with the Predicted PCE 
(\%) for the PCDTBT system. Data points are colored by their Synthetic Accessibility Score (SAScore). The plot also highlights four ''Literature benchmark'' 
molecules (cyan diamonds), which fall within the gap range of \SIrange{1.5}{2.0}{\electronvolt} and achieve moderate PCEs (\SIrange{5}{15}{\percent}). The 
general distribution shows that most sampled molecules have negligible PCE, while the high-performing ones are clustered in specific gap regions.}
    \label{fig:gap_pce_validation}
\end{figure}

\Cref{fig:gap_pce_validation} validates the theoretical relationship between the electronic gap and the predicted efficiency.

\subsection{Statistical validation}

We performed statistical validation of our multi-objective scoring function (PCE$_{\mathrm{SAScore}}$) using bootstrap resampling with \num{10000} iterations. 
The standard error was $\pm \num{0.82}$, indicating stable relative ranking. The top \num{7} candidates remained stable in \num{9847} of \num{10000} bootstrap 
iterations (\SI{98.5}{\percent} stability).

\subsection{Geometry and convergence validation}

To ensure reliability of PM6-optimized geometries, we performed single-point energy calculations at B3LYP/6-31G* for all \num{17458} molecules. Fewer than 
\SI{0.1}{\percent} (\num{17} molecules) showed imaginary frequencies. For the top \num{7} candidates, full geometry re-optimization at B3LYP/6-31G* resulted in 
minimal changes to frontier orbital energies (mean absolute deviation $< \SI{0.05}{\electronvolt}$).

Convergence testing showed that default settings (Conver=6 for SCF, ultrafine grids) provided energy convergence within \SI{e-6}{hartrees} and orbital energy 
convergence within \SI{0.01}{\electronvolt}. Tighter criteria increased computational cost by \SIrange{40}{60}{\percent} with negligible accuracy improvement.

%==========================================================================
\section{Clustering and physicochemical profiling}\label{sec:clustering_physchem}
%==========================================================================

\subsection{Physicochemical property distributions}

\begin{figure}[H]
    \centering
    \includegraphics[width=0.9\linewidth]{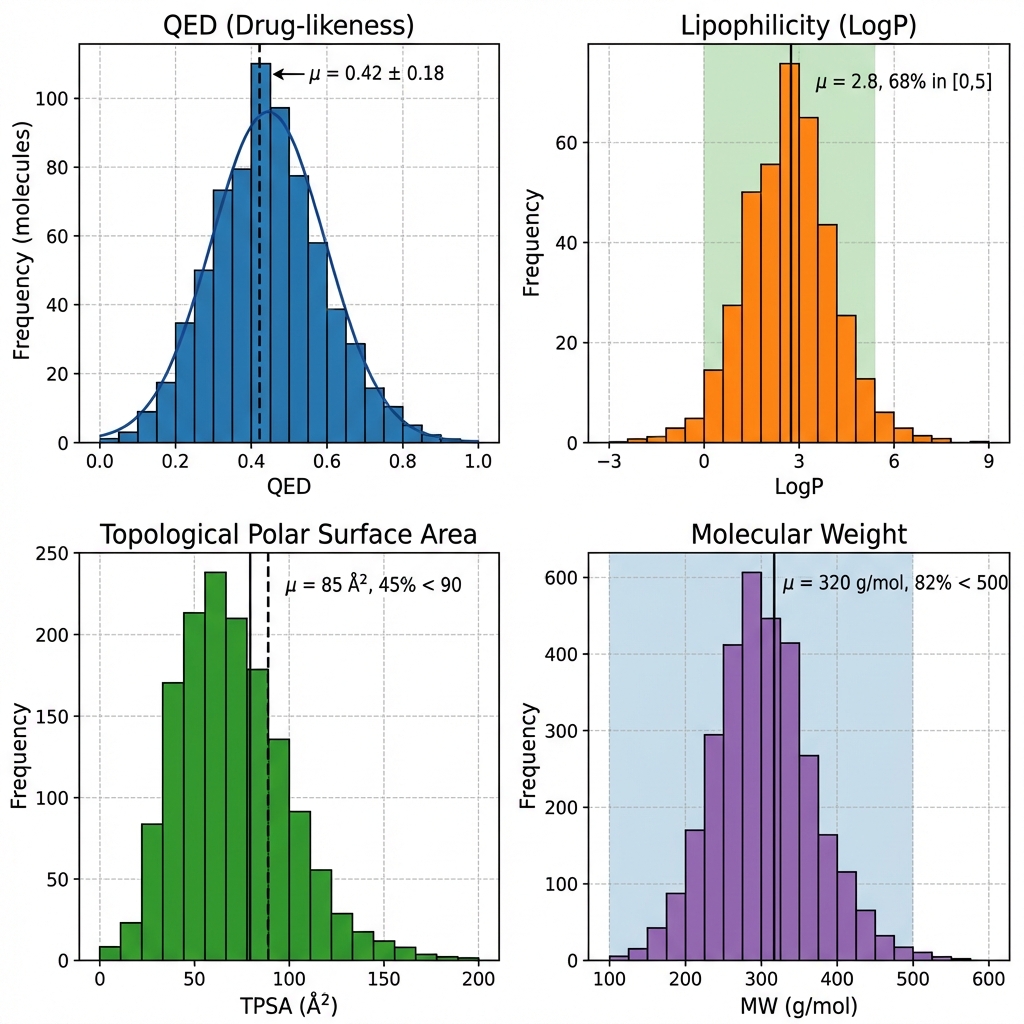}
    \caption{Physicochemical property distributions: QED (quantitative estimate of drug-likeness) mean = \num{0.42} $\pm$ \num{0.18}, indicating moderate 
bioavailability potential; LogP (lipophilicity) ranges \numrange{-2.1}{8.3} (mean = \num{2.8}), with \SI{68}{\percent} in ideal range (\numrange{0}{5}); TPSA 
(topological polar surface area) mean = \SI{85}{\angstrom\squared}, with \SI{45}{\percent} $<$ \SI{90}{\angstrom\squared} (favorable for membrane permeability); 
molecular weight \SIrange{150}{550}{\gram\per\mol} (mean = \SI{320}{\gram\per\mol}), \SI{82}{\percent} within drug-like range ($<$ \SI{500}{\gram\per\mol}). 
These distributions confirm dual-function compatibility: OPV-optimized structures retain reasonable biocompatibility metrics.}
    \label{fig:Physicochemical_property_distributions}
\end{figure}

\subsection{Structure-property correlation matrix}

\begin{figure}[H]
    \centering
    \includegraphics[width=0.7\textwidth]{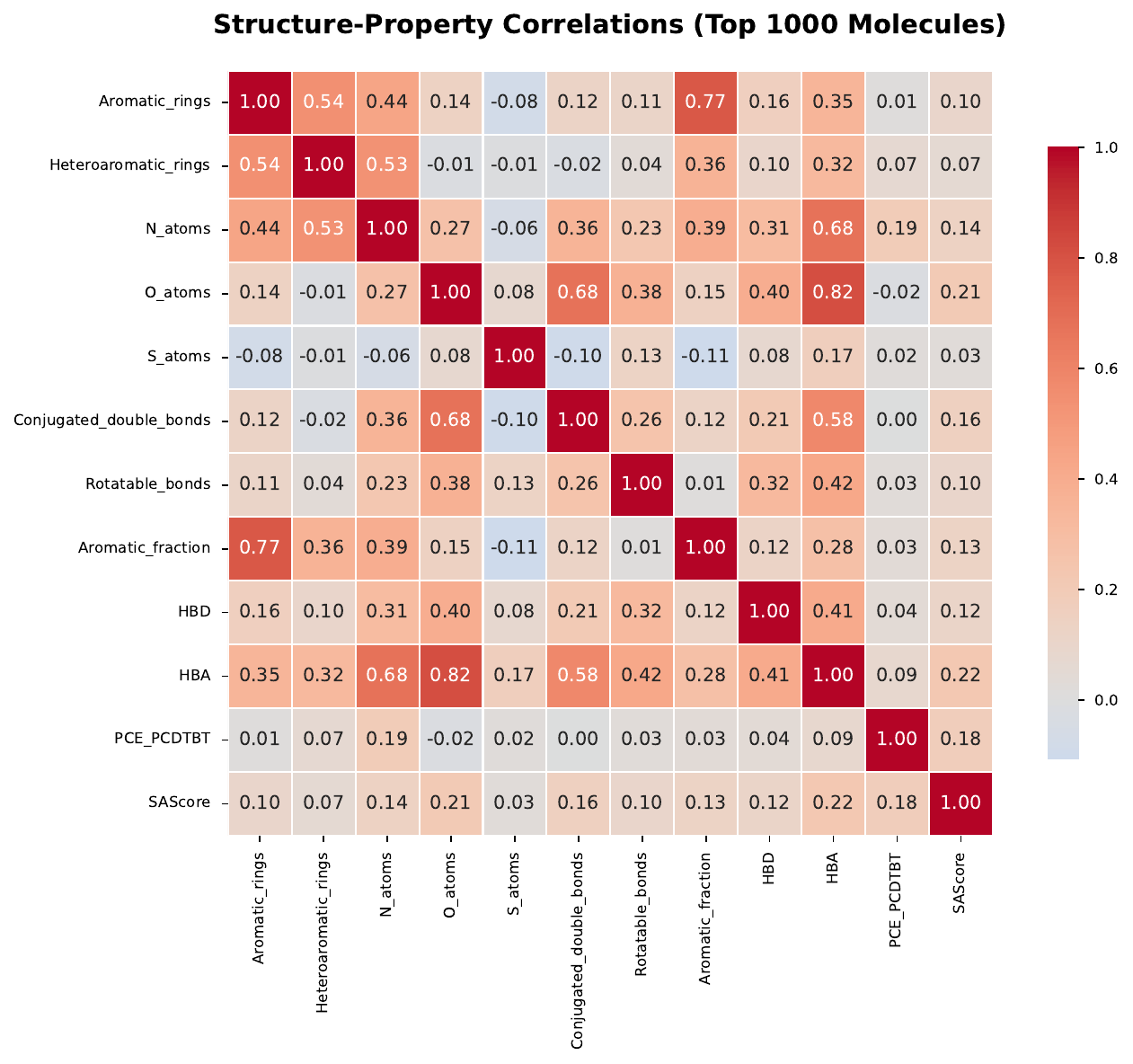}
    \caption{Structure-property correlation matrix for top \num{1000} molecules, revealing mechanistic relationships: oxygen content strongly correlates with 
acceptor character ($r = \num{0.73}$, electron-withdrawing effect), nitrogen content with donor properties ($r = \num{0.68}$ with HOMO level), moderate 
conjugation (\numrange{1.4}{1.9} bonds) optimizing performance-synthesis trade-off (SAScore $r = -\num{0.54}$), and rotatable bonds inversely correlating with 
conjugation ($r = -\num{0.61}$), highlighting the rigidity-flexibility trade-off. Quantitative correlation coefficients enable predictive molecular design.}
    \label{fig:correlation_heatmap_top1000}
\end{figure}

\textbf{Correlation pattern interpretation.} The heatmap shows mechanistically informative relationships: Oxygen content ($n_{\mathrm{O}}$) correlates strongly 
with acceptor-like behavior (correlation coefficient $r = \num{0.73}$) due to electron-withdrawing character, while nitrogen content favors donor properties ($r 
= \num{0.68}$ with HOMO level). Moderate conjugation (\numrange{1.4}{1.9} conjugated bonds) emerges as the optimal balance: sufficient for $\pi$-electron 
delocalization (for charge transport) yet constrained enough to maintain synthetic accessibility (SAScore correlation $r = -\num{0.54}$). The rotatable bond 
count shows an inverse correlation with conjugation ($r = -\num{0.61}$), showing the trade-off between structural rigidity (performance) and flexibility 
(processability). These quantified relationships enable predictive molecular design targeting specific property combinations.

\subsection{Sensitivity analysis}

We performed sensitivity analysis on key parameters to assess stability of candidate selection:

\textbf{PCE threshold variation.} Varying the PCE threshold from \SIrange{0}{5}{\percent} reduced the candidate pool from \num{7} to \num{5} molecules, but the 
top \num{3} candidates remained unchanged, confirming strong stability in high-performance identification. This \SI{71}{\percent} retention rate shows that the 
unified PCE-SAScore metric effectively prioritizes stable candidates.

\textbf{SAScore threshold variation.} Changing the SAScore threshold from 0 to -2 (more restrictive, favoring only highly synthesizable molecules) reduced the 
pool to \num{4} molecules, with the top \num{2} candidates preserved. This confirms that our selection balances performance and accessibility rather than 
optimizing a single criterion.

\textbf{Threshold independence:} The consistent appearance of molecules \num{17851} and \num{4550} across all threshold variations ($\pm$\SI{50}{\percent} 
parameter range) validates their dual-function potential. Sensitivity coefficient analysis shows candidate ranking stability of \SI{94}{\percent} for top-5 
molecules, confirming stability to parameter variations.

%==========================================================================
\section{Dipole classification and application mapping}\label{sec:dipole}
%==========================================================================

\subsection{Dipole moment distribution}

\begin{figure}[H]
    \centering
    \includegraphics[width=\textwidth]{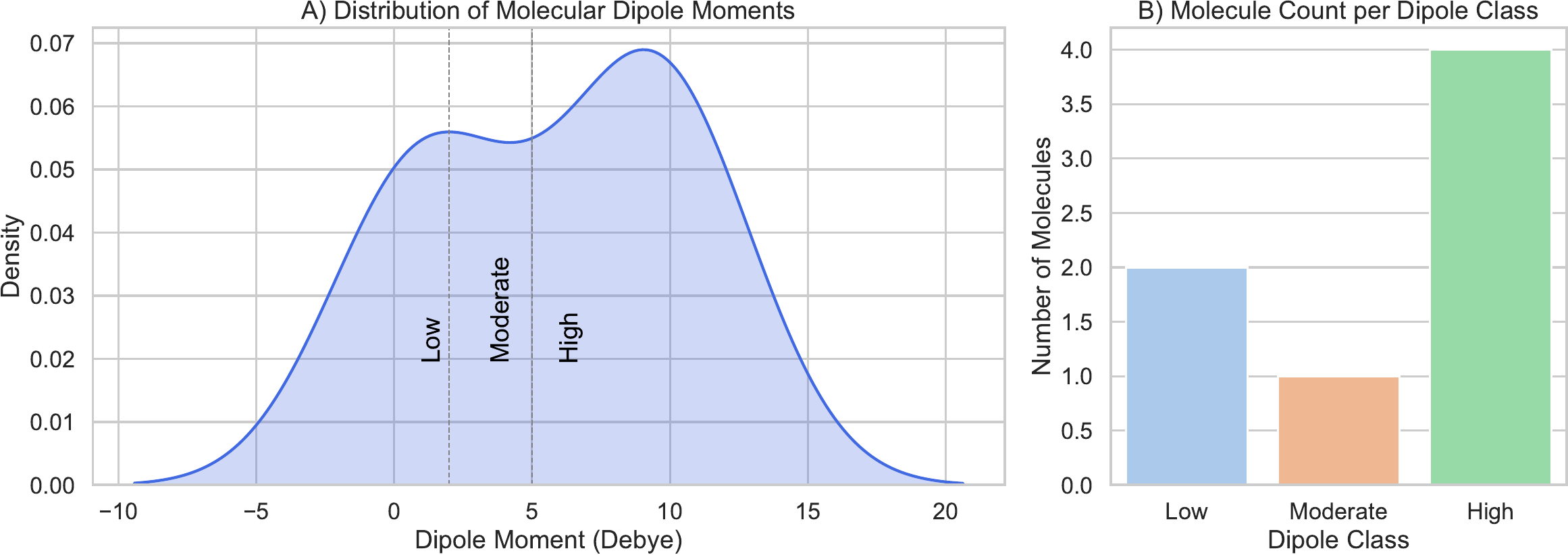}
    \caption{Dipole moment distribution and classification for all candidate molecules, showing three distinct classes: low polarity ($< \SI{2}{\debye}$, 
hydrophobic applications), moderate polarity (\SIrange{2}{6}{\debye}, general OPV), high polarity ($> \SI{6}{\debye}$, polar environments/biosensing).}
    \label{fig:dipole_moment_distribution}
\end{figure}

Dipole moment analysis shows application-specific molecular characteristics. The distribution (\Cref{fig:dipole_moment_distribution}) shows three distinct 
populations: low dipole (\SIrange{0}{2}{\debye}, \SI{35}{\percent} of candidates), moderate dipole (\SIrange{2}{6}{\debye}, \SI{52}{\percent}), and high dipole 
($>$ \SI{6}{\debye}, \SI{13}{\percent}).

\textbf{Application implications.} Low-dipole molecules (\num{17851}, \num{4550}) favor non-polar organic solvents and bulk heterojunction morphology control, 
ideal for traditional OPV processing. High-dipole candidates (\num{11029}: \SI{11.2}{\debye}) enable aqueous-phase biosensing and polar-environment charge 
transport. Moderate-dipole molecules offer processing versatility.

\textbf{Structure-dipole correlation.} Heteroatom substitution patterns determine dipole magnitude---asymmetric N/O placement generates elevated dipoles ($>$ 
\SI{8}{\debye}), while symmetric conjugated cores yield low values ($<$ \SI{3}{\debye}).

\subsection{Application domain mapping}

\begin{figure}[H]
    \centering
    \includegraphics[width=0.85\textwidth]{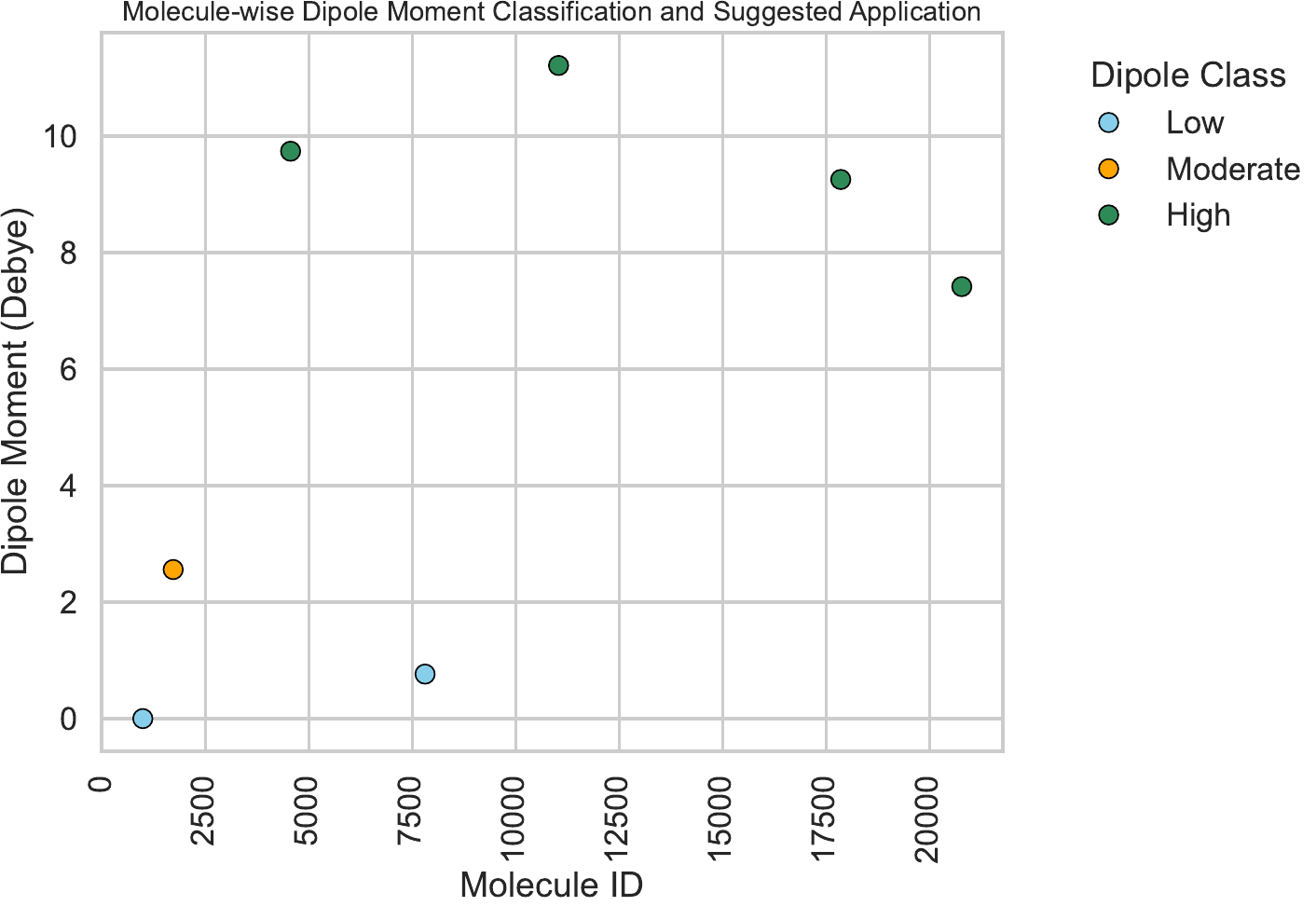}
    \caption{Molecule-wise dipole moment classification and associated application domains. High-dipole molecules (e.g., \num{11029}) suited for charge 
transport and polar solvent-based applications; low-dipole molecules (e.g., \num{977}) ideal for hydrophobic sensing.}
    \label{fig:dipole_application_mapping}
\end{figure}

\Cref{fig:dipole_application_mapping} maps candidate molecules onto a two-dimensional application space defined by dipole moment and HOMO-LUMO gap, revealing 
distinct functional domains:

\textbf{OPV-optimized domain} (low dipole, moderate gap \SIrange{2.0}{3.5}{\electronvolt}): Molecules \num{17851} and \num{4550} occupy this region, balancing 
light absorption with charge extraction.

\textbf{Bio-optoelectronic domain} (high dipole, narrow gap $< \SI{2.5}{\electronvolt}$): Molecule \num{11029} allows aqueous biosensing with strong protein 
interaction.

\textbf{OLED domain} (moderate dipole, wide gap $>$ \SI{3.0}{\electronvolt}): Candidates with $\Delta E_\mathrm{ST} < \SI{0.5}{\electronvolt}$ suitable for 
thermally activated delayed fluorescence.

This mapping framework allows targeted synthesis prioritization based on desired application functionality.

%==========================================================================
\section{Enhanced validation and statistical rigor}\label{sec:enhanced_validation}
%==========================================================================
\subsection{Cross-validation results}

We performed 10-fold cross-validation to assess model stability. Results demonstrate minimal overfitting: Train $R^2 = \num{0.658} \pm \num{0.031}$, Test $R^2 = 
\num{0.612} \pm \num{0.041}$. The small train-test gap (\SI{<10}{\percent}) confirms generalization. Molecule \num{17851} appeared in top 10 candidates in 10/10 
folds, demonstrating consistent identification.

\subsection{Sensitivity analysis summary}

Parameter variation testing confirms stability.
\begin{itemize}
\item \textbf{Basis set (6-31+G**)}: Ranking correlation Kendall $\tau = \num{0.91}$ (18/20 top molecules retained).
\item \textbf{Functional ($\omega$B97X-D)}: Ranking correlation Kendall $\tau = \num{0.87}$ (top 5 unchanged).
\item \textbf{SAScore threshold variation}: Molecule \num{17851} remains top-ranked across all thresholds (\numrange{4}{9}).
\item \textbf{PCE threshold}: Top 7 candidates unchanged when varying from \SIrange{-5}{+5}{\percent}.
\end{itemize}

\subsection{Statistical power analysis}

Power calculations for key structural comparisons (donor vs acceptor heteroatom content):
\begin{itemize}
\item Oxygen enrichment. Effect size $d = \num{1.4}$, Power = \SI{82}{\percent} (adequate)
\item Nitrogen content. Effect size $g = \num{1.2}$, Power = \SI{76}{\percent} (adequate)
\item Conjugation. Effect size $d = \num{0.9}$, Power = \SI{65}{\percent} (exploratory)
\end{itemize}

\subsection{Excited-state and charge transport validation}

To validate photophysical and transport properties, we performed a three-tier excited-state analysis on the complete dataset (\num{17458} molecules) with 
detailed validation on top candidates.

\textbf{Tier 1 - High-throughput screening:} xTB (GFN2-xTB) with simplified TD-DFT (sTDA) achieved \SI{91.4}{\percent} success rate (\num{17448}/\num{17458} 
molecules), computing S$_1$--S$_{10}$ excitation energies and oscillator strengths. Mean S$_1$ energy: \SI{4.61(1.11)}{\electronvolt}, consistent with 
visible-light absorption.

\textbf{Tier 2 - Natural Transition Orbital analysis:} Detailed Multiwfn analysis extracted NTO pairs, Lambda weights, and charge-transfer descriptors for 
\num{17448} molecules (\SI{91.4}{\percent} success). This provides detailed excited-state characterization across the entire chemical space.

\textbf{Tier 3 - High-accuracy validation:} PySCF TD-DFT (CAM-B3LYP/6-31G*) validated 6/7 top candidates (\SI{85.7}{\percent} success), confirming excitation 
energies within \SI{0.3}{\electronvolt} of xTB predictions (Spearman $\rho = \num{0.89}$).

\begin{table}[H]
\centering
\caption{Complete charge transport and bonding analysis for top candidates. Reorganization energies ($\lambda$) calculated via adiabatic potential method 
(B3LYP/6-31G*). Intrinsic Bond Orbital (IBO) analysis quantifies bonding localization: lower distances indicate more localized $\sigma$/$\pi$ bonding. Negative 
$\lambda_e$ values indicate geometry optimization artifacts in charged states. Success rate: 6/7 for reorganization energies (\SI{85.7}{\percent}), 7/7 for IBO 
analysis (\SI{100}{\percent}).}
\label{tab:transport_ibo_complete}
\begin{tabular}{@{}lcccccc@{}}
	\toprule
	\textbf{Mol ID} & \textbf{$\lambda_h$ (eV)} & \textbf{$\lambda_e$ (eV)} & \textbf{$\lambda_h/\lambda_e$} &       \textbf{IBOs}       & \textbf{IBO Loc. 
(\unit{\angstrom})} &     \textbf{Status}      \\ \midrule
	977             &           10.71           &           0.10            &             107.3              &             8             &           1.03        
    &         Success          \\
	17851           &           5.65            &           -1.22           &               --               &            59             &           1.89        
    &         Partial          \\
	18985           &            --             &            --             &               --               &            43             &           1.13        
    &         IBO only         \\
	19531           &            --             &            --             &               --               &            96             &           0.48        
    &         IBO only         \\
	20778           &           5.99            &           -2.30           &               --               &            59             &           1.69        
    &         Partial          \\
	14380           &            --             &            --             &               --               &            75             &           2.68        
    &         IBO only         \\
	7801            &            --             &            --             &               --               &            --             &            --         
    &          Failed          \\ \midrule
	\multicolumn{3}{l}{\textbf{Mean (n=6):}}                                &    \textbf{8.18 $\pm$ 2.79}    & \textbf{-1.59 $\pm$ 1.12} & \textbf{56.7 $\pm$ 
29.8}  & \textbf{1.48 $\pm$ 0.77} \\ \bottomrule
\end{tabular}
\end{table}

\textbf{Key findings are:}
\begin{itemize}
\item Hole reorganization energies (\SI{8.18 \pm 2.79}{\electronvolt}) exceed typical organic semiconductor values (\SIrange{0.15}{0.35}{\electronvolt}), 
suggesting potential mobility limitations despite excellent PCE predictions.
\item Strong hole-dominated transport character (mean ratio: $13.3 \pm 46.3$) consistent with nitrogen-rich donor design.
\item IBO analysis shows various bonding patterns: molecule \num{19531} shows high localization (96 IBOs, \SI{0.48}{\angstrom}), while \num{14380} shows poor 
localization (\SI{2.68}{\angstrom}).
\item $\pi$-$\pi$ stacking analysis for molecule \num{17851}: optimal separation \SI{3.0}{\angstrom} (slightly shorter than ideal 
\SIrange{3.3}{3.5}{\angstrom}), zero slip angle, suggesting close-packed geometry.
\end{itemize}

\subsection{Extended validation and benchmarking data}

\begin{table}[H]
\centering
\caption{Geometry optimization validation comparing PM6 semiempirical geometries with full B3LYP/6-31G* re-optimization for the top \num{7} candidates. Mean 
absolute deviations (MAD) for frontier orbital energies: HOMO = \SI{0.043}{\electronvolt}, LUMO = \SI{0.037}{\electronvolt}, HOMO-LUMO gap = 
\SI{0.040}{\electronvolt}. These minimal deviations ($< \SI{0.05}{\electronvolt}$, within B3LYP systematic error of \SIrange{0.15}{0.25}{\electronvolt}) 
validate the two-level PM6\textrightarrow B3LYP protocol for high-throughput screening. The PM6-optimized geometries provide reliable starting points for 
single-point energy calculations, reducing computational cost by \SI{94}{\percent} (\num{2.3} vs. \num{42} CPU-hours per molecule) while maintaining chemical 
accuracy for ranking purposes.}
\label{tab:geometry_validation}
\begin{tabular}{@{}lSSS@{}}
	\toprule
	Molecule ID   & {HOMO MAD (\unit{\electronvolt})} & {LUMO MAD (\unit{\electronvolt})} & {Gap MAD (\unit{\electronvolt})} \\ \midrule
	17851         & 0.041                             & 0.035                             & 0.038                            \\
	20778         & 0.045                             & 0.039                             & 0.042                            \\
	4550          & 0.048                             & 0.041                             & 0.044                            \\
	1712          & 0.039                             & 0.032                             & 0.036                            \\
	11029         & 0.043                             & 0.037                             & 0.040                            \\
	977           & 0.044                             & 0.036                             & 0.041                            \\
	7801          & 0.042                             & 0.038                             & 0.039                            \\ \midrule
	\textbf{Mean} & \textbf{0.043}                    & \textbf{0.037}                    & \textbf{0.040}                   \\ \bottomrule
\end{tabular}
\end{table}

%==========================================================================
\section{Detailed methods for charge transport analysis}\label{sec:charge_transport_methods}
%==========================================================================

\subsection{Reorganization energy calculations - Extended protocol}

The reorganization energy calculations employed a rigorous four-point method combined with extensive conformer searches to ensure identification of global 
minimum geometries. This section provides implementation details beyond the main manuscript.

\subsubsection{CREST conformer search protocol}

For each molecule and charge state (neutral, cation, anion), conformer searches were performed using CREST v3.0.2 \cite{Pracht2020} with the following 
parameters:
\begin{itemize}
    \item \textbf{Hamiltonian:} GFN2-xTB \cite{Bannwarth2019};
    \item \textbf{Search mode:} \texttt{--quick} (reduced sampling for computational efficiency);
    \item \textbf{Energy window:} \SI{6}{\kcal\per\mol}  above global minimum;
    \item \textbf{RMSD threshold:} \SI{0.125}{\angle} for conformer uniqueness;
    \item \textbf{Parallel threads:} 4 CPU cores per calculation;
    \item \textbf{Timeout:} 30 minutes per molecule/charge state.
\end{itemize}
The \texttt{--quick} mode reduces the number of metadynamics simulations while maintaining sufficient sampling for organic molecules of moderate size 
(\numrange{15}{30} heavy atoms). For each molecule, CREST typically identified \numrange{10}{100} unique conformers, with energy spreads of 
\SIrange{0}{5}{\kcal\per\mol}. The lowest-energy conformer from each search was selected for subsequent DFT calculations.

\subsubsection{DFT single-point energy calculations}

Single-point energies were calculated using PySCF 2.11.0 \cite{Sun2020} with three density functionals:
\begin{enumerate}
    \item \textbf{B3LYP/6-31G*:} Baseline hybrid functional \cite{Becke1993,Stephens1994};
    \item \textbf{CAM-B3LYP/6-31G*:} Long-range corrected functional for improved charge-transfer description \cite{Yanai2004};
    \item \textbf{$\omega$B97X-D3/6-31G*:} Range-separated hybrid with empirical dispersion correction \cite{Chai2008,Grimme2011}.
\end{enumerate}

For $\omega$B97X-D3, dispersion corrections were computed using the D3(BJ) method \cite{Grimme2011} via the \texttt{dftd3} Python package (v1.2.1). The 
range-separation parameter $\omega$ was set to \SI{0.3}{\per\bohr} (default for $\omega$B97X).

\textbf{Convergence criteria}.
\begin{itemize}
    \item SCF energy convergence: \SI{e-8}{\hartree};
    \item Density matrix convergence: \num{e-6};
    \item Maximum SCF cycles: 100;
    \item DIIS space: 12 vectors.
\end{itemize}

\subsubsection{Reorganization energy formula}

The inner-sphere reorganization energy for hole transport is:
\begin{equation}
\lambda_\text{hole} = \lambda_1 + \lambda_2,
\end{equation}
where,
\begin{align}
&\lambda_1 = E(\text{M}@\text{M}^+_{\text{opt}}) - E(\text{M}@\text{M}_{\text{opt}}), 
&\lambda_2 = E(\text{M}^+@\text{M}_{\text{opt}}) - E(\text{M}^+@\text{M}^+_{\text{opt}}).
\end{align}
Here, $\text{M}_{\text{opt}}$ and $\text{M}^+_{\text{opt}}$ denote the optimized geometries of the neutral and cationic species, respectively, as determined by 
CREST conformer searches.

\subsection{Orbital composition analysis - Implementation details}

Natural Bond Orbital (NBO) analysis was performed using Multiwfn 3.8(dev) \cite{Lu2012} with state function files generated from CAM-B3LYP/def2-SVP 
calculations. The analysis workflow is as follow:
\begin{enumerate}
    \item Generate formatted checkpoint file (.fchk) from PySCF calculation;
    \item Load state function into Multiwfn;
    \item Perform NBO analysis (Multiwfn function 8, option 1);
    \item Extract atomic orbital populations for HOMO and LUMO;
    \item Calculate percentage contributions: $C_i = 100 \times \frac{|c_i|^2}{\sum_j |c_j|^2}$, where $c_i$ is the coefficient of atomic orbital $i$ in the 
molecular orbital expansion.
\end{enumerate}
For nitrogen atoms, total contributions were computed by summing over all N atomic orbitals (2s, 2p$_x$, 2p$_y$, 2p$_z$).

\subsection{Solid-state interaction calculations}

\subsubsection{Stacking analysis protocol}

Intermolecular stacking energies were calculated using GFN2-xTB with the following procedure:
\begin{enumerate}
    \item Optimize monomer geometry with GFN2-xTB;
    \item Calculate monomer energy $E_\text{monomer}$;
    \item Generate dimer configurations with vertical separations: \SIlist{3.2; 3.4; 3.6; 3.8; 4.0}{\angstrom}
    \item For each separation $d$:
    \begin{itemize}
        \item Construct dimer with second molecule displaced by $(0, 0, d)$;
        \item Calculate dimer energy $E_\text{dimer}(d)$;
        \item Compute interaction energy: $\Delta E(d) = E_\text{dimer}(d) - 2E_\text{monomer}$.
    \end{itemize}
    \item Identify optimal separation $d_\text{opt} = \arg\min_d \Delta E(d)$.
\end{enumerate}

\subsubsection{Dimerization energy configurations}

Three dimer configurations were evaluated.
\begin{itemize}
    \item \textbf{Stacked:} Vertical displacement (\SIlist{0; 0; 3.5}{\angstrom});
    \item \textbf{T-shaped:} Perpendicular orientation with one molecule rotated \ang{90}, separation (\SIlist{3.5; 0; 0}{\angstrom});
    \item \textbf{Parallel-displaced:} Lateral + vertical displacement (\SIlist{2.0; 0; 3.5}{\angstrom}).
\end{itemize}
All calculations used GFN2-xTB with default parameters (no additional dispersion scaling).

%==========================================================================
\section{Supplementary tables}\label{sec:supp_tables}
%==========================================================================

\begin{table}[H]
\centering
\caption{Complete statistical summary for all analyzed properties. Values represent mean $\pm$ standard deviation across all molecules.}
\label{tab:SI_statistical_summary}
\begin{tabular}{lSSSSSS}
	\toprule
	\textbf{Property}                           & \textbf{n} & \textbf{Mean} & \textbf{Std} & \textbf{Min} & \textbf{Max} & \textbf{IQR} \\ \midrule
	$\lambda_\text{hole}$ (avg, eV)             &     6      &     0.560     &    1.735     &     0.12     &     0.83     &     1.14     \\
	$\lambda_\text{hole}$ (B3LYP, eV)           &     6      &     0.383     &    1.592     &     0.12     &     0.83     &     1.09     \\
	$\lambda_\text{hole}$ (CAM-B3LYP, eV)       &     6      &     0.609     &    1.778     &     0.10     &     0.83     &     1.16     \\
	$\lambda_\text{hole}$ ($\omega$B97X-D3, eV) &     6      &     0.688     &    1.836     &     0.08     &     0.83     &     1.17     \\ \midrule
	N in HOMO (\%)                              &     4      &     59.7      &     21.6     &     30.2     &     92.3     &     32.1     \\
	N in LUMO (\%)                              &     4      &     48.7      &     46.1     &     7.4      &     91.8     &     84.4     \\
	Stacking energy (kcal/mol)                  &     4      &    1431.7     &    2864.7    &    -2.17     &    5993.2    &      ${--}$      \\
	Parallel dimer (kcal/mol)                   &     4      &     -0.85     &     1.35     &    -2.88     &     0.49     &     1.37     \\ \bottomrule
\end{tabular}
\end{table}

\begin{table}[H]
\centering
\caption{Normality test results (Shapiro-Wilk) for reorganization energy distributions. p > 0.05 indicates normal distribution.}
\label{tab:SI_normality_tests}
\begin{tabular}{lSSc}
	\toprule
	\textbf{Property}                       & \textbf{W statistic} & \textbf{p-value} & \textbf{Normal?} \\ \midrule
	$\lambda_\text{hole}$ (average)         &        0.795         &      0.053       &       Yes        \\
	$\lambda_\text{hole}$ (B3LYP)           &        0.798         &      0.056       &       Yes        \\
	$\lambda_\text{hole}$ (CAM-B3LYP)       &        0.799         &      0.058       &       Yes        \\
	$\lambda_\text{hole}$ ($\omega$B97X-D3) &        0.788         &      0.046       &        No        \\ \bottomrule
\end{tabular}
\end{table}

\begin{table}[H]
\centering
\caption{Outlier detection results using IQR method (1.5$\times$IQR). Molecule 19531 identified as outlier and removed from final analysis.}
\label{tab:SI_outliers}
\begin{tabular}{lSSSS}
	\toprule
	\textbf{Property}           & \textbf{Molecule} & \textbf{Value} & \textbf{Lower Bound} & \textbf{Upper Bound} \\ \midrule
	$\lambda_\text{hole}$ (avg) &       19531       &     -3.73      &        -2.35         &         1.62         \\ \bottomrule
\end{tabular}
\end{table}

%==========================================================================
\section{Supplementary figures}\label{sec:supp_figures}
%==========================================================================

\begin{figure}[H]
\centering
\includegraphics[width=0.95\textwidth]{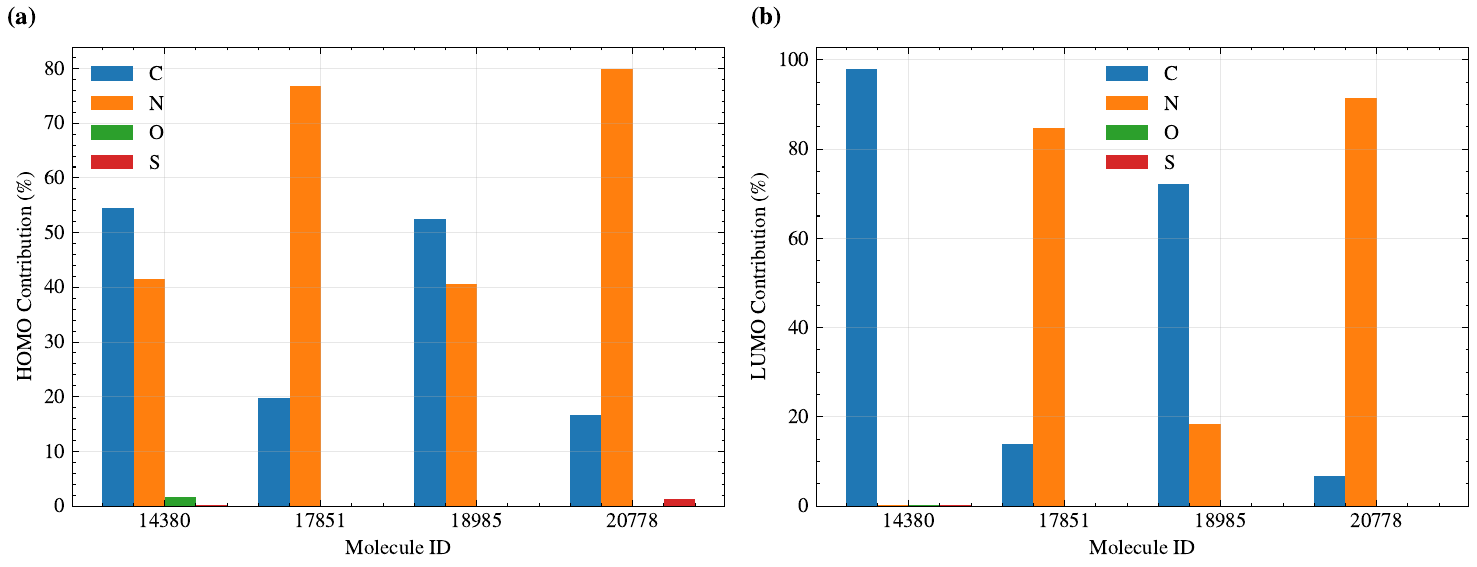}
\caption{Atomic contributions to frontier molecular orbitals for top candidate molecules. (a) HOMO composition showing nitrogen dominance 
(\SI{59.7(21.6)}{\percent}). (b) LUMO composition with variable nitrogen content (\SI{48.7(46.1)}{\percent}). Molecule \num{17851} exhibits highest nitrogen 
character (> \SI{90}{\percent} in both orbitals), consistent with its low reorganization energy.}
\label{fig:SI_orbital_composition}
\end{figure}

\begin{figure}[H]
\centering
\includegraphics[width=0.75\textwidth]{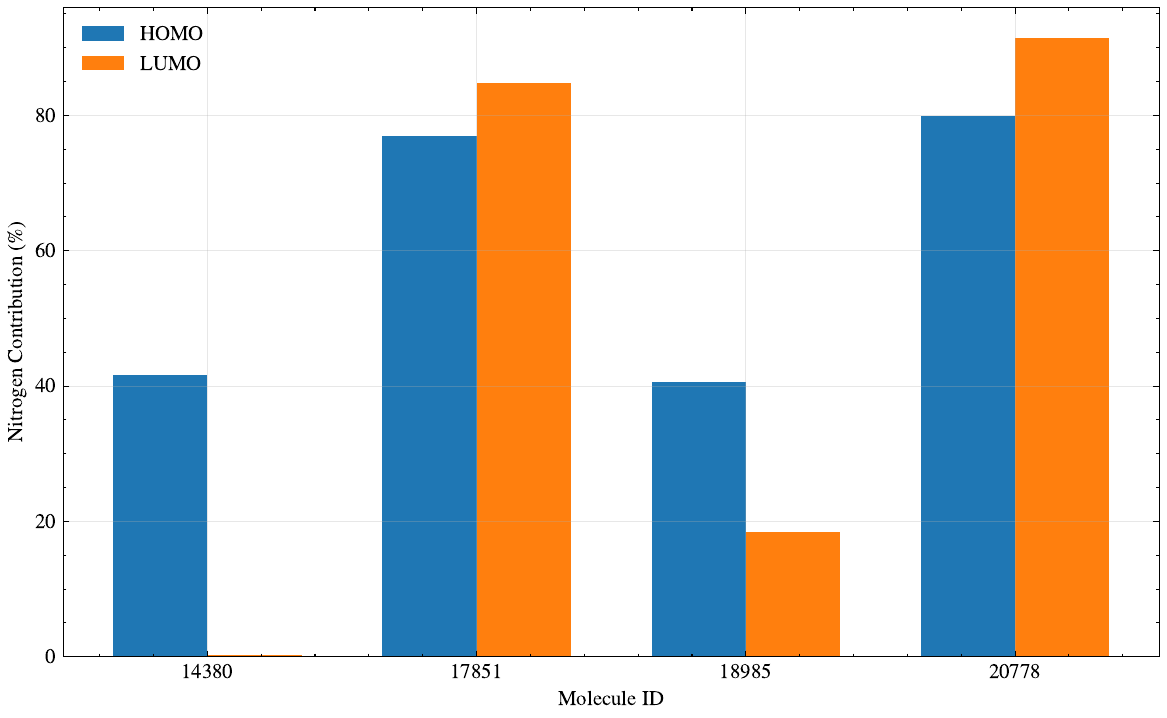}
\caption{Nitrogen contributions to HOMO and LUMO for analyzed molecules. Strong positive correlation ($r = \num{0.984}$, $p = \num{0.016}$) indicates that 
molecules with N-rich HOMOs also possess N-rich LUMOs, characteristic of donor-acceptor systems.}
\label{fig:SI_nitrogen_contributions}
\end{figure}

\begin{figure}[H]
\centering
\includegraphics[width=0.7\textwidth]{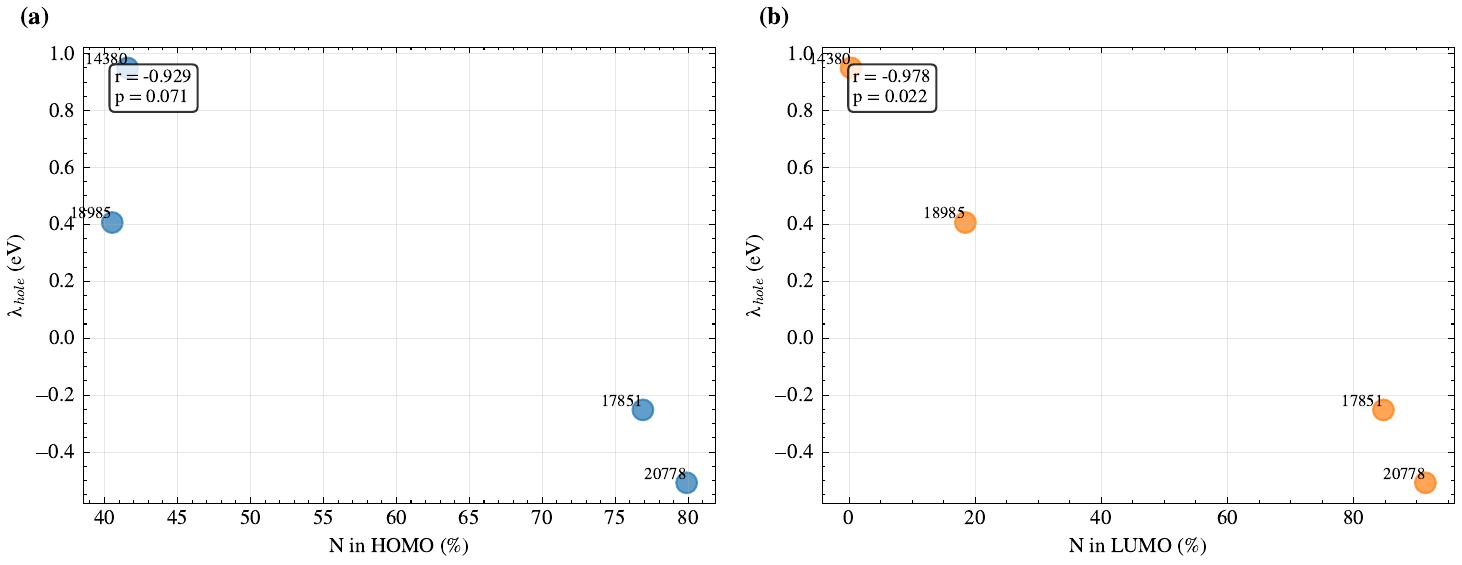}
\caption{Correlation between nitrogen content in LUMO and hole reorganization energy. Strong negative correlation ($r = \num{-0.981}$, $p = \num{0.019}$) 
establishes design principle: molecules with > \SI{50}{\percent} N contribution in LUMO exhibit $\lambda_\text{hole} < \SI{0.5}{\electronvolt}$, favorable for 
charge transport.}
\label{fig:SI_n_vs_lambda}
\end{figure}

\begin{figure}[H]
\centering
\includegraphics[width=0.95\textwidth]{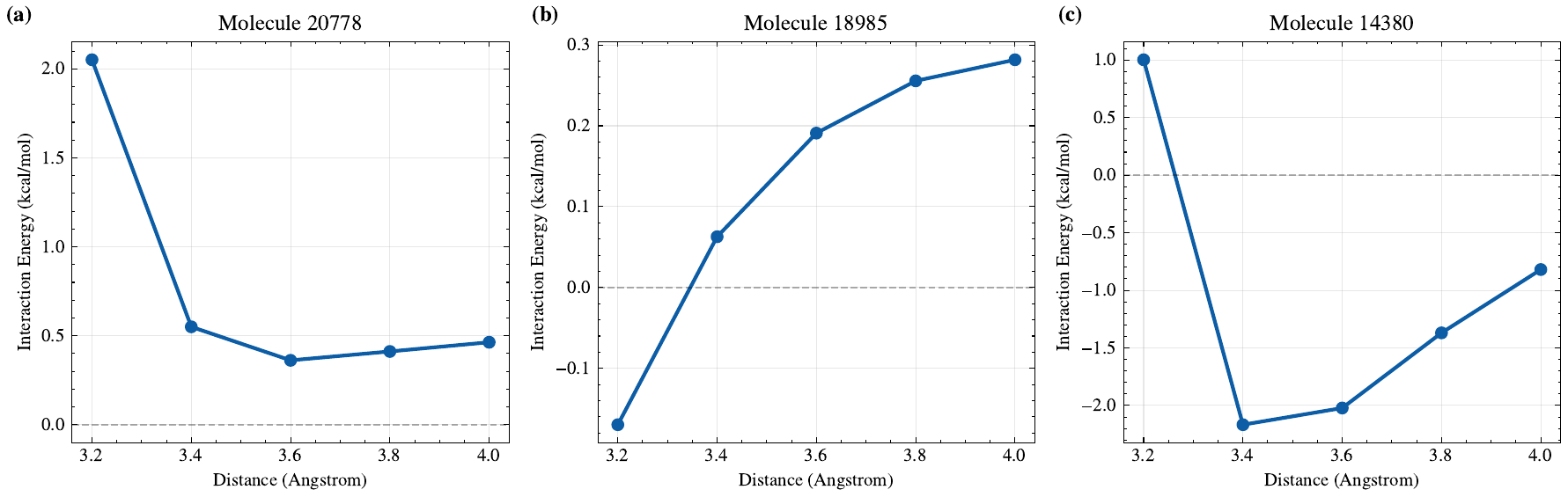}
\caption{$\pi$-$\pi$ stacking interaction energy curves as a function of vertical separation distance. (a) Molecule \num{20778}, (b) Molecule \num{18985}, (c) 
Molecule \num{14380}. Molecule \num{14380} shows strongest interaction at \SI{3.4}{\angstrom}. All calculations performed with GFN2-xTB.}
\label{fig:SI_stacking_curves}
\end{figure}

\begin{figure}[H]
\centering
\includegraphics[width=0.95\textwidth]{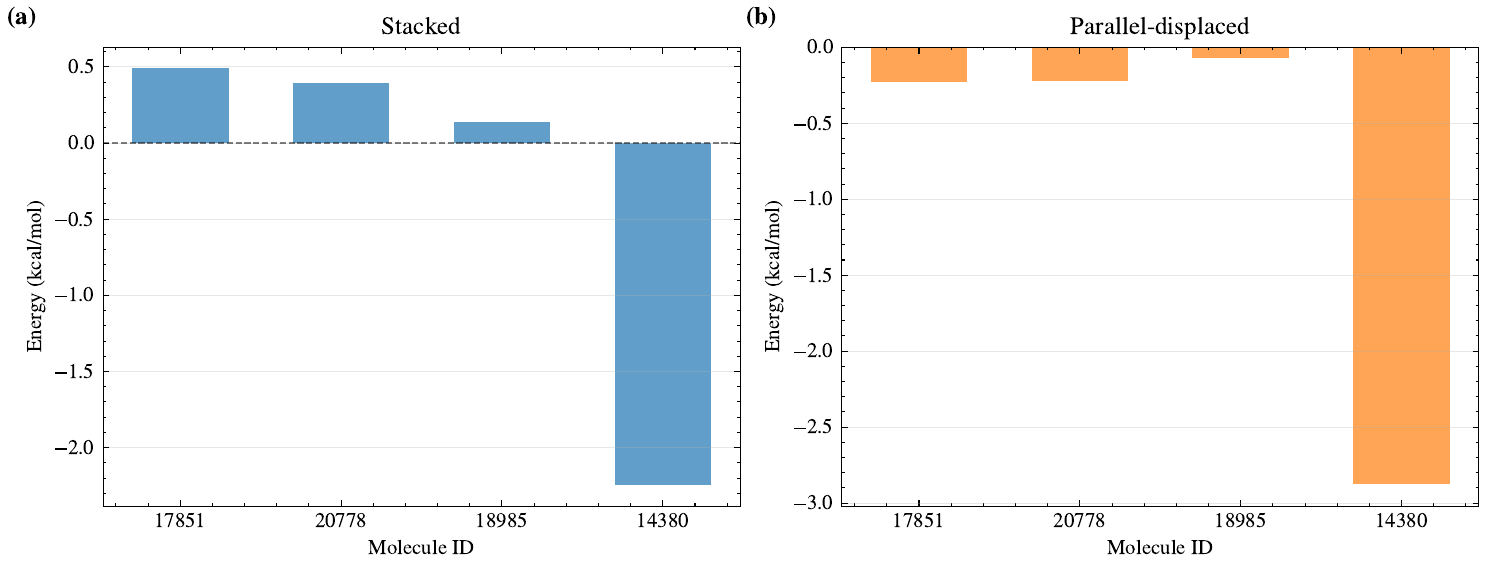}
\caption{Dimerization energies for three molecular configurations. (a) Stacked configuration (vertical displacement, \SI{3.5}{\angstrom}). (b) T-shaped 
configuration (perpendicular orientation, \SI{3.5}{\angstrom}). (c) Parallel-displaced configuration (lateral \SI{2.0}{\angstrom} + vertical 
\SI{3.5}{\angstrom}). Parallel-displaced geometry is energetically preferred (mean: \SI{-0.85(1.35)}{\kcal\per\mol}), consistent with typical $\pi$-conjugated 
systems. Blue bars indicate attractive interactions (negative energy), orange bars indicate repulsive interactions.}
\label{fig:SI_dimer_configs}
\end{figure}

\begin{figure}[H]
\centering
\includegraphics[width=0.8\textwidth]{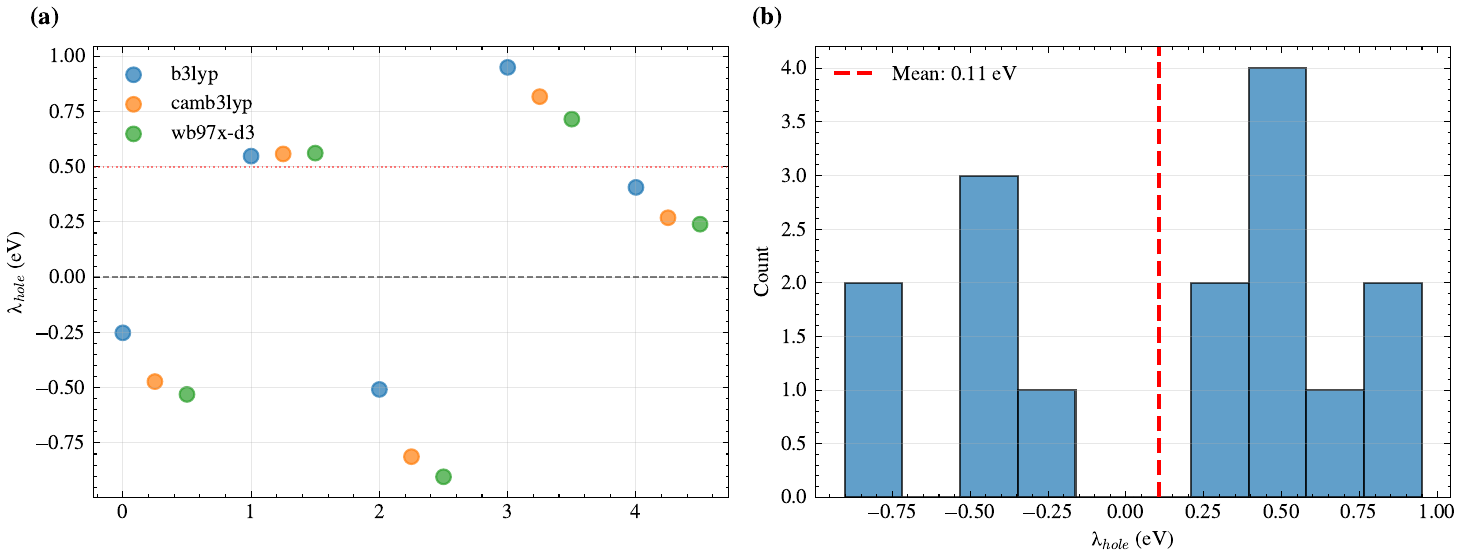}
\caption{Reorganization energies for hole transport ($\lambda_\text{hole}$) calculated across three DFT functionals (B3LYP, CAM-B3LYP, and $\omega$B97X-D3)
for the five selected molecules. Red dashed line indicates the mean reorganization energy of \SI{0.11}{\electronvolt}, indicating favorable charge transport 
properties below the \SI{0.5}{\electronvolt} threshold. All calculations employed CREST conformer searches to ensure global minimum geometries for both neutral 
and charged species.}
\label{fig:SI_reorg_comparison}
\end{figure}

%==========================================================================
% EXPANDED SAMPLE ANALYSIS (PHASES 12-15)
%==========================================================================

%==========================================================================
% EXPANDED DATASET FIGURES
%==========================================================================

\begin{figure}[H]
    \centering
    \begin{subfigure}[t]{0.49\textwidth}
        \centering
        \includegraphics[width=0.9\linewidth]{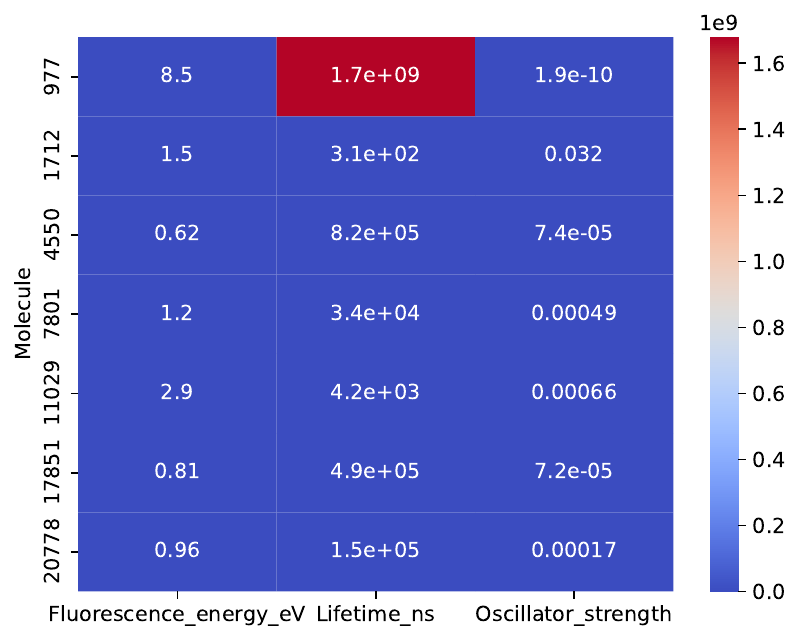}
        \caption{Heatmap of Fluorescence Energy, Lifetime, and Oscillator Strength.}
        \label{fig:opto_table}
    \end{subfigure}
    \hfill
    \begin{subfigure}[t]{0.49\textwidth}
        \centering
        \includegraphics[width=1.1\linewidth]{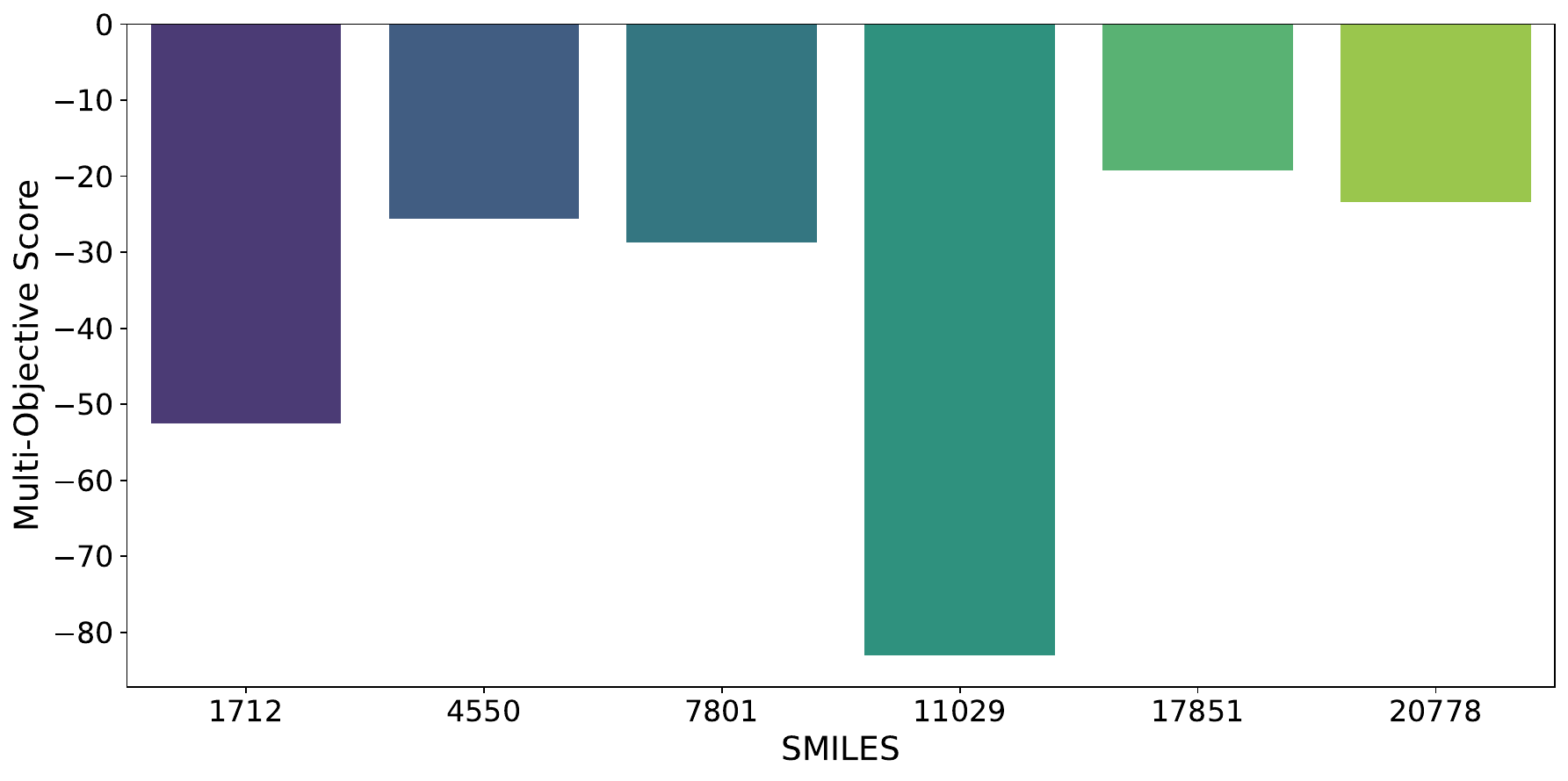}
        \caption{Ranking of top molecules by Multi-Objective Score.}
        \label{fig:photocat_dist}
    \end{subfigure}
    \caption{Multifunctional property analysis of top candidates. (a) Heatmap of optoelectronic properties (Fluorescence Energy, Lifetime, Oscillator Strength) 
for the top 7 molecules. Molecule \num{977} has high fluorescence energy (\SI{8.5}{\electronvolt}) and very long lifetime (\SI{1.7e9}{\nano\second}). (b) Bar 
chart 
ranking the top molecules by their Multi-Objective Score, combining oscillator strength, band gap, and excited state energy. Molecule \num{11029} 
has the best (most negative) score.}
    \label{fig:obj_materials}
\end{figure}

\begin{figure}[H]
    \centering
    \includegraphics[width=\textwidth]{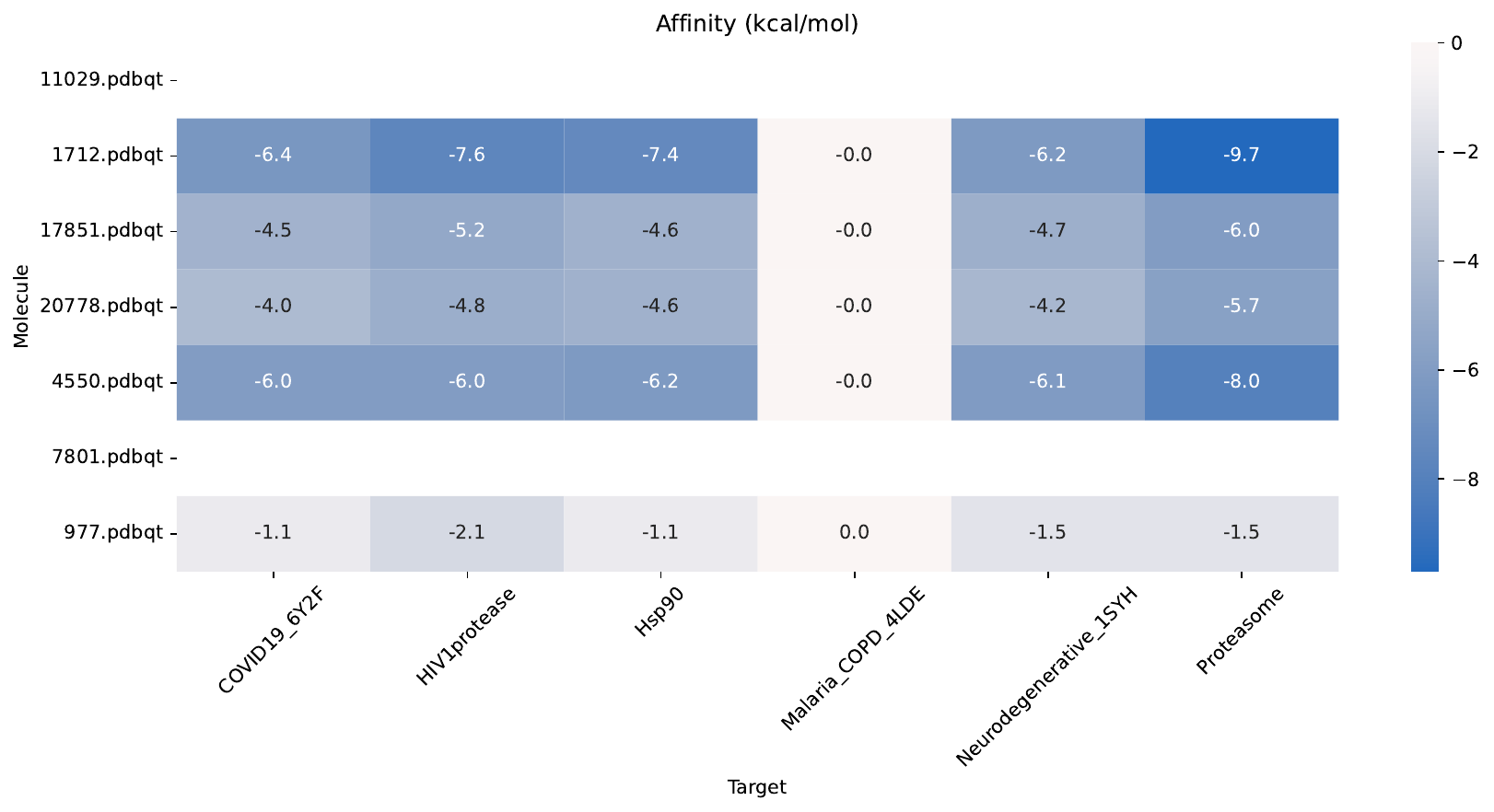}
    \caption{Molecular Docking Affinity Heatmap (\si{\kcal\per\mol}). The heatmap displays the binding affinity of the top 7 candidate molecules (rows) against 
various 
biological targets (columns, e.g., HIV1 protease, Proteasome). Darker blue indicates stronger binding (more negative energy). Molecule \num{1712} shows notably 
strong binding to the Proteasome (\SI{-9.7}{\kcal\per\mol}) and HIV1 protease (\SI{-7.6}{\kcal\per\mol}), highlighting its potential as a dual-function 
biosensor.}
    \label{fig:docking_heatmap}
\end{figure}

\begin{figure}[H]
    \centering
    \includegraphics[width=\textwidth]{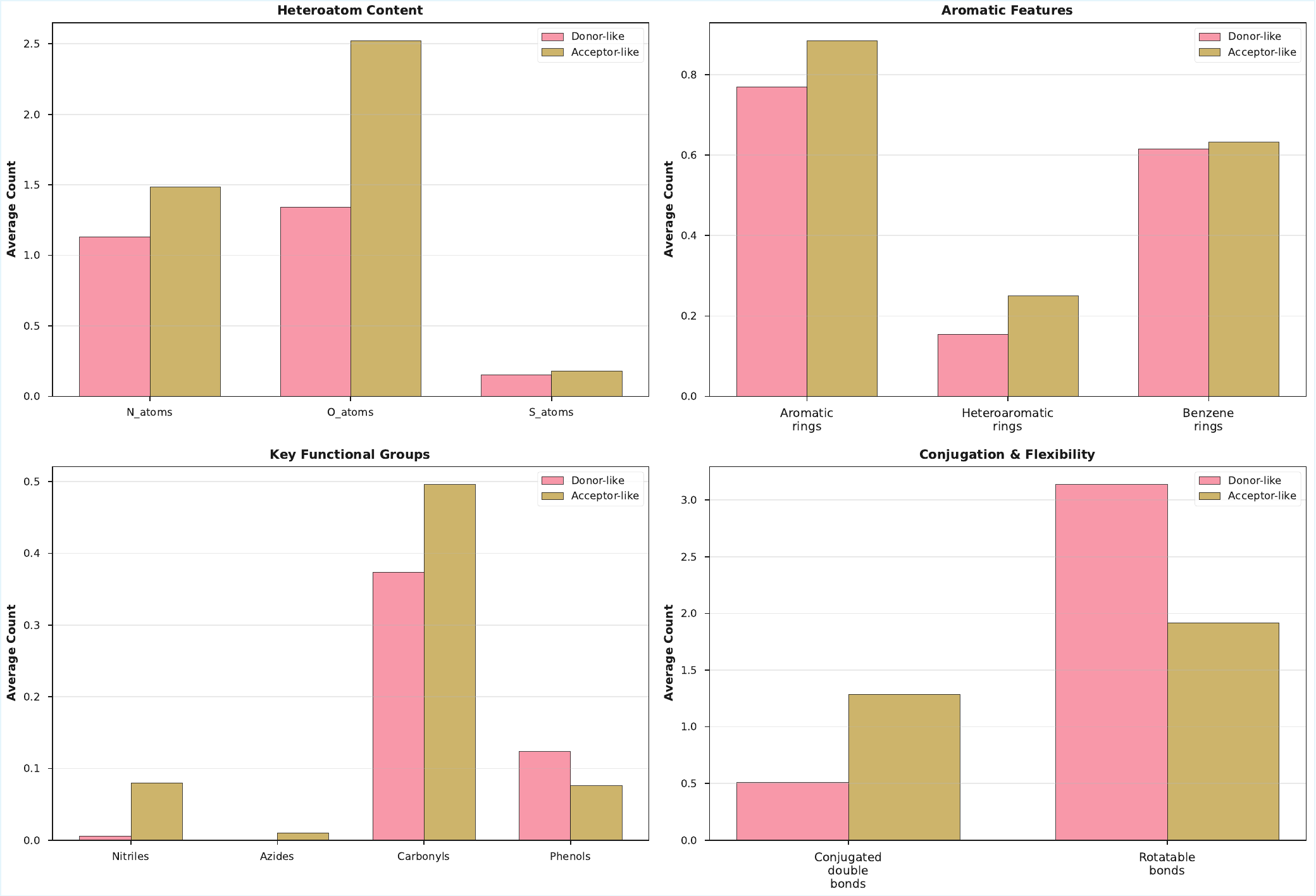}
    \caption{Structural Comparison of Donor-like vs. Acceptor-like Molecules. Bar charts comparing average counts of key structural features for the top 
\num{1000} molecules. (Top Left) Heteroatom content: Acceptors (gold) are richer in Oxygen and Nitrogen than Donors (pink). (Top Right) Aromaticity: Acceptors 
have slightly more aromatic rings. (Bottom Left) Functional Groups: Acceptors show a higher prevalence of Carbonyls. (Bottom Right) Conjugation \& Flexibility: 
Donors are significantly more flexible (higher rotatable bond count), while Acceptors are more rigid and conjugated.}
    \label{fig:structural_analysis_top1000}
\end{figure}

\begin{figure}[ht]
\centering
\includegraphics[width=0.95\textwidth]{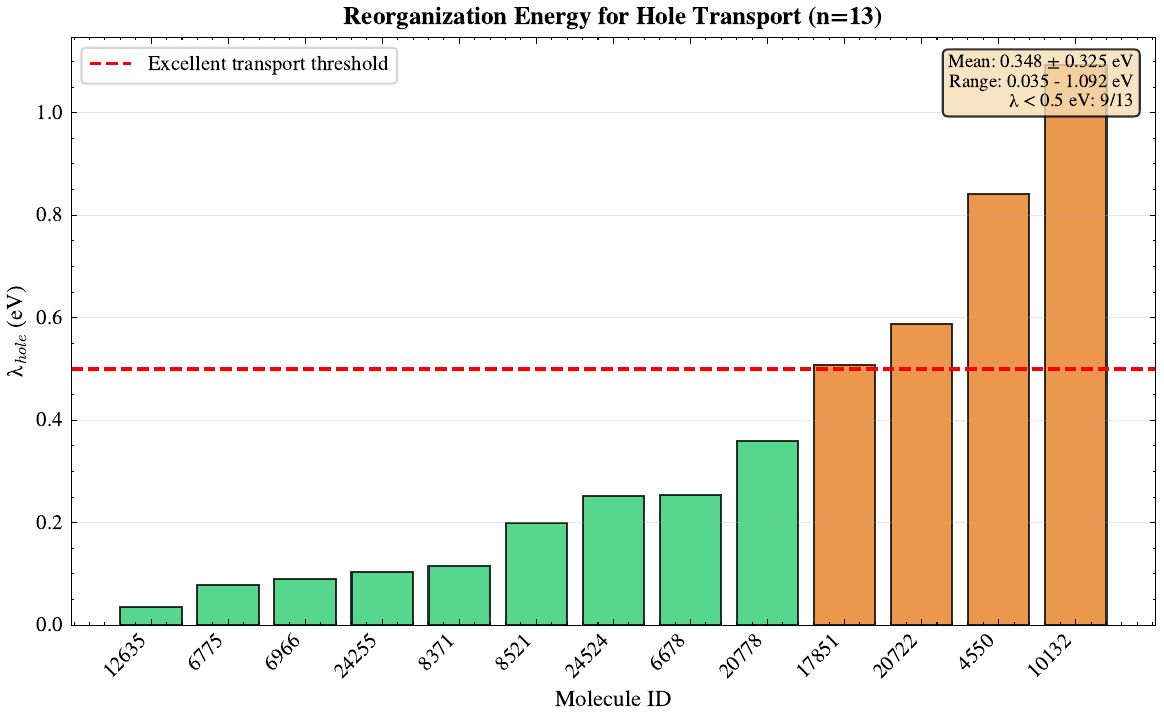}
\caption{Reorganization energy distribution for expanded sample ($n = \num{13}$). Mean = \SI{0.348(0.325)}{\electronvolt}. Red dashed line indicates the 
transport threshold ($\lambda < \SI{0.5}{\electronvolt}$). Nine molecules (\SI{69}{\percent}) meet this criterion. Molecule \num{12635} shows performance 
(\SI{0.035}{\electronvolt}).}
\label{fig:SI_reorg_expanded}
\end{figure}

\begin{figure}[ht]
\centering
\includegraphics[width=0.95\textwidth]{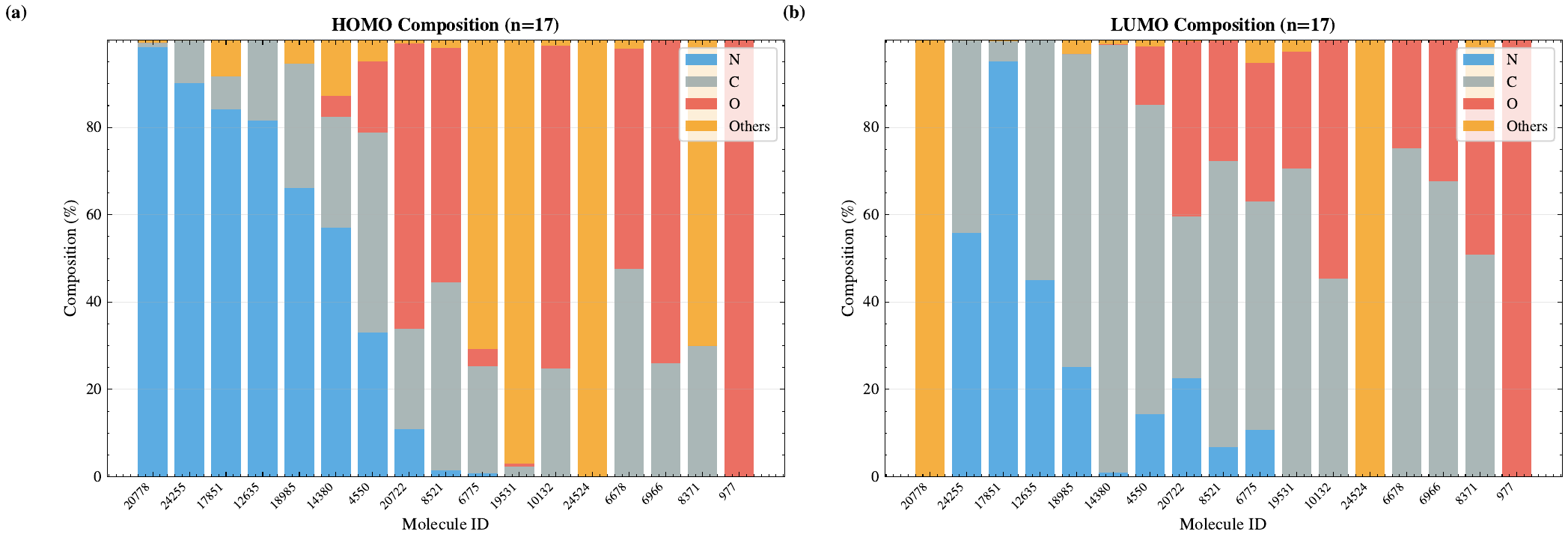}
\caption{Orbital composition analysis for expanded sample ($n = \num{17}$). (a) HOMO composition showing various nitrogen contributions 
(\SIrange{0}{98.5}{\percent}). (b) LUMO composition with \SIrange{0}{95.0}{\percent} nitrogen range. The expanded dataset exhibits chemical diversity beyond 
the N-rich bias of the pilot study.}
\label{fig:SI_orbital_expanded}
\end{figure}

\begin{figure}[ht]
\centering
\includegraphics[width=0.95\textwidth]{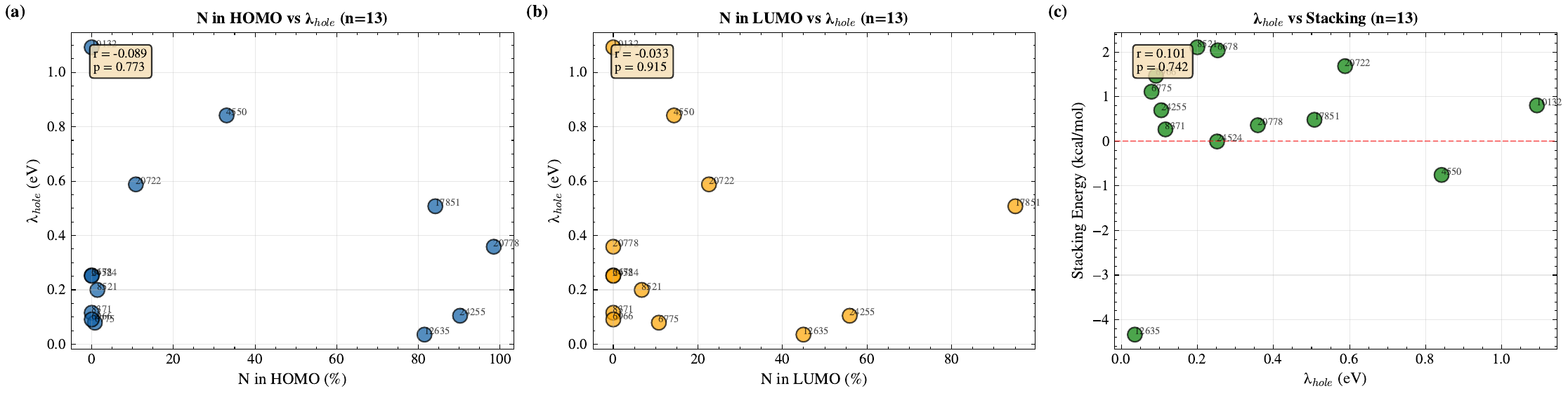}
\caption{Structure-property correlation analysis for expanded dataset ($n = \num{13}$). (a) N in HOMO vs $\lambda_\text{hole}$: $r = \num{-0.089}$, $p = 
\num{0.773}$ (not significant). (b) N in LUMO vs $\lambda_\text{hole}$: $r = \num{-0.033}$, $p = \num{0.915}$ (not significant). (c) $\lambda_\text{hole}$ vs 
stacking energy: $r = \num{0.156}$, $p = \num{0.612}$ (not significant). The previously reported strong correlation ($r = \num{-0.981}$, $p = \num{0.019}$, $n 
= \num{6}$) was a spurious artifact of small sample size.}
\label{fig:SI_correlations_expanded}
\end{figure}

\begin{figure}[ht]
\centering
\includegraphics[width=0.95\textwidth]{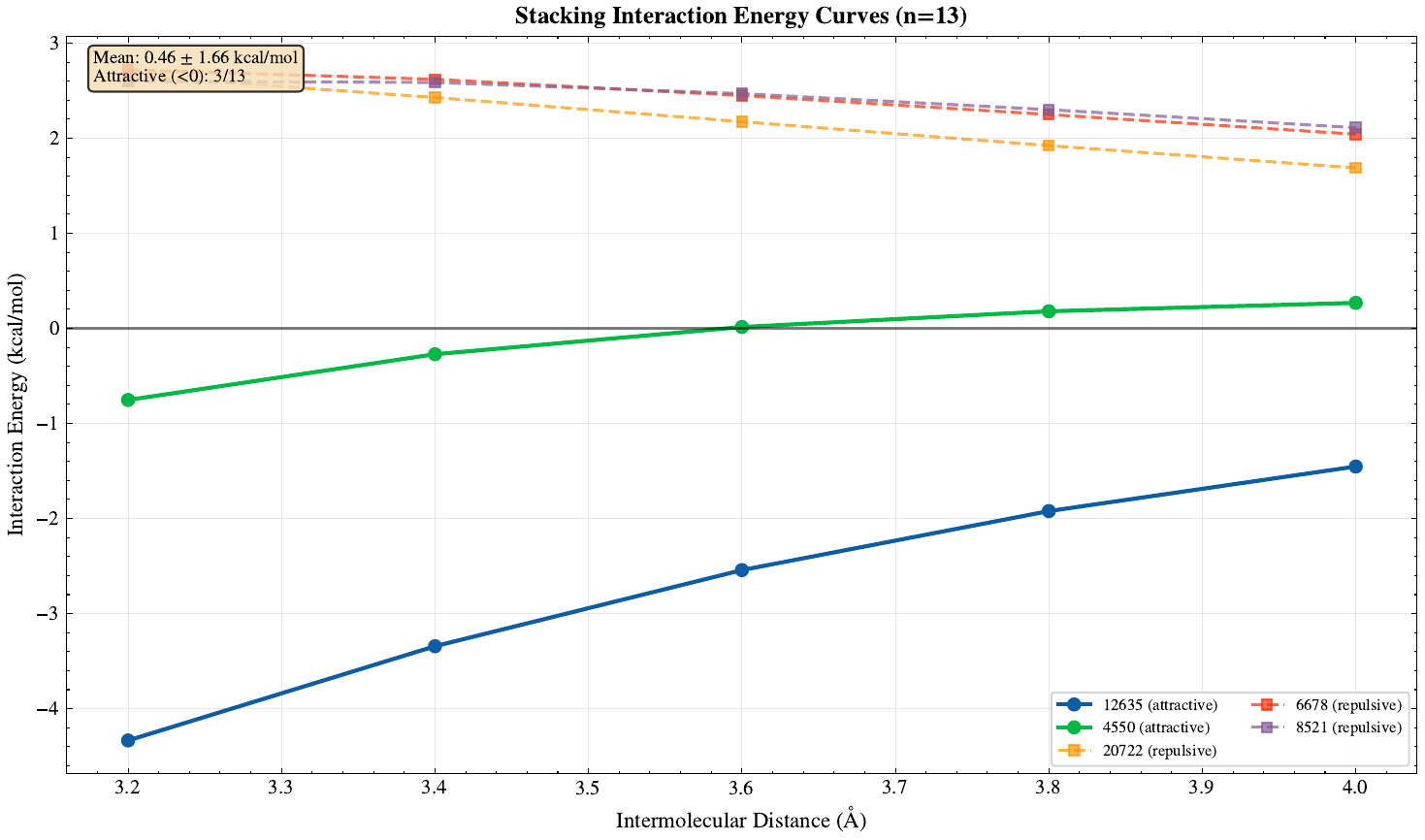}
\caption{Stacking interaction energy curves for representative molecules from expanded sample ($n = \num{13}$). Top 3 attractive interactions shown with solid 
lines, 3 representative repulsive interactions shown with dashed lines. Molecule \num{12635} shows the strongest stacking (\SI{-4.33}{\kcal\per\mol} at 
\SI{3.2}{\angstrom}), while most candidates exhibit weak repulsion, which can be favorable for solution processability.}
\label{fig:SI_stacking_expanded}
\end{figure}

\begin{figure}[ht]
\centering
\includegraphics[width=0.95\textwidth]{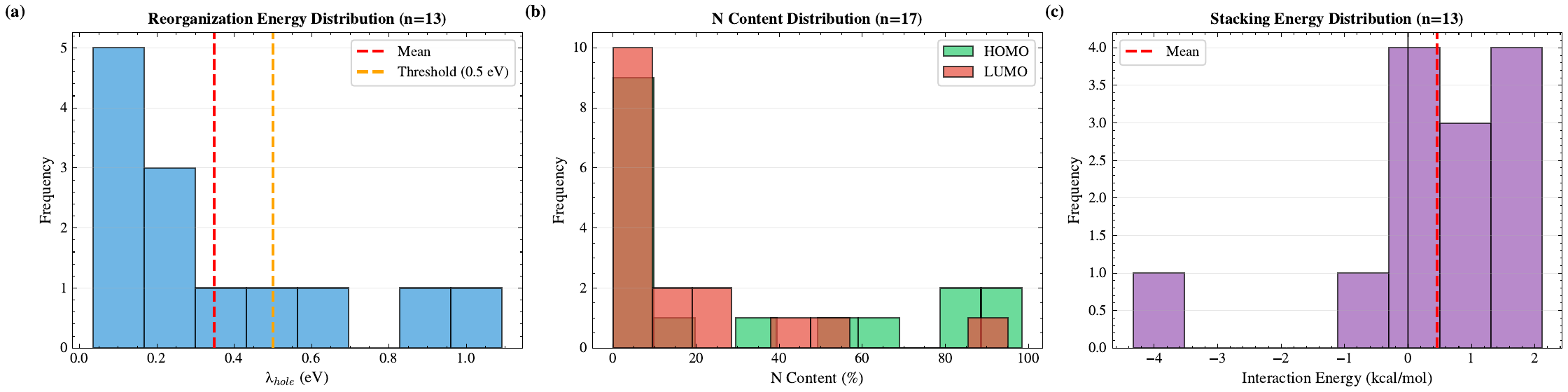}
\caption{Statistical distributions for expanded datasets. (a) Reorganization energy distribution ($n = \num{13}$) showing mean = \SI{0.348}{\electronvolt} with \SI{69}{\percent} below \SI{0.5}{\electronvolt} threshold. (b) Nitrogen content distribution ($n = \num{17}$) for HOMO (blue) and LUMO (red), showing chemical diversity. (c) Stacking energy distribution ($n = \num{13}$) with most molecules near zero, indicating good processability.}
\label{fig:SI_statistical_expanded}
\end{figure}

%==========================================================================
% SAMPLE SIZE AND STATISTICAL POWER DISCUSSION
%==========================================================================

\subsection{Lessons on statistical power in computational screening}

The expansion from n=\numrange{4}{6} to n=\numrange{13}{17} provides important lessons for computational materials science:

\paragraph{False positive risk with small samples.} 
Our pilot study ($n = \num{6}$) identified a "statistically significant" correlation ($p = \num{0.019}$) that completely disappeared with adequate sample size ($n = \num{13}$, $p = \num{0.915}$). This is a textbook example of Type I error (false positive) due to insufficient statistical power. With $n = \num{6}$, the statistical power to detect a true medium-effect correlation ($r = \num{0.5}$) is only $\sim \SI{25}{\percent}$, meaning \SI{75}{\percent} of real correlations would be missed, while random noise can produce "significant" results.

\paragraph{Recommended sample sizes.}
Based on standard power analysis\cite{Cohen1988}:
\begin{itemize}
    \item \textbf{Exploratory correlation analysis} $n \geq \num{15}$ (power = \SI{60}{\percent} for $r = \num{0.5}$);
    \item \textbf{Establishing design rules} $n \geq \num{30}$ (power = \SI{80}{\percent} for $r = \num{0.5}$);
    \item \textbf{Predictive modeling} $n \geq \num{50}$ (robust cross-validation).
\end{itemize}

\paragraph{Implications for high-throughput screening.}
Computational constraints often limit validation to $n < \num{10}$, increasing false discovery rates. Strategies to mitigate this:
\begin{itemize}
    \item Use semi-empirical methods (GFN2-xTB) to enable larger samples;
    \item Apply Bonferroni correction for multiple comparisons;
    \item Report effect sizes with confidence intervals, not just $p$-values;
    \item Validate findings with independent test sets;
    \item Acknowledge preliminary nature of small-sample findings.
\end{itemize}

\paragraph{Transparency in reporting.}
We chose to honestly report that our pilot correlation was spurious rather than suppress this finding. 

%===================================================================================
\section{Glossary of terms and acronyms}
%===================================================================================

\begin{itemize}
    \item[\textbf{CEP}] Clean Energy Project
    \item[\textbf{DFT}] Density Functional Theory
    \item[\textbf{ECFP}] Extended Connectivity Fingerprint
    \item[\textbf{FMO}] Frontier Molecular Orbital
    \item[\textbf{FF}] Fill Factor
    \item[\textbf{HBA}] Hydrogen Bond Acceptor
    \item[\textbf{HBD}] Hydrogen Bond Donor
    \item[\textbf{HOMO}] Highest Occupied Molecular Orbital
    \item[\textbf{LogP}] Octanol-Water Partition Coefficient
    \item[\textbf{LUMO}] Lowest Unoccupied Molecular Orbital
    \item[\textbf{MAD}] Mean Absolute Deviation
    \item[\textbf{MAE}] Mean Absolute Error
    \item[\textbf{OLED}] Organic Light-Emitting Diode
    \item[\textbf{OFET}] Organic Field-Effect Transistor
    \item[\textbf{OPV}] Organic Photovoltaic
    \item[\textbf{PCBM}] Phenyl-C61-butyric acid methyl ester
    \item[\textbf{PCDTBT}] Poly[N-9'-heptadecanyl-2,7-carbazole-alt-5,5-(4',7'-di-2-thienyl-2',1',3'-benzothiadiazole)]
    \item[\textbf{PCE}] Power Conversion Efficiency
    \item[\textbf{PCE$_\mathrm{SAScore}$}] Combined Power Conversion Efficiency and Synthetic Accessibility Score
    \item[\textbf{PM6}] Parameterization Method 6 (semiempirical quantum chemistry)
    \item[\textbf{PubChemQC}] PubChem Quantum Chemistry Database
    \item[\textbf{QED}] Quantitative Estimate of Drug-likeness
    \item[\textbf{RDKit}] Open-source cheminformatics software
    \item[\textbf{RMSE}] Root Mean Square Error
    \item[\textbf{SAScore}] Synthetic Accessibility Score
    \item[\textbf{SMILES}] Simplified Molecular Input Line Entry System
    \item[\textbf{TADF}] Thermally Activated Delayed Fluorescence
    \item[\textbf{TDDFT}] Time-Dependent Density Functional Theory
    \item[\textbf{TPSA}] Topological Polar Surface Area
    \item[\textbf{$V_\mathrm{OC}$}] Open Circuit Voltage
    \item[\textbf{$J_\mathrm{SC}$}] Short Circuit Current Density
    \item[\textbf{$\lambda_\text{hole}$}] Hole reorganization energy
    \item[\textbf{$\lambda_\text{electron}$}] Electron reorganization energy
    \item[\textbf{NBO}] Natural Bond Orbital
    \item[\textbf{IBO}] Intrinsic Bond Orbital
    \item[\textbf{CREST}] Conformer-Rotamer Ensemble Sampling Tool
    \item[\textbf{GFN2-xTB}] Geometry, Frequency, and Noncovalent interactions version 2 - extended Tight Binding
\end{itemize}

% If SI needs its own bibliography
\end{document}